\documentclass[12pt]{article}

\usepackage{soul}

\usepackage{tikz,pgfplots}
\pgfplotsset{compat=newest}
\usetikzlibrary{decorations.pathmorphing}
\usepackage{epsfig,latexsym,amsfonts,amsmath,amsthm,amssymb,amsbsy,multirow,slashed,wasysym,textcomp,wrapfig,datetime,comment,mathtools,cancel,cite,mathrsfs,tensor}
\usepackage{graphicx}
\usepackage{booktabs}
\usepackage{subcaption}
\definecolor{darkblue}{rgb}{0,0,0.7}
\definecolor{darkred}{rgb}{0.7,0,0}
\usepackage[unicode, colorlinks, citecolor=darkblue, linkcolor=darkred, urlcolor=blue]{hyperref}
\usepackage[font={footnotesize}]{caption}

\def\ee{\end{equation}}
\def\be{\begin{equation}}
\def\bea{\begin{eqnarray}}
\def\eea{\end{eqnarray}}
\newcommand{\beq}{\begin{eqnarray}}
\newcommand{\eqq}{\end{eqnarray}}
 \newcommand{\badat}{\begin{alignedat}}
 \newcommand{\eadat}{\end{alignedat}}

\newcommand{\eal}[1]{\be \begin{aligned} #1 \end{aligned}\end{equation}} 
\newcommand{\eqn}[1]{\be #1 \end{equation}} 
\newcommand{\eqa}[1]{\bea  #1\end{eqnarray}}

\long\def\new#1\endnew{{\bf #1}}		
\long\def\del#1\enddel{}

\def\del{\partial}

\usepackage{color}

\newcommand{\pink}[1]{\textcolor{\pink}{#1}}

\definecolor{dblue}{rgb}{0.2,0.50,0.80}

\newcommand{\ab}[1]{\left|#1\right|}

\newcommand{\br}[1]{\left[#1\right]}

\newcommand{\pa}[1]{\left(#1\right)}

\newcommand{\dt}{\mathop{}\!\delta}

\newcommand{\pd}{\mathop{}\!\partial}

\newcommand{\ts}{\textsuperscript}

\def\l({\left(}
\def\r){\right)}

\numberwithin{equation}{section} 

\begin{document}
\begin{titlepage}

  \thispagestyle{empty}
  \begin{flushright}
  
    \end{flushright}
  \bigskip
  \begin{center}
	 \vskip2cm
  \baselineskip=13pt {\LARGE \scshape{
  \vspace{0.5em}pp-Waves and the Hidden Symmetries \\ \vspace{.1 in}of Black Hole Quasinormal Modes}}

	 \vskip2cm
   \centerline{Daniel Kapec and Ahmed Sheta}
    
 \vskip.5cm
 \noindent{\em Center for the Fundamental Laws of Nature,}
  \vskip.1cm
\noindent{\em  Harvard University,}
{\em Cambridge, MA 02138, USA}
\bigskip
  \vskip1cm
  \end{center}
  \begin{abstract}
There are two interesting classes of trapped null geodesics in any black hole spacetime: those that lie  on the photon ring and those that generate the horizon. Recent work introduced a ``near-ring'' scaling limit  that exhibits the emergent symmetries of the eikonal quasinormal mode (QNM) spectrum associated to the photon ring. This analysis was reformulated geometrically by Fransen using the Penrose limit, which produces pp-waves from geodesics.  We elaborate on and generalize various aspects of this construction for the Schwarzschild black hole.  We also discuss the Penrose limit onto the horizon generators. This second limit, although technically simpler, also displays emergent near-horizon symmetries that explain the equally-spaced overtones of the highly-damped QNM spectrum. In both examples, symmetry considerations distinguish the QNM solutions from the scattering states and produce overtones as  descendants.
\end{abstract}

\end{titlepage}
\tableofcontents
\newpage

\section{Introduction}

Most of what is currently understood about the quantum mechanics of black holes and black branes can be traced back in one way or another to  the low-energy, near-horizon scaling limit for near-extremal black holes. This limit has led to progress because it makes precise the otherwise ambiguous (and non-covariant) notion of the stretched horizon. Unfortunately, this demarcation between the ``black hole part'' and the ``far part'' of the spacetime, marked in the extremal case by the conformal boundary of the AdS throat (appropriately regulated), is not available for generic black holes far from extremality. Finding the replacement or generalization of this concept is key to understanding holography for asymptotically flat black holes, and the failure to do so has hindered progress away from extremality.   

The near-extremal scaling limit actually involves three simultaneous limits, each of which is responsible for a different effect. For a Kerr black hole of mass $M$ and angular momentum $J=aM$, the limit corresponds to taking
\begin{align}\label{eq:nearNHEK}
    \text{Near-NHEK Limit:}\qquad
    \begin{cases}
         \displaystyle M-a\ll {M}&\qquad\text{near-extremal},\\
        \ab{r-M}\ll M&\qquad\text{near-horizon},\\
        \ab{\omega_R-m\Omega_H}\ll M^{-1}&\qquad\text{near-superradiant}.
    \end{cases}
\end{align}
The third limit has a different character than the first two. The first limit involves a sequence of spacetimes and ensures that the region outside the black hole develops a large warped throat. The second limit is geometric, zooming in on a particular region of \textbf{configuration} space created by the first limit. The third limit is performed in \textbf{momentum} space for fields propagating in the Kerr background. It is a constraint on the way we probe the black hole. So the full limit really involves a combined limit in phase space.

Taken together, the first two limits result in a new, simpler, more symmetric spacetime, usually called NHEK (near-horizon extreme Kerr). Because the second limit actually involves a singular scaling of Boyer-Lindquist coordinates $(t,r,\phi)$ relative to NHEK coordinates $(T,R,\Phi)$
\begin{equation}
    t=\frac{T}{\varepsilon \Omega_H} \; , \qquad r=r_+(1+\varepsilon R)\; , \qquad \phi=\Phi + T \;, \qquad \varepsilon\to 0 \; ,
\end{equation}
not every perturbation of Kerr has a non-singular limit into NHEK. Indeed, since for finite $\varepsilon$
\begin{equation}\label{eq:singularLimitNHEK}
    \Omega_H\partial_T = \frac{1}{\varepsilon}(\partial_t + \Omega_H\partial_\phi) \; , 
\end{equation}
it is clear that waves with finite energy with respect to NHEK time $\partial_T$ must have $\omega_R-m\Omega_H\sim \varepsilon$ in Kerr. For waves of these frequencies the absorption cross section is small, so that they do not cause significant disturbances in the throat when they impinge on the black hole.\footnote{Above these frequencies, the absorption cross section is positive, and below this bound the wave undergoes superradiant amplification with negative cross section. $\omega_R=m\Omega_H$ corresponds to the zero of the cross section.} The third limit therefore selects perturbations which have a non-singular limit into the NHEK region.

The emergent conformal symmetries in the NHEK throat mean that the solutions to the wave equation in extremal Kerr which are non-singular in the limit \eqref{eq:nearNHEK} are related to each other by $SL(2,\mathbb{R})$ transformations in the near-horizon region. Since these solutions in the near-horizon region extend to global solutions in the full Kerr geometry, there is a sense in which a subset of solutions to the full Teukolsky equation are constrained by the enhanced symmetries present in the limit \eqref{eq:nearNHEK}. In other words, the solutions fit into $SL(2,\mathbb{R})$ families even if the wavefunctions are not related by geometric symmetry transformations far from the black hole.

Although the limit \eqref{eq:nearNHEK} is extremely powerful, it only encodes information about the low temperature behavior of the quantum system dual to the Kerr black hole. As the temperature goes up the throat shrinks and the separation between the black hole region and the far region becomes more ambiguous. Since the black hole is just curvature and the whole space is curved, it is difficult to say where the black hole ``is'' covariantly and quantum mechanically. For instance, absorption cross sections depend on the geometry several Schwarzschild radii from the horizon.\footnote{Greybody factors depend on the peak of the radial potential which occurs at the photon ring at $r\sim 3M$.}

Recently, a different scaling limit \cite{Hadar:2022xag,Kapec:2022dvc} was introduced for non-extremal black holes which also encodes some universal features of the black hole system.  The limit bears some similarities to (and was inspired by) the limit \eqref{eq:nearNHEK}, but differs in important ways. In particular, while the limit \eqref{eq:nearNHEK} essentially captures universal low-energy (large red-shift) physics, the ``near-ring" limit studied in \cite{Hadar:2022xag,Kapec:2022dvc} is designed to isolate the high-energy universality for generic black holes. It involves a single limit in configuration space and two limits in the momentum space for fields in the black hole background. For Schwarzschild it takes the simple form
\begin{align}\label{eq:NearRing}
    \text{Near-Ring Limit:}\qquad
    \begin{cases}
        \ab{r-3M}\ll M&\qquad\text{near-peak},\\
        \displaystyle\ab{\frac{\ell}{\omega_R}-3\sqrt 3M}\ll M&\qquad\text{near-critical},\vspace{3pt}\\
        \displaystyle\frac{1}{\omega_R}\ll {M}&\qquad\text{high-frequency}.
    \end{cases}
\end{align}
The near-extremal constraint on the geometry is replaced by the near-critical constraint on the wave, and the low energy (near-superradiant) condition is replaced by a high frequency limit.  
The conditions \eqref{eq:NearRing} essentially ensure the validity of the geometric optics approximation, with the near-critical condition ensuring that the waves satisfy quasinormal boundary conditions.

In familiar physical systems, symmetry enhancement often  occurs both at low energies and at high energies. In non-gravitational systems in flat spacetime, the renormalization group flow between conformal fixed points captures this symmetry enhancement, but the symmetry structure for dynamics in black hole backgrounds is slightly different. The low-energy near horizon limit for near-extremal black holes \eqref{eq:nearNHEK} has long been known to display enhanced conformal symmetry, but in \cite{Hadar:2022xag,Kapec:2022dvc}
it was pointed out that the harmonic oscillator algebra emerges naturally in the high-energy near-ring limit \eqref{eq:NearRing} and controls the overtone spectrum of eikonal QNMs. 
As we will describe, not every solution to the wave equation in Schwarzschild  has a nonsingular near-ring limit, but the ones that do are organized into representations of the enhanced symmetry algebra of the near-ring region. These families extend into the far region even though they are not related by geometric symmetry transformations there. 

Shortly after the limit \eqref{eq:NearRing}  was introduced, Fransen \cite{Fransen:2023eqj} pointed out that the near-ring scaling limit can be geometrized using the Penrose limit \cite{Penrose1976}. This limit previously played a crucial role in establishing stringy tests of the AdS/CFT correspondence \cite{Berenstein:2002jq}, and one hopes that it will similarly shed light on the quantum Kerr black hole. For instance, while we cannot solve string theory in the Kerr background, string theory in the Penrose limit corresponding to \eqref{eq:NearRing} is exactly solvable, as is the wave equation. Much of the first half of this paper is a technical elaboration and generalization of the work \cite{Fransen:2023eqj}, and we believe there are still many ideas to be explored in this area.

For many purposes, the Penrose limit can be thought of as a covariant geometrization of the  geometric optics approximation. Given a null geodesic congruence, the Penrose limit scales into the vicinity of the central geodesic, retaining only properties of the geodesic deviation.  The resulting geometry is a plane  wave with parallel propagation (pp-wave for short), and the hidden symmetries identified in \cite{Hadar:2022xag} become isometries of the limiting spacetime.
As we will see, the solutions to the black hole wave equation with smooth limits into the pp-wave  are high frequency QNMs, so the limit \eqref{eq:NearRing} really captures universal information about the  high-frequency Ruelle spectrum of the black hole independent of its temperature. 

There is yet another family of well-studied QNMs whose spectrum is understood semi-analytically and which seem to organize into symmetric families. These are the highly-damped QNMs, which for Schwarzschild take the form
\begin{equation}\label{eq:IntroDamped}
    \omega = \frac{\kappa}{2\pi}\log 3 -  i \kappa \left(n+\frac12\right) \; , 
\end{equation}
where $\kappa$ is the surface gravity of the horizon and $n\gg1$. The presence of the  Matsubara frequency appearing in this formula suggests that these modes should be related to the near-horizon Rindler-like region of the black hole, and the equal spacing suggests a potential emergent symmetry in the limit of large damping. In other words, when the real part of $\omega_{\text{QNM}}$ becomes large, the equal spacing of the spectrum is realized by the emergent inverted harmonic oscillator (IHO) algebra, which is an exact symmetry in the Penrose limit of the photon ring. Is a similar mechanism at play when the imaginary part of $\omega_{\text{QNM}}$ becomes large? 

It turns out that the equal spacing in \eqref{eq:IntroDamped} can be understood in terms of symmetries, but the relevant algebra is not that of the IHO. Instead of a near-ring limit, one studies the wave equation locally in the near-horizon region, where it reduces to the massive Rindler equation. The 2d Poincar\'e algebra emerges naturally as a symmetry of the near-horizon wave equation and generates the equally-spaced overtones of \eqref{eq:IntroDamped}. In analogy with the photon ring, these emergent symmetries can be geometrized by performing a scaling limit onto the horizon. This can in turn be viewed as the Penrose limit onto the past horizon generators. The resulting pp-wave is of course flat (Rindler space), and the spectrum-generating Poincar\'e symmetries  become isometries.
We summarize the similarities between the two symmetry structures in the table below.

\begin{table}[h!]
\centering
\begin{tabular}{c|c|c} 
  QNM branch: & Eikonal & Highly Damped \\ \midrule
 Emergent Symmetry: & Inverted Harmonic Oscillator & 2d Poincar\'e Algebra \\
 & $b_-,b_+,H$ & $\partial_U, \partial_V, U\partial_U-V\partial_V$ \\ \midrule
 Associated Geodesic: & Photon Ring     &  Generator of the Past Horizon \\ \midrule
Penrose Limit: & Vacuum pp-Wave & Flat pp-Wave (Rindler) \\ \midrule
Fundamental Mode & $b_-\Psi_0=0 \;,$ & $(U\partial_U-V\partial_V) \Psi_0$ \\ 
(pp-Wave Ground State) & $H\Psi_0=-\frac{i}{2} \gamma_L\Psi_0$ & \quad$=\left(\frac12+ \frac{i}{2\pi}\log 3 \right) \Psi_0$ \\ \midrule
Overtone Descendants & $\Psi_n=(b_+)^n\Psi_0 $ & $\Psi_n=(\partial_V)^n \Psi_0 $ \\ 
\end{tabular} 
\end{table}

The 2d Poincar\'e algebra is distinct from the IHO algebra of the near-ring limit, but it shares several important properties. In general, an equally-spaced overtone structure in the spectrum arises from a symmetry algebra with the ladder structure
\begin{equation}
    [W, R_+] = R_+ \;, \qquad [W, R_-] = -R_- \;.
\end{equation}
Here $W$ measures the frequency, while $R_{\pm}$  raise and lower the frequency by one unit. In the context of black hole physics, this algebra closes in  3 different ways depending on the situation
\begin{equation}
    [R_-,R_+] = 0 \;, \qquad [R_-, R_+] \propto 1 \;, \qquad [R_-, R_+] = 2 W \; .
\end{equation}
The first option is the 2d Poincar\'e algebra, which is realized in Rindler-like systems (such as the horizons of non-extremal black holes). It is a symmetry of the highly damped Schwarzschild QNM spectrum. The second option, in which the algebra closes on a central element, gives the SHO/IHO algebra.  This symmetry is realized in the near-ring limit for black hole photon rings. The third possibility is the $SL(2)$ algebra,\footnote{There have been previous attempts to interpret the modes \eqref{eq:IntroDamped} geometrically \cite{Moretti:2002mp,Solodukhin:2003mp,Padmanabhan:2003fx,Medved:2003rga,Medved:2003pr,Kim:2012ax,Kim:2012mh,Raffaelli:2013ih,Ponglertsakul:2018rot,raffaelli2022}, many of which attempted to construct an $SL(2,\mathbb{R})$ symmetry near the horizon of non-extremal black holes. These constructions differ from the spectrum-generating Poincar\'e symmetry that we identify.} which is known to emerge near the horizon of an extremal black hole. It seems interesting that these 3 possibilities are all realized in the vicinity of bound null geodesics in a black hole geometry.

Throughout the paper, we restrict attention to Penrose limits in the Schwarzschild black hole. The equatorial photon ring trajectories in Kerr (for which the Schwarzschild discussion carries over) were treated by \cite{Fransen:2023eqj}, but the generic complicated polar motion in the Kerr photon ring does pose a technical hurdle. We take concrete steps towards addressing it in section \ref{sec:Tilt} while also noting some intrinsic drawbacks of the pp-wave approach.  We expect the Schwarzschild discussion to carry over to non-equatorial Kerr trajectories modulo some subtleties. On the other hand, the discussion of the highly-damped QNMs is expected to change qualitatively for Kerr and Reissner-Nordstrom, as we discuss in section \ref{sec:future}.

We should also note that the discussion in this paper does not generalize straightforwardly to black holes in AdS. The AdS radial potential is confining at spatial infinity. Since the asymptotics of the radial potential are completely different, the long-lived QNM spectrum also exhibits different properties (for example, the QNMs controlled by the photon sphere in AdS-Schwarzschild occur at large \textit{imaginary} angular momentum \cite{Dodelson:2023nnr}).  Indeed, the emergent symmetries that we identify are sensitive to the asymptotic structure of the radial potential, so they will play a different role in AdS. 

The outline of this paper is as follows. \textbf{Section \ref{sec:first}} reviews the derivation of the eikonal QNMs within the WKB and geometric optics approximations. We discuss the near-ring limit, its hidden symmetries and associated conservation laws, and their relation to the equally-spaced overtones of the eikonal spectrum. We conclude with the scaling limit into the pp-wave geometry of the equatorial Schwarzschild photon ring. In \textbf{section \ref{sec:plane wave}} we analyze the wave equation in the equatorial pp-wave geometry. We describe  how solutions in the pp-wave lift to a subset of the eikonal QNM wavefunctions in Schwarzschild, and we discuss how the hidden near-ring symmetries become isometries that generate overtones. \textbf{Section \ref{sec:Tilt}} analyzes the Penrose limit of a tilted (i.e. non-equatorial) photon-ring. 
We use this example to compare general features of how the Penrose limit geometrizes the WKB approximation with first-order vs. second-order turning points. We also address the complications introduced by nontrivial polar motion and reproduce more general eikonal QNM wavefunctions from the tilted pp-wave. \textbf{Section \ref{sec:highDamp}} treats the highly-damped QNMs of the Schwarzschild black hole. We consider the near-horizon limit of the wave equation and identify the emergent symmetries that control the equally-spaced overtone structure. These considerations lead us to consider a generalized geometric optics approximation based on the geodesics that generate the past horizon. The construction shares many similarities with the near-ring limit. \textbf{Section \ref{sec:future}} concludes with a list of open questions. Appendix \ref{App:SCHpenrose} constructs the two Penrose limits used in section \ref{sec:Tilt}. Appendix \ref{App:monodromy} reviews the monodromy argument that determines the highly-damped QNMs and clarifies why it is most naturally interpreted in the vicinity of the past horizon. Appendix \ref{App:rindler} discusses various aspects of the massless and massive Rindler wave equation relevant to the discussion in section \ref{sec:highDamp}.

\section{Eikonal Quasinormal Modes and the Near-Ring Region}\label{sec:first}

The eikonal QNM spectrum of Schwarzschild is well known and takes the form
\begin{equation} \label{eq:eikonal}
    \omega = \l( l + \frac{1}{2} \r) \Omega_R - i \l( n + \frac{1}{2} \r) \gamma_L \;, \qquad \mathrm{as } \quad l \to \infty \;,
\end{equation}
where $\Omega_R = \gamma_L = \l(3 \sqrt{3} M\r)^{-1}$ are the orbital frequency and Lyapunov exponent of the bound null geodesic at $r=3M$. 
In \cite{Hadar:2022xag} it was noted that the overtone structure of this spectrum can be reproduced using symmetries by zooming into the ``near-ring region'' in the phase space of solutions to the wave equation, defined by
\begin{align} 
    \text{Near-Ring Limit:}\qquad
    \begin{cases}
        \ab{r-3M}\ll M&\qquad\text{near-peak},\\
        \displaystyle\ab{\frac{\ell}{\omega_R}-3\sqrt 3M}\ll M&\qquad\text{near-critical},\vspace{3pt}\\
        \displaystyle\frac{1}{\omega_R}\ll {M}&\qquad\text{high-frequency}.
    \end{cases}
\end{align}
In this region of phase space, an emergent inverted harmonic oscillator (IHO) algebra acts dynamically on the phase space and generates the symmetry of the resonance spectrum \eqref{eq:eikonal}. 

In this section, we review the WKB derivation of the eikonal QNM spectrum \eqref{eq:eikonal} and its relation to the near-ring limit of the wave equation. Using geometric optics, we discuss how this limit can also be understood from an analogous near-ring limit in the phase space of massless geodesics. Following Fransen \cite{Fransen:2023eqj}, this second limit is then reformulated in terms of the Penrose limit of the photon ring, which we explore in more detail in section \ref{sec:plane wave}.

\subsection{Quasinormal Modes}\label{sec:QNMsubsec1}
Linearized scalar perturbations on a black hole background obey the wave equation
\begin{equation}
    \Box \Psi(x) = 0\; .
\end{equation}
Using $\partial_t$ to normalize frequencies and spherical symmetry to separate the wave equation
\begin{equation} \label{eq:full wavefunction}
    \Psi_{\omega l m}(x) = e^{-i \omega t} Y_{lm} (\theta, \phi) \frac{\psi_{\omega l}(r)}{r}\; ,
\end{equation}
one finds that $\psi(r)$ is governed by the radial ODE
\begin{equation} \label{eq:radial}
    \left[ \partial_{r_*}^2 + \omega^2 - V_l(r_*) \right] \psi_{\omega l}(r_*) = 0 \;.
\end{equation}
Here $r_* = r+ 2M \log \l( \frac{r}{2M} - 1\r)$ is the tortoise coordinate, and the radial wave potential is
\begin{equation} \label{eq:radial_potential}
    V_l(r_*) = f(r) \left[ \frac{l(l+1)}{r^2} + \frac{2M}{r^3} \right] \;
\end{equation}
with $f(r)=1-\frac{2M}{r}$. QNMs are solutions to \eqref{eq:radial} with outgoing boundary conditions at spatial infinity   and infalling boundary conditions at the horizon  
\begin{equation}\label{eq:QNMbc}
    \psi \sim e^{i \omega r_*} \quad \text{as} \quad r_* \to \infty \; , \qquad 
    \psi \sim e^{-i \omega r_*}\quad \text{as} \quad r_* \to -\infty \; .
\end{equation}
These are dissipative boundary conditions, so the differential operator has a complex spectrum
\begin{equation}\label{eq:complexSpec}
    \omega = \omega_R + i \omega_I \;.
\end{equation}
The Schwarzschild black hole is known to be mode-stable ($\omega_I < 0$). Due to continual scattering off of the long-range Coulombic potential, perturbations decay polynomially at late times (the Price tails \cite{Price:1971fb,Price:1972pw}) rather than exponentially as the mode analysis \eqref{eq:complexSpec} might suggest. Nevertheless, the QNMs are expected to control the initial ringdown phase of gravitational mergers and are a prime target for upcoming gravitational wave experiments. 

\subsection{WKB and the Near-Ring Limit} \label{subsec:near-ring dynamical symmetry}
The boundary conditions \eqref{eq:QNMbc} can be viewed as an extreme case of the scattering problem for waves with  reflected and transmitted components much larger than the incident wave itself. The typical approach to   estimate the frequencies \eqref{eq:complexSpec} for which this holds is the WKB approximation \cite{schutz,Iyer:1986np,Iyer:1986nq}. In order to have a large transmission amplitude, the energy of the incident wave must be tuned to the top of the potential barrier (otherwise WKB would predict exponentially small transmission). The turning point is therefore second order, and is  controlled by the inverted harmonic oscillator rather than the Airy equation. 

WKB matching about this turning point implies that the QNM frequencies are given by
\begin{equation} \label{eq:eikonal_WKB}
    \omega \approx \sqrt{V(\tilde{r}_*)} -i \sqrt{\frac{V''(\tilde{r}_*)}{2 V(\tilde{r}_*)}} \l( n + \frac12 \r) \; ,
\end{equation}
where $\tilde{r}_*$ is the location of the peak of the potential $V(r_*)$ (derivatives are taken with respect to $r_*$). This approximation works best
for $l \to \infty$, in which case the peak of the potential is just the photon ring $\tilde{r} =r_0 + O\l(l^{-1}\r)=3M+O\l(l^{-1}\r)$, and $\omega_R = \sqrt{V(\tilde{r}_*)} \approx \frac{1}{3 \sqrt{3} M} \l( l + \frac{1}{2} \r) \to \infty$.
In that case \eqref{eq:eikonal_WKB} is simply  \eqref{eq:eikonal} and we see that the eikonal QNM spectrum is completely controlled by the photon ring.

In particular, the WKB first-order ansatz assumes a phase of the form
\begin{equation}
    \psi_{\pm} (r_*) \sim e^{\pm i \int^{r_*} \sqrt{\omega^2 - V(r_*)} dr_*}  = e^{\pm i \int^r \sqrt{\omega^2 - V(r)} \frac{dr}{f(r)}} \; .
\end{equation}
These functions have the asymptotic form
\begin{equation}\label{eq:phaseSign}
    \psi_{\pm} (r) \sim \begin{cases}
        e^{ \mp i \omega r_*} &\qquad r \to 2M \\
        e^{ \pm i \omega \frac{\sqrt{3}}{2 M} (r-3M)^2 } &\qquad r \to 3M \\
        e^{ \pm i \omega r_*} &\qquad r \to \infty
    \end{cases}
\end{equation}
Hence, to impose the QNM boundary conditions \eqref{eq:QNMbc}, it's sufficient to study the wave equation in the vicinity of the peak of the potential, and impose the boundary condition that
\begin{equation} \label{eq:near-ring IHO bc}
    \psi(r) \sim e^{ i \omega \frac{\sqrt{3}}{2 M} (r-3M)^2 }
\end{equation}
in that region.

The near-ring limit \eqref{eq:NearRing} isolates the eikonal 
QNMs in a region of  phase  space where they are controlled by the inverted harmonic oscillator potential. Since this is precisely the region that determines the QNM spectrum through WKB matching, the eikonal overtone spectrum follows from the enhanced symmetries  present in this limit. 
The near-ring wave equation \cite{Hadar:2022xag} is
\begin{equation} \label{eq:wave_approx_IHO}
    \br{\pd_{r_*}^2+\frac{\omega_R^2}{3M^2}\dt r^2+2i\omega_R\omega_I}\psi(\dt r)=0\;,
\end{equation}
where $\delta r \equiv r - 3M $, $\partial_{r_*} \approx f(3M) \partial_{\delta r} = \frac{1}{3} \partial_{\delta r}$, and $\omega_R = \frac{1}{3 \sqrt{3} M} \l( l + \frac{1}{2} \r)$. The eigenvalue problem that computes the imaginary parts of the spectrum at fixed $\omega_R$ is therefore given by 
\begin{align}\label{eq:ImEigen}
    \mathcal{H}\psi=i\omega_I\psi\;,\qquad
    \mathcal{H}=-\frac{1}{2\omega_R}\br{\pd_x^2+\gamma_L^2\omega_R^2x^2}\;,\qquad
    x=r_*-r^0_* 
    =\frac{\dt r}{f(3M)}\;,
\end{align}
with $\gamma_L \equiv \frac{1}{3\sqrt{3}M}$. This ODE has two qualitatively different classes of solutions. The eigenvalue problem \eqref{eq:ImEigen} is a version of the Weber equation and its solutions for generic eigenvalue are parabolic cylinder functions \cite{schutz,Bender:1999box} 
\begin{equation}\label{eq:PCF}
\psi(x) =     c_1 D_{-\frac12 -\omega_I/\gamma_L}\left(e^{3i\pi/4}\sqrt{2\gamma_L\omega_R}x\right) +c_2 D_{-\frac12 +\omega_I/\gamma_L}\left(e^{i\pi/4}\sqrt{2\gamma_L\omega_R}x\right)\;. 
\end{equation}
A generic linear combination of these two solutions contains waves that are incoming and outgoing on both sides of the potential barrier and which therefore do not match onto quasinormal modes. 
The behavior \eqref{eq:near-ring IHO bc} of the WKB approximation in the region far from the turning point is $e^{\frac{i}{2} \gamma_L\omega_R x^2}$ on both sides while the asymptotics for the cylinder functions $D_\nu(z)$ as $z\to \infty$ are
\begin{align}
    &D_\nu(z)\sim z^\nu e^{-z^2/4} \; , \qquad\qquad \qquad \qquad \qquad \qquad \qquad |\text{arg} \;z|<\frac{3\pi}{4}   \; ,\\
    &D_\nu(z)\sim z^\nu e^{-z^2/4} - \frac{(2\pi)^{1/2}}{\Gamma(-\nu)}e^{i\pi \nu}z^{-\nu -1}e^{z^2/4} \; , \qquad \frac{\pi}{4}<\text{arg} \;z < \frac{5\pi}{4} \; .
\end{align}
Focusing only on the exponentials, the behavior at large positive $x$ is therefore
\begin{equation}
    \psi(x)\sim c_1\left[e^{i\frac12 \gamma_L\omega_R x^2}-\frac{(2\pi)^{1/2}}{\Gamma(\frac12+\omega_I/\gamma_L)}e^{-i\frac12 \gamma_L\omega_R x^2}\right] +c_2 e^{-i\frac12 \gamma_L\omega_R x^2}
\end{equation}
so matching dictates that
$c_2=0$ as well as
\begin{equation}
    \omega_I = - \gamma_L \l(n+\frac12\r) \;.
\end{equation}
For these values of $\omega_I$ the parabolic cylinder function $D_n$ degenerates into the analytic continuation of the eigenstates of the normal harmonic oscillator \cite{IHO_eigenfunctions}. The wavefunctions in the near-ring region can therefore be constructed algebraically using 
the inverted harmonic oscillator algebra
\begin{equation} \label{eq:symm_dynamical}
    b_\pm=\frac{e^{\mp\gamma_Lt}}{\sqrt{2\gamma_L\omega_R}}\pa{\pm i\pd_x-\gamma_L\omega_Rx}\;,\qquad
    \frac{1}{\gamma_L}\mathcal{H}=-\frac{1}{2}\pa{b_+b_-+b_-b_+}
	\;.
\end{equation}
The fundamental mode $\psi_0$ satisfies a highest-weight condition 
\begin{equation}\label{eq:groundstate}
    b_-\psi_0 =0 \; , \qquad  \; \psi_0 (\delta r)=e^{i\frac{\sqrt{3}}{2M}\omega_R \delta r^2}
\end{equation}
and the overtones are descendants
\begin{equation}\label{eq:nearRingOvertone}
    \psi_n(\delta r) = b_+^n \psi_0(\delta r)\sim\delta r^n e^{-n\gamma_L   t} \psi_0\; .
\end{equation}

The advantage of this discussion is that the quasinormal boundary conditions and the uniform overtone spacing both arise from symmetry considerations.  

\subsection{Geometric Optics and Geodesics}\label{sec:QNMsubsec3}
Next we consider massless geodesics in the Schwarzschild geometry. For simplicity we consider motion in the equatorial plane since all other geodesics can be obtained by an $SO(3)$ symmetry transformation. In terms of the conserved quantities
\begin{equation}
    E = f(r) \frac{dt}{d\lambda} \;, \qquad L = r^2 \frac{d \phi}{d \lambda} \;,
\end{equation}
the radial motion is governed by the equation
\begin{equation}
    \l( \frac{dr}{d\lambda} \r)^2 + V_g(r) = E^2 \;,
\end{equation}
where $\lambda$ is affine time and the geodesic radial potential is
\begin{equation} \label{eq:Schw_geo_pot}
    V_g(r) = f(r) \frac{L^2}{r^2}\;.
\end{equation}
At leading order in the eikonal limit ($\omega_R \sim l \to \infty$), the wave potential \eqref{eq:radial_potential} reduces to the geodesic radial potential  if we make the identifications $E = \omega_R$ and $L = \l( l+\frac12\r)$
\begin{equation}
    V(r) \approx V_g(r) \;. 
\end{equation}
The WKB analysis of the radial wave equation is therefore equivalent to the geometric optics approximation based on the unstable bound orbit at the photon ring $r_0 = 3M$. The wavefronts of the QNM $\psi(r)$ are locally propagated by the radial motion of high-energy massless geodesics, while the decay of the wave is controlled by the expansion of the geodesic congruence. The energy of geodesics at the photon ring obeys 
\begin{equation}\label{eq:impact}
    E =  \frac{L}{3 \sqrt{3} M} = L \Omega_R \;,
\end{equation}
where
\begin{equation}
    \Omega_R = \left. \frac{d \phi}{dt} \right|_{r_0}
\end{equation}
is the angular velocity of the photon ring. This part of the geodesic phase space can be isolated through the near-ring geodesic scaling limit
\begin{align}\label{eq:NearRing_geo}
    \text{Near-Ring Limit of Geodesics:}\qquad
    \begin{cases}
        \ab{r-3M}\ll M&\qquad\text{near-peak},\\
        \displaystyle\ab{\frac{L}{E}-\frac{1}{\Omega_R}}\ll M&\qquad\text{near-critical},\vspace{3pt}\\
        \displaystyle\frac{1}{E}\ll {M}&\qquad\text{high-energy}.
    \end{cases}
\end{align}
Geodesics with $r=3M$ and $E=L\Omega_R$ orbit the black hole forever, but those which are slightly perturbed $r=3M+\delta r$ diverge from the photon ring exponentially fast in Schwarzschild time  
\begin{equation}\label{eq:divergence}
    \frac{d \delta r}{dt} = \frac{f(r_0)}{E} \sqrt{\frac{V_g''(r_0)}{2}} \delta r \equiv \gamma_L \delta r \;.
\end{equation}
This geodesic divergence controls the decay of the wave as well as the overtone structure. The geometric optics approximation builds solutions to the wave equation using the ansatz
\begin{equation}
    \Psi(x) = A(x) e^{i S(x)} \; .
\end{equation}
It assumes that the phase varies faster than the amplitude. Taking derivatives one finds
\begin{equation}\label{eq:terms}
    -p_\mu p^\mu \cdot A +i(2p^\mu \partial_\mu A + A \cdot\nabla_\mu p^\mu) + \nabla_\mu \nabla^\mu A = 0 \;, \qquad p_\mu\equiv \partial_\mu S(x) \; .
\end{equation}
At first order this is simply the equation for a null geodesic
\begin{equation}
    p_\mu p^\mu =0 \; , \qquad p^\mu\nabla_\mu p^\nu =0 \; .
\end{equation}
At second order the expansion of the congruence $\theta=\nabla^\mu p_\mu$ controls the transport of the amplitude   
\begin{equation}\label{eq:amp}
    p^\mu\partial_\mu A=-\frac12 \theta A \; .
\end{equation}
If the expansion is positive, the wave decays along the geodesic. The near-ring geodesics have the approximate Hamilton-Jacobi function
\begin{equation}
    S(x) \approx -E t + L \phi + \frac{\sqrt{3} E}{2M} \delta r^2 \; , \qquad L=3\sqrt{3}M E
\end{equation}
and one can check using \eqref{eq:divergence} that the expansion along the geodesic congruence is
\begin{equation}
    \theta=3\gamma_L \omega_R \; .
\end{equation}
The fundamental solution to the amplitude equation \eqref{eq:amp} is therefore
\begin{equation}\label{eq:a0}
    A_0(x) = e^{-\frac{1}{2} \gamma_L t} \; .
\end{equation}
Combined with the phase, this reproduces the wavefunction \eqref{eq:groundstate} in the near-ring region
\begin{equation}
    \Phi_0(x)=e^{-\frac12\gamma_L t}e^{-i\omega_R t}e^{il\phi}e^{i\frac{\sqrt{3}}{2M} \omega_R\delta r^2}
\end{equation}
and corresponds to the $n=0$ mode in the eikonal QNM spectrum
\begin{equation}\label{eq:eikonal2}
    \omega_R = \l( l + \frac{1}{2} \r) \Omega_R \;, \qquad \omega_I = -\l( n + \frac{1}{2} \r) \gamma_L \;.
\end{equation}

\subsection{Overtones From Conservation Laws}\label{subsec:overtoneSym}

Given the $n=0$ solution, there is a simple geometric trick to generate the overtones \cite{Hadar:2022xag}. Equation \eqref{eq:amp} says that if $A_0$ is a solution to the amplitude equation and $Q(x,p)$ is a quantity conserved along the geodesic (so that $p^\mu \nabla_\mu Q=0$) then $f(Q)A_0e^{iS}$ is also an approximate solution to the wave equation. Taylor expanding $f(Q)$ one encounters a family of solutions   
\begin{equation}\label{eq:geometricOvertones}
    \Phi_n(x)=\left[Q(x,p)\right]^nA_0e^{iS}.
\end{equation} 
There are essentially two ways to find a quantity that is conserved along a geodesic. If there is a Killing symmetry generated by the vector field $K^\mu(x)$, then the quantity $Q(x,p)=K^\mu(x) p_\mu$ is automatically conserved along every geodesic in the spacetime. The action of the symmetry on wavefunctions in the geometric optics approximation is particularly simple at leading order
\begin{equation}\label{eq:KillingOvertone}
    \mathcal{L}_K A_0e^{iS} \sim K^\mu \partial_\mu S \cdot A_0e^{iS}=(K^\mu p_\mu)A_0e^{iS} \; .
\end{equation}
So the symmetry transformation simply multiplies the seed solution by powers of the conserved quantity. This construction is not interesting for the set of Killing fields which commute and can be simultaneously diagonalized. In that case, \eqref{eq:KillingOvertone}  amounts to multiplying the seed solution by a position-independent constant that does not produce a new solution. If the symmetry group is non-abelian and $K^\mu$ is not diagonalized, then $K^\mu p_\mu$ often contains position dependence and \eqref{eq:KillingOvertone} is a genuinely new waveform.

In the absence of a Killing symmetry, it is still sometimes possible to produce solutions \eqref{eq:geometricOvertones}  for individual geodesic congruences. If one is able to explicitly solve for the coordinate embedding of the geodesic, dynamical invariants can be constructed. These invariants  differ for each geodesic and may be nonlinear in coordinates and momenta.  For instance, according to \eqref{eq:divergence} the quantity
\begin{equation}\label{eq:b}
    Q=\delta r \cdot e^{-\gamma_L t}
\end{equation}
is conserved along the perturbed photon ring geodesic in the near-ring region.  The amplitudes
\begin{equation}\label{eq:overtoneAmp}
    A_n=Q^nA_0 =\left(\delta r\right) ^ne^{-(n+\frac12)\gamma_L t}
\end{equation}
are therefore all approximate solutions to \eqref{eq:amp}. They fill out the rest of the overtone spectrum \eqref{eq:eikonal2} and agree with the near-ring wavefunctions \eqref{eq:nearRingOvertone}. As we will see in the following sections, taking a Penrose limit onto the geodesic congruence often converts the dynamically conserved quantity \eqref{eq:b} into a kinematically conserved quantity associated to an emergent Killing symmetry in the plane wave geometry.

To understand how this occurs, note that the conservation of a quantity like \eqref{eq:b} is equivalent to the existence of a (congruence dependent) conserved current
\cite{Misner:1973prb}
\begin{equation}
    j_{(A,p)}^\mu=A_0^2 \, p^\mu \; .
\end{equation}
This vector field is divergence free in the vicinity of the congruence since according to \eqref{eq:amp}
\begin{equation}
    \nabla_\mu j_{(A,p)}^\mu = A_0\left(2p^\mu \nabla_\mu A_0 + A_0 \nabla^\mu p_\mu\right)=0 \; ,
\end{equation}
which implies that its integrated charge is conserved.
In the case at hand with $A_0$ given by \eqref{eq:a0}, the quantity   conserved along the congruence is 
\begin{equation}
Q(x,p)=    \int d^3x  j_t\sim 4\pi E e^{-\gamma_L t} \int_{r_0}^{r_0+\delta r} r^2 dr\sim 4\pi r_0^2E e^{-\gamma_L t}  \delta r
\end{equation}
which is conservation of \eqref{eq:b}. $Q$ can be thought of as the conserved number of photons.

Often times, when there is an emergent symmetry, the conserved quantity can be written as the contraction of a vector field (specific to the congruence and not a Killing vector) with the tangent to the congruence
\begin{equation}
  Q(x,p)=  V^\mu(x,p)p_\mu \; .
\end{equation}
Typically the components of $V(x,p)$ will be nonlinear in $p$ (this is the case for Kerr, when the Lyapunov exponent involves elliptic integrals with arguments determined by $p$). In that case the overtone structure in geometric optics arises as in \eqref{eq:KillingOvertone}, with
\begin{equation}\label{eq:hiddenVect}
    \mathcal{L}_{V} A_0e^{iS} \sim \left(V^\mu \partial_\mu S\right) \cdot A_0e^{iS}=Q(x,p)A_0e^{iS} \; ,
\end{equation}
although the construction only works for the single congruence. When we take the Penrose limit, we isolate the congruence,  $V^\mu$ becomes an isometry, and \eqref{eq:hiddenVect} becomes \eqref{eq:KillingOvertone}.

\subsection{Penrose Limit \& pp-Wave Spacetime} \label{subsec:scaling_limit}
The geometric optics approximation reduces the wave problem in a complicated geometry to a simpler geodesic calculation in the same geometry. As emphasized by Fransen \cite{Fransen:2023eqj}, there is an equivalent way to perform this approximation that instead reduces the wave problem in the complicated geometry to a wave calculation in a much simpler geometry. The construction proceeds by taking the ``Penrose limit'' of the relevant geodesic congruence \cite{Penrose1976, Blau:notes}. This limit zooms into the congruence, retaining only information about the geodesic deviation. Since it is precisely this data which controls the first two orders of the geometric optics approximation, solving for the spectrum and waveforms in this geometry is equivalent to the procedure reviewed in sections 
\ref{subsec:near-ring dynamical symmetry}-\ref{subsec:overtoneSym}. This approach effectively geometrizes and covariantizes the near-ring limit. In particular, the emergent symmetries of the eikonal spectrum are transformed into geometric isometries of the limiting plane wave spacetime.

\subsubsection*{Scaling Limit of the Geometry}

According to \eqref{eq:impact} equatorial photon ring geodesics are parameterized by affine time $\lambda$ with
\begin{equation}
    \l( t(\lambda), r(\lambda), \theta (\lambda), \phi (\lambda) \r) = \l( 3 E \lambda, 3M, \frac{\pi}{2}, 3E \Omega_R \lambda \r) \;, \qquad \Omega_R^{-1} = 3 \sqrt{3} M\; .
\end{equation}
For the photon ring, the first step amounts to performing the coordinate transformation\footnote{ This coordinate transformation is obtained by constructing a pseudo-orthonormal frame that's adapted to this geodesic, as in \cite{Fransen:2023eqj}, and integrating the legs of the co-frame. \label{fn:coframe}} ${\l(t,r,\theta,\phi \r) \to \l( u,v,\rho,\psi \r)}$
\begin{equation} \label{eq:coord_transf}
    \begin{aligned}
        & t = 3 u \; ,\\
        & r = 3M + \varepsilon \frac{\rho}{\sqrt{3}} \; ,\\
        & \theta = \frac{\pi}{2} + \varepsilon \frac{\psi}{3M}\;, \\
        & \phi =  3 \Omega_R u + \varepsilon^2 \Omega_R v \;.
    \end{aligned}
\end{equation}
Note that an affine transformation of $\lambda$ can always be used to scale away the dependence on $E$. Expanding the line element about $\varepsilon=0$, the Schwarzschild metric becomes 
\begin{equation}
    ds^2 = \varepsilon^2 \l( 2 du dv + \frac{1}{3M^2} \l( \rho^2 - \psi^2  \r) du^2 + d \rho^2 + d \psi^2 \r) + O(\varepsilon^3) \;.
\end{equation}
To perform the Penrose limit, we combine the singular diffeomorphism $\varepsilon\to 0$ with a singular conformal transformation, resulting in the plane wave metric  
\begin{equation} \label{eq:PR_Pen_metric}
    ds^2_{\mathrm{Pen}} = \lim_{\varepsilon \to 0} \varepsilon^{-2} ds^2 = 2 du dv + \frac{1}{3M^2} \l( \rho^2 - \psi^2  \r) du^2 + d \rho^2 + d \psi^2 \;.
\end{equation}
Note that unlike the near-extremal near-horizon limit \eqref{eq:nearNHEK}, we must perform a conformal rescaling of the metric as we zoom into the photon ring. That is because the photon ring has vanishing volume in the black hole spacetime, in contrast to the extremal throat whose analogous expansion begins at order $\varepsilon^0$. 

\subsubsection*{Scaling Limit of the Solutions}
We would now like to understand which solutions to the Schwarzschild wave equation are nonsingular in the Penrose limit. In other words, if we start with a wavefunction of the form
\begin{equation}
    f_{\omega, m} (x) = e^{-i \omega t + i m \phi}
\end{equation} 
in the original Schwarzschild geometry, we would like to know how it behaves under the singular diffeomorphism \eqref{eq:coord_transf} with $\varepsilon \to 0$. For finite $\varepsilon$ we can simply perform the coordinate transform  
\begin{equation}
    f_{\omega, m} (x) \propto e^{-3 i u (\omega - \Omega_R m)} e^{i \Omega_R m \varepsilon^2 v} \equiv e^{-i \tilde{\omega} u + i p v}\; .
\end{equation} 
The frequency and momenta in the two coordinate systems are therefore related by
\begin{equation}\label{eq:momentaMatch}
    \tilde{\omega} = 3 (\omega - \Omega_R m) \;, \qquad p = \varepsilon^2 \Omega_R m \; .
\end{equation}
For finite $\varepsilon$ the periodicity of the coordinate $v \sim v + \frac{1}{\varepsilon^2} \frac{2 \pi}{\Omega_R}$ imposes the quantization condition $\frac{1}{\varepsilon^2} \frac{p}{\Omega_R} \in \mathbb{Z}$, which is equivalent to integrality of $m$. The limit $\varepsilon\to 0$ is therefore a decompactification limit. Since the pp-wave spacetime is insensitive to $\varepsilon$, it is natural to keep the quantities $\tilde{\omega},p$ finite in the limit. Inverting \eqref{eq:momentaMatch} one finds
\begin{equation}\label{eq:matching}
    \qquad m = \frac{1}{\varepsilon^2} \frac{p}{\Omega_R} \;, \qquad \omega = \frac{1}{\varepsilon^2} p + \frac{\tilde{\omega}}{3} \;, \qquad \frac{\omega}{m} = \Omega_R + \varepsilon^2 \frac{\Omega_R \tilde{\omega}}{3 p} \;. 
\end{equation}
In the scaling limit as $\varepsilon \to 0$ we therefore have 
\begin{equation}
   m \sim \frac{1}{\varepsilon^2} \gg 1 \;, \qquad \omega \sim \frac{1}{\varepsilon^2} \frac{1}{M} \gg \frac{1}{M} \;, \qquad \frac{\omega}{m} - \Omega_R \sim \varepsilon^2 \frac{1}{M} \ll \frac{1}{M} \; .
\end{equation}
This is precisely the near-ring limit \eqref{eq:NearRing} of the Schwarzschild wave equation. 
In other words, only modes with $m$ and $\omega$ that scale in this way have non-singular limits (finite $\tilde{\omega}, p$) as we scale into the plane wave spacetime. Of course, these are also precisely the conditions that single out the eikonal QNM spectrum, so we expect the solutions to the wave equation in the geometry \eqref{eq:PR_Pen_metric} to reproduce the spectrum and the enhanced symmetries encountered in sections \ref{subsec:near-ring dynamical symmetry}-\ref{subsec:overtoneSym}. In the next section we discuss general features of this geometry.

\section{Plane Wave Limit of the Schwarzschild Photon Ring} \label{sec:plane wave}

The Penrose limit of an arbitrary null geodesic in Schwarzschild takes the form \cite{Blau:notes}
\begin{equation} \label{eq:Schw_general penrose limit}
    ds^2 = 2dudv + \frac{3 M b^2}{r(u)^5} \l( x^2 - y^2 \r) du^2 + dx^2 + dy^2 \;,
\end{equation}
where $r(u)$ is the radial motion of the geodesic with affine parameter $\lambda=u$, and the affine parameter has been scaled to set $E=1$, such that $b \equiv L/E = L$. In the case of photon ring geodesics ($r(u)=r_0=3M$ and $b=b_0=3\sqrt{3}M$), this geometry becomes $u$-independent and is simply
\begin{equation}\label{eq:plane_wave_metric}
    ds^2 = 2du dv + \frac{1}{3M^2} \l( x^2 - y^2 \r) du^2 + dx^2 + dy^2 \;.
\end{equation}
This is a nontrivial solution to the vacuum Einstein equation presented in what are typically called ``Brinkmann coordinates.'' The coefficient of the $\l(x^2 - y^2\r) du^2$ term can be rescaled by the transformation  ${u \to \alpha u}$, $v \to \frac{1}{\alpha} v$.  The covariant geometric information\footnote{All curvature scalars for pp-wave geometries vanish identically.} is instead contained in the geodesic deviation, which is controlled by  the two independent nonvanishing components of the Riemann tensor (appropriately contracted with the 4-velocity and deviation vector) 
\begin{equation}
    R_{uxux} =-\frac{1}{3M^2} \; , \qquad R_{uyuy} = \frac{1}{3M^2} \; .
\end{equation}
The factor $1/(3M^2)$ captures the geodesic deviation along the photon ring, with the positive sign signaling Lyapunov instability in the radial direction and the negative sign signaling bound oscillations in the polar direction. The geometry \eqref{eq:plane_wave_metric} represents a solution to the Einstein equation in its own right. However, performing a large diffeomorphism that rescales $u$ (and thus the harmonic oscillator coefficient) changes the way that it attaches to the exterior Schwarzschild geometry and amounts to performing the limit with a rescaled affine time along the geodesic.

\subsubsection*{Symmetries} \label{subsec:symmetries}
The isometry algebra of \eqref{eq:plane_wave_metric} is generated by
\begin{equation}\label{eq:ppISO}
\begin{split}
    & a_+ = \frac{i}{\sqrt{2 \beta}} e^{-i \beta u} \l(i \beta y \partial_v +  \partial_y \r) \; , \qquad a_- = \frac{-i}{\sqrt{2 \beta}} e^{i \beta u} \l(-i \beta y \partial_v +  \partial_y \r) \\
    & b_+ = \frac{1}{\sqrt{2 \beta}} e^{- \beta u} \l( -\beta x \partial_v - \partial_x \r) \; , \qquad b_- = \frac{1}{\sqrt{2 \beta}} e^{\beta u} \l( -\beta x \partial_v + \partial_x \r) \\
    & V = i \partial_v \; , \qquad H = \frac{i}{\beta} \partial_u \; ,
\end{split}
\end{equation}
where $\beta^2 = 1/(3M^2)$. Note that $a_+ = (a_-)^*$. The only nonvanishing Lie brackets are
\begin{equation}
    [a_-, a_+] = V \; , \quad [b_-, b_+] = i V \; , \quad [H, a_{\pm}] = \pm a_{\pm} \; , \quad [H, b_{\pm}] = \mp i b_{\pm} \; .
\end{equation}
The $a_{\pm}$ are SHO ladder operators, while the $b_{\pm}$ are IHO resonance creation/annihilation operators.

\subsection{Wave Equation, Boundary Conditions \& Stability} \label{Sec:pp_wave_eq}
The scalar wave equation in the geometry \eqref{eq:plane_wave_metric} is
\begin{equation}\label{eq:BOX}
 \Box=   2\partial_u\partial_v - \frac{1}{3M^2}(x^2-y^2)\partial_v^2+\partial_x^2+\partial_y^2\; . 
\end{equation}
Separating variables using the ansatz
\begin{equation}
    \Psi(u,v,x,y)=e^{-i\tilde{\omega} u +i p v}R(x)P(y)
\end{equation}
leads to an equation for the radial and polar parts of the wavefunction
\begin{equation}\label{eq:dispersion}
    [2\tilde{\omega} p +p^2\beta^2(x^2-y^2)+(\partial_x^2 + \partial_y^2)]R(x)P(y)=0 \; ,\qquad  \beta^2= \frac{1}{3M^2} \; .
\end{equation}
The polar direction is controlled by a harmonic oscillator potential while the radial wavefunction is subject to the IHO 
\begin{align}\label{eq:QHM}
    &-\left(-\partial_y^2+ \beta^2 p^2 y^2\right)P(y)=E_yP(y) \; ,\\
    &-\left(-\partial_x^2- \beta^2 p^2 x^2\right)R(x)=E_xR(x) \; .\label{eq:IHM}
\end{align}
To proceed further we have to discuss boundary conditions. 

\subsubsection*{Stable Direction (Angular)}
Equation \eqref{eq:QHM} has solutions for any value of the separation constant $E_y$, but only the discrete set $E_y(n_1)=-2\beta p(n_1+1/2)$ correspond to parabolic cylinder functions which are $L^2$ normalizable. Those solutions are    
\begin{equation}\label{eq:penroseAngular}
    P_{n_1}(y)=\frac{1}{\sqrt{2^{n_1} n_1!}} \left(\frac{\beta p}{\pi}\right)^{1/4}e^{-\beta py^2/2}H_{n_1}(\sqrt{\beta p}\; y) \; , \qquad E_y=-2\beta p\left(n_1+\frac12\right)\; .
\end{equation}
It may seem bizarre that the polar part of the groundstate wavefunction is peaked about a particular angle without oscillations (this is certainly not generic behavior for spherical harmonics). Indeed, this feature  is  peculiar to the Penrose limit of the equatorial geodesic.  As we will discuss in section \ref{sec:Tilt}, generic photon ring geodesics oscillate between two polar turning points $\theta_+$ and $\pi-\theta_+$. These geodesics have nontrivial polar momentum $p_\theta$ and they traverse the full  range $[\theta_+,\pi-\theta_+]$. On the other hand, the equatorial geodesic has $\theta_+=\pi-\theta_+=\frac{\pi}{2}$ and $p_\theta=0$, so one can think of it as the limit where the two turning points become very close together near the bottom of the angular potential (which is always located at $\theta=\frac{\pi}{2}$). This is precisely the limit in which the usual WKB approximation (with first order turning points and an Airy approximation) breaks down and must be replaced by a second-order turning point modeled on the harmonic oscillator. That harmonic oscillator is precisely the one appearing in \eqref{eq:QHM}.

\begin{figure}[h]
\includegraphics[scale=.38]{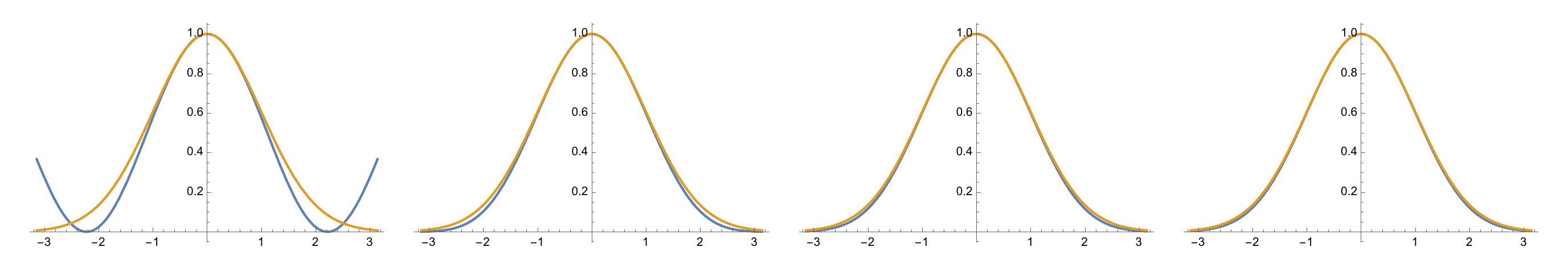}
\caption{Plots of $\frac{1}{(2l-1)!!}P_l^l\left(\cos\left(\frac{\pi}{2}-\frac{y}{\sqrt{l}}\right)\right)$   and $e^{-\frac12 y^2}$ for $l=2,6,10,14$}\label{fig:Gaussian_Legendre}
\end{figure}

The equatorial geodesic is related to the solutions of the wave equation with $m=l\gg1$ whose support is also concentrated on the equator. To understand the connection to \eqref{eq:QHM}
we substitute the coordinate transformation \eqref{eq:coord_transf} into the associated Legendre function (with $\psi=y$)
\begin{align}
    P_m^m(\cos \theta)&\equiv(-1)^m(2m+1)!! (1-\cos^2\theta)^{m/2}
    =(-1)^m(2m+1)!! \left(1-\cos^2\left(\frac{\pi}{2}+\varepsilon\frac{y}{3M}\right)\right)^{m/2}\notag\\
    &\sim (-1)^m(2m+1)!! \left(1-\varepsilon^2 \frac{y^2}{9M^2} \right)^{m/2} \; .
\end{align}
Now using \eqref{eq:matching} we can identify
\begin{equation}
    \varepsilon^2=\frac{p}{m}3\sqrt{3}M \; .
\end{equation}
Taking the limit $m\to \infty$ (equivalently $\varepsilon \to 0$) using
\begin{equation}
    \lim_{m\to \infty}\left(1- \frac{1}{m}\frac{py^2}{\sqrt{3}M} \right)^{m/2}=e^{-\frac12 \frac{p}{\sqrt{3}M}y^2}
\end{equation}
one recovers the the groundstate \eqref{eq:penroseAngular}. As one can see from figure \ref{fig:Gaussian_Legendre}, this is already a very good approximation even for rather small values of $l$.

\begin{figure}[h]
\includegraphics[scale=.37]{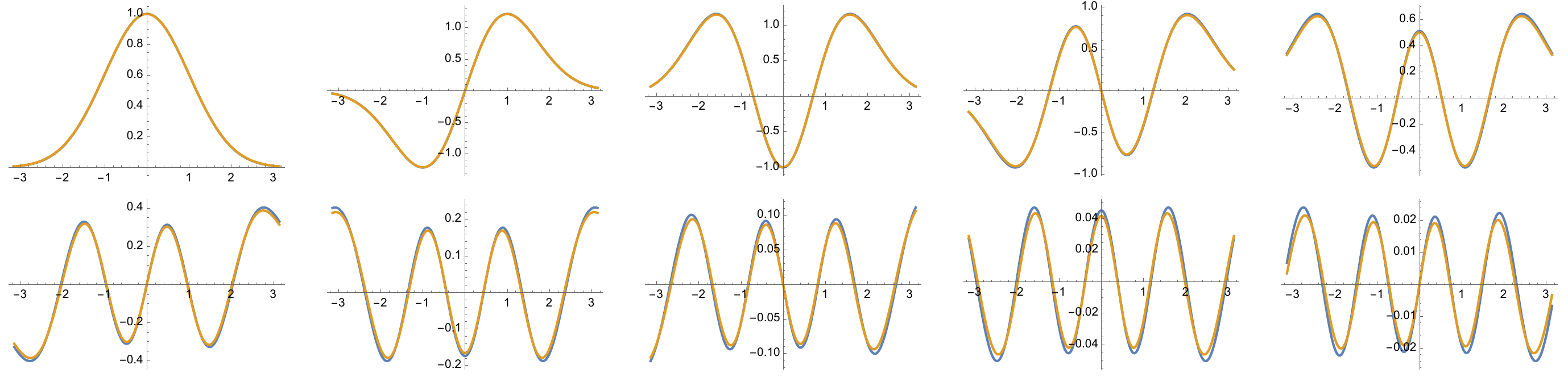}
\caption{Plots of $\frac{m^{-n/2}}{(2m-1)!!}P_{m+n_1}^m\left(\cos\left(\frac{\pi}{2}-\frac{y}{\sqrt{m+n_1}}\right)\right)$ and $\frac{1}{n_1!}H_{n_1}(y)e^{-\frac12 y^2}$ for $m=100$ and $n_1=0,1,\dots, 9$.}\label{fig:Hermite_Legendre}
\end{figure}

It is also possible to obtain the relationship between the excited SHO states \eqref{eq:penroseAngular} and the associated Legendre polynomials with $l-m\ll l$ using the ladder operator
\begin{equation}
    L_-=e^{-i\phi}(-\partial_\theta + i\cot \theta \partial_\phi) \; .
\end{equation}
Acting $n_1$ times with the lowering operator and using the relation \eqref{eq:coord_transf} gives
\begin{align}
    Y_{m+n_1}^m&\propto e^{-in_1\phi}(-\partial_\theta + i\cot \theta \partial_\phi)^{n_1}e^{i(m+n_1)\phi} P_{m+n_1}^{m+n_1}(\theta)\\
    &=e^{-in_1\phi}\left(-\frac{3M}{\varepsilon}\partial_y -i\frac{\varepsilon}{3M}y \partial_\phi\right)^{n_1}e^{i(m+n_1)\phi} P_{m+n_1}^{m+n_1}(\theta) \; .
\end{align}
Then using \eqref{eq:matching} and recalling that the wavefunction also contains a factor $e^{ipv}$ we have
\begin{align}
     Y_{m+n_1}^m&\sim e^{-in_1\phi}\left(-\frac{3M}{\varepsilon}\partial_y +\frac{p}{\varepsilon}\sqrt{3}y\right)^{n_1}e^{i(m+n_1)\phi} P_{m+n_1}^{m+n_1}(\theta)\\
    &= \left(\frac{3M}{\varepsilon}\right)^{n_1} e^{-im\phi}\left(-\partial_y +\frac{p}{\sqrt{3}M}y\right)^{n_1} e^{-\frac12 \beta p y^2}\\
    &\propto m^{n_1/2} e^{-im\phi}\left(\partial_y +i\beta y\partial_v\right)^{n_1} e^{-\frac12 \beta p y^2} \; .
\end{align}
This is proportional to the action of the SHO raising isometry $a_+$ given in \eqref{eq:ppISO}, which produces the Hermite polynomials when acting repeatedly on the ground state. As can be seen in figure \ref{fig:Hermite_Legendre} this is a good approximation for large $m$ and small $n_1 \ll m$. 

From this construction it is clear that the pp-wave SHO excitation number $n_1$ lifts to 
\begin{equation}
    n_1=l-m
\end{equation}
in the full Schwarzschild geometry.

\subsubsection*{Boundary Conditions in the Unstable Direction (Radial) and Mode Stability}
We next turn to the radial equation \eqref{eq:IHM}, which also requires a discussion of boundary conditions mirroring that of section \ref{subsec:near-ring dynamical symmetry}. 
The inverted harmonic oscillator has solutions for any value of the separation constant $E_x$ in \eqref{eq:IHM}. However, for generic values the resulting eigenfunction will be a linear combination of incoming and outgoing waves on both sides of the barrier. These solutions are limits of scattering states in Schwarzschild, not limits of QNMs. 
 
If we choose the outgoing QNM boundary conditions   so that the ``groundstate''  looks like $e^{i\beta px^2/2}$, then the separation constant is 
\begin{equation}\label{eq:Ex}
    E_x(n_2)=2i\beta p\left( n_2 + \frac12 \right) \; .
\end{equation}
The separation constants in \eqref{eq:penroseAngular} and \eqref{eq:Ex} determine the dispersion relation according to \eqref{eq:dispersion} 
\begin{equation}
    2\tilde{\omega} p +E_x(n_2)+E_y(n_1)=0 \; .
\end{equation}
 Since the separation constants are linear in $p$ this simply gives
\begin{equation}\label{eq:PPspec}
\tilde{\omega}=\beta\left(n_1+\frac12\right)  -i\beta\left(n_2+\frac12\right) \; .
\end{equation}
Relating this spectrum to the Schwarzschild quantities using \eqref{eq:momentaMatch}, we find
\begin{equation}
\begin{split}
    \omega  &= \Omega_R m  + \frac{1}{3} \tilde{\omega} \\
    &= \l( m  + n_1 + \frac{1}{2} \r) \Omega_R  - i \l( n_2   + \frac{1}{2} \r) \gamma_L \;,
\end{split}
\end{equation}
where $\Omega_R = \gamma_L = \frac{1}{3}\beta$. Recalling that $n_1=l-m$ and identifying $n_2$ with the Schwarzschild overtone number we exactly reproduce the eikonal QNM spectrum
\begin{equation}
    \omega_{ln} = \l( l + \frac{1}{2} \r) \Omega_R  - i \l( n   + \frac{1}{2} \r) \gamma_L \; .
\end{equation}

Note that the choice of the AQNM groundstate would have produced modes with positive imaginary part, while the choice of a real separation constant in \eqref{eq:Ex} would have resulted in a purely real spectrum for $\tilde{\omega}$. 
In other words, the choice of boundary conditions for the inverted harmonic oscillator determines whether the frequencies $\tilde{\omega}$ are real or not. There are questions as to whether or not the geometry \eqref{eq:plane_wave_metric} is classically stable since potentials which are unbounded from below often signal instabilities. Equation \eqref{eq:PPspec} makes it clear that the system  is mode stable since the resonance spectrum has purely negative imaginary parts. Of course, mode stability does not necessarily imply linear (or non-linear) stability, but the photon ring pp-wave is mode stable in the same way that Schwarzschild itself is mode stable: 
the inverted harmonic oscillator potential represents \textbf{unstable} trapping. The resonant boundary conditions do lead to complex frequencies, but they have the correct sign so that sensible initial data decays with time. Since we expect the Penrose limit to capture the near-ring geometric optics solutions, it would  be surprising to find an unstable resonance since it should lift to a high frequency solution in the black hole geometry.

\subsection{Overtones from Symmetry}
Recall the scalar wave operator was given by
\begin{equation} 
 \Box=   2\partial_u\partial_v - \beta^2(x^2-y^2)\partial_v^2+\partial_x^2+\partial_y^2 \; .
\end{equation}
In terms of the isometry generators \eqref{eq:ppISO}, this operator can be written as
\begin{equation}
    \frac{1}{\beta} \Box = - 2 H V + \l(a_+ a_- + a_- a_+ \r) - \l( b_+ b_- + b_- b_+ \r)
\end{equation}
which is the quadratic Casimir of the algebra. This means that, given one solution to the wave equation, we can act on it with an isometry generator to generate another solution. To solve the wave equation, we diagonalize a maximally commuting set of isometries, which we choose to be $H$ and $V$. Since the other isometries do not commute with these vector fields, acting with them  will generate new solutions to the wave equation. An interesting aspect of the Penrose limit is the ability to specify boundary conditions using symmetry alone. In order to implement outgoing boundary conditions for the unstable direction and normalizable boundary conditions for the stable direction, we simply look for ``lowest-weight" solutions $\phi_0$ that are annihilated by $a_-$ and $b_-$:
\begin{equation}
    H \phi_0 = \frac{\tilde{\omega}}{\beta} \phi_0 \;, \qquad V \phi_0 = - p\, \phi_0 \;, \qquad a_- \phi_0 = b_- \phi_0 = 0 \; .
\end{equation}
This solution will have the form
\begin{equation}
    \phi_0 = e^{-i \tilde{\omega} u + i p v} e^{i\beta p x^2/2}e^{-\beta p y^2/2}
\end{equation}
and we can generate new solutions by repeatedly acting with the ladder operators
\begin{equation} \label{eq:pp overtones}
    \phi_{n_1, n_2} = (a_+)^{n_1} (b_+)^{n_2} \phi_0 \; .
\end{equation}
This geometrizes the construction \eqref{eq:nearRingOvertone} of \cite{Hadar:2022xag} as noted first  by Fransen \cite{Fransen:2023eqj}. 

Note that one could also pick an arbitrary wavefunction $\psi(x)$ and begin acting with the resonance creation operator $b_+$. However, if the separation constant of the seed function is real (so that the state satisfies scattering boundary conditions rather than QNM boundary conditions) then the separation constant of the descendant will be complex. When lifted to the Schwarzschild geometry, these wavefunctions correspond to analytic continuations of scattering states to a particular set of complex frequencies and are of no particular significance. 

\subsection{pp-Wave Isometries Lifted to Schwarzschild}\label{subsec:plane wave_eikonal qnm}

The isometries \eqref{eq:ppISO} of the  plane wave geometry can be lifted (non-uniquely) to vector fields in the full Schwarzschild geometry using the coordinate transformation \eqref{eq:coord_transf}. Their form far from the photon ring is irrelevant since they only constrain dynamics in the near-ring region. There, they take the form (near the equator)
\begin{alignat}{2} \label{eq:symmetries_Schw_coords}
        & a_+ = \varepsilon \frac{i}{3M\sqrt{2 \beta}} e^{-i \beta t/3} \l(i \l( \theta - \pi/2 \r) \partial_\phi +  \partial_\theta \r) \;, \qquad 
    &&a_- = \varepsilon \frac{-i}{3M\sqrt{2 \beta}} e^{i \beta t/3} \l(-i  \l( \theta - \pi/2 \r) \partial_\phi + \partial_\theta \r)\;,  \notag\\
    & b_+ = \varepsilon \frac{1}{\sqrt{6 \beta}} e^{- \beta t/3} \l( -\beta^2  \l(r-3M \r) \partial_\phi -  \partial_r \r) \; , \qquad 
    &&b_- =  \varepsilon \frac{1}{\sqrt{6 \beta}} e^{\beta t/3} \l( -\beta^2  \l(r-3M \r) \partial_\phi +  \partial_r \r) \; , \notag\\
    & V = i\Omega_R\varepsilon^2  \partial_\phi \; , \qquad \qquad \qquad &&H = \frac{3i}{\beta} \l( \partial_t + \Omega_R \partial_\phi \r) \;.
\end{alignat}
As explained in section \ref{subsec:scaling_limit}, only wavefunctions with $\omega  \sim m  \sim \frac{1}{\varepsilon^2}$ and $\frac{m}{\omega} = 3\sqrt{3}M + \mathcal{O}(\varepsilon^2)$ survive the scaling limit. From the perspective of the full Schwarzschild geometry, the action of the generators of \eqref{eq:symmetries_Schw_coords} on these wavefunctions is finite. 

Under the scaling \eqref{eq:coord_transf}, the Schwarzschild $SO(3)$  generators $L_{\pm} = L_x \pm i L_y$ limit to
\begin{equation}
    \varepsilon \frac{1}{\sqrt{2 \cdot 27^{1/2} M}} L_{\pm} \to i a_{\mp} \;, \qquad \varepsilon^2 \Omega_R L_z \to - V \;, \qquad \text{as} \quad \varepsilon \to 0 \; .
\end{equation}
This describes the contraction of the $SO(3)$ algebra to the Heisenberg algebra, pointed out in \cite{Fransen:2023eqj}. This contraction underlies the scaling limit of the Legendre polynomials onto the Hermite polynomials as explained in section \ref{Sec:pp_wave_eq}.

On the other hand, the IHO algebra generated by $b_-,b_+$ captures the geometrization of the (hidden) phase space symmetry \eqref{eq:symm_dynamical} that acts on the eikonal QNMs. In particular, acting on the eikonal QNM wavefunctions, we get
\begin{equation}
    i \varepsilon   \sqrt{\omega_R} b_{\pm}^{\textrm{hidden}} \to b_\pm^{\textrm{iso}} \;.
\end{equation}

\section{Tilted Photon Ring}\label{sec:Tilt}
The $SO(3)$ symmetry of Schwarzschild ensures that the QNM spectrum does not depend on the azimuthal number $m$. In particular,  computing the frequencies for the special case $m=l$   correctly reproduces the entire spectrum. This observation allowed us to  focus on the Penrose limit of the equatorial photon orbit in section \ref{sec:plane wave} while sidestepping any discussion of complicated polar motion. Unfortunately, this simplification obscures an important subtlety in the relationship between the WKB approximation and the Penrose limit. This distinction becomes crucial in Kerr where most photon orbits execute complicated polar motion.

In order to illustrate the issue and warm up for the Kerr problem,  in this section we generalize the analysis of section \ref{sec:plane wave} to handle tilted (non-equatorial) photon orbits in Schwarzschild. 
\begin{figure}[h]
    \centering
    \includegraphics[width=7cm]{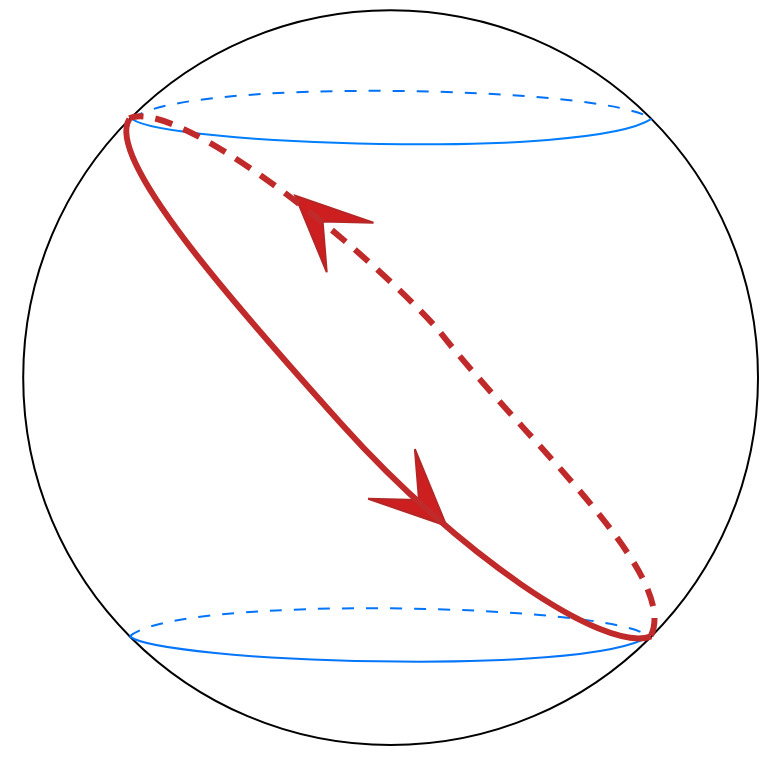}
    \caption{The tilted photon ring geodesic}
    \label{fig:Tilted PR}
\end{figure}

\subsection{WKB, Geometric Optics and the Penrose Limit}

The original, and most transparent, way to take the Penrose limit is via adapted coordinates. Given a geodesic congruence, one chooses the affine time $U=\lambda$ as a first coordinate, and the Hamilton-Jacobi function $V = S(x)$ of the congruence as the second coordinate \cite{Patricot:2003dh,Blau:2004yi,Blau:notes}. Together with the transverse coordinates $(R,\Psi)$, this defines a nonsingular diffeomorphism ${(t,r,\theta,\phi)\to(U,V,R,\Psi)}$. One then makes the singular coordinate transformation
\begin{equation}
    (U,V,R,\Psi) \to (u,\varepsilon^2v,\varepsilon \rho,\varepsilon \psi) \;, \qquad \varepsilon \to 0 \;,
\end{equation}
and rescales the metric to obtain a regular solution. The resulting plane wave always has a $\partial_v$ isometry, which means that solutions to the wave equation will always have a factor of
\begin{equation}\label{eq:Phase}
    \exp \pa{i p v} = \exp \pa{i \frac{p}{\varepsilon^2} S} \; .
\end{equation}
As $\varepsilon$ becomes small, the function in the original geometry has precisely the form of a leading-order geometric optics wavefunction.\footnote{Our convention is to normalize $S(x)$ by setting $E=1$ with a rescaling of affine parameter. The geometric optics ansatz is then $\exp(i\omega S)$, with $\omega = p/\varepsilon^2$ large for small $\varepsilon$. Factors of $L$ in the geodesic equation are then replaced by $b=L/E$. }

The geometric optics ansatz \eqref{eq:Phase} builds an approximate solution to the wave equation using a \textit{single} geodesic with tangent vector $p_\mu=\partial_\mu S$.
When the geodesic motion is separable the Hamilton-Jacobi function can be written 
\begin{equation}\label{eq:HJfun}
    S=-t + L_z\phi \pm \int^\theta_{\theta_0} p_\theta \, d\theta' \pm \int^r_{r_0}p_r \, dr'\; .
\end{equation}
Importantly, the geodesic equation fixes the form of $p_\theta$ and $p_r$ but not their sign, which flips when the motion reaches a turning point.
For eikonal QNMs, the quasinormal mode boundary condition makes use of a second order turning point in the radial motion. The  geodesic underlying the geometric optics approximation is taken to be slightly perturbed from $r=3M$ with ${L=3\sqrt{3}ME}$ so that $p_r$ vanishes linearly as $r\to 3M$. This correctly produces the near-ring phase \eqref{eq:phaseSign} using the last term in \eqref{eq:HJfun}. It amounts to performing a modified WKB approximation to the second order turning point in the radial motion (both WKB matching conditions are imposed simultaneously using the IHO approximation to the potential).

\begin{figure}[h]
    \centering
    \includegraphics[width=15cm]{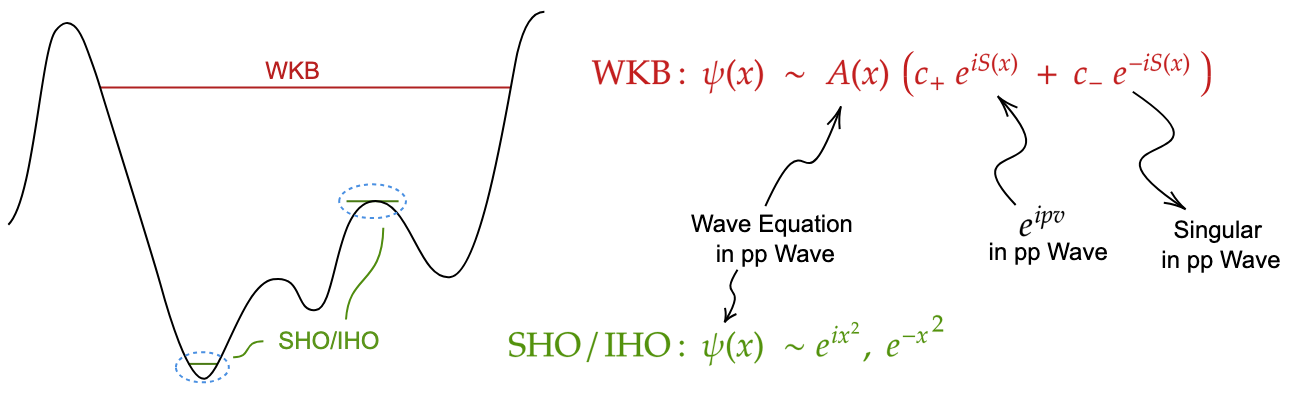}
    \caption{The two asymptotic approximations of a Schrodinger-like 2\ts{nd}-order ODE are \textbf{either} WKB (labeled in red) or the SHO/IHO (labeled in green). Note that WKB really refers to a first-order turning point WKB approximation, whereas the SHO/IHO refers to a second-order turning point WKB approximation. In the figure, we denote which factors in the asymptotic approximation are produced by the wave equation in the pp-wave.}
    \label{fig:WKBvsSHO}
\end{figure}

Approximating the phase \eqref{eq:HJfun} using the wave equation involves first-order WKB for the some of the terms, and modified WKB based on second-order turning points (SHO/IHO) for the other terms. Below, we discuss differences in how the Penrose limit geometrizes these two cases. The discussion is summarized in figure \ref{fig:WKBvsSHO}.

\textbf{IHO/SHO:} When we reproduce the near-ring wavefunctions \eqref{eq:PCF} using the Penrose limit, we use the geodesic with  $p_r=0$ identically for all times (rather than one with $p_r$ vanishingly small in the far past but $O(1)$ later). For that geodesic
\begin{equation} \label{eq:HJ equatorial}
    S=(-t + b_0 \phi )=\varepsilon^2 v 
\end{equation}
so that the factor of $e^{ipv}$ in the plane wave geometry reproduces the first two terms of \eqref{eq:HJfun} with $p=\varepsilon^2\omega$ held fixed. The last term of \eqref{eq:HJfun} is then obtained using the ground state of the inverted harmonic oscillator in the unstable direction. As far as the Penrose limit is concerned, when there is a second-order turning point, the phase \eqref{eq:Phase} does not contain dependence on the unstable direction. Instead, the IHO in the pp-wave wave equation  reproduces the rest of the geometric optics wavefunction, including the leading phase and overtones.

\begin{figure}[h]
\includegraphics[scale=.35]{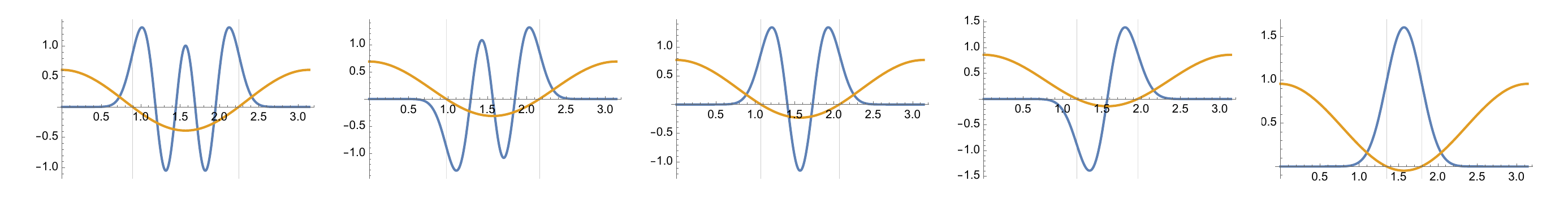}\caption{Plots of $P_l^m(\cos \theta)$ (in blue) and the corresponding angular potential (in orange) for $l=20$ and $m=16,\dots 20$ with the classical turning points marked.}\label{fig:PolarPotential}
\end{figure}

Something similar happens for the polar coordinates in the special case of the equatorial photon orbit.
In section \ref{Sec:pp_wave_eq} we found that the behavior of the $P_l^l(\cos\theta)$ was well-approximated by the Gaussian ground state of the harmonic oscillator associated to the angular transverse direction. In particular, there was no oscillatory phase in the solution, only damping away from the equator.
This behavior is highly non-generic, and arises precisely because the ordinary WKB approximation of the associated Legendre functions breaks down when $m\sim l$. The angular potential for the polar motion has turning points when $\sin \theta \sim \frac{m}{l+\frac12}$. When $m\ll l$ these turning points are well-separated, and the usual WKB matching can be imposed separately on both. However, as $m\to l$, the turning points come together near the bottom of the potential (illustrated in figure \ref{fig:PolarPotential}), and the ordinary WKB matching has to be imposed simultaneously. This amounts to approximating the turning points with the harmonic oscillator potential, and it is that SHO which shows up in the Penrose limit of the equatorial photon ring. The third term in \eqref{eq:HJfun} is zero, and the nontrivial angular dependence comes from the harmonic oscillator of the pp-wave, as was the case for the radial direction.

\textbf{1\ts{st}-order WKB:} The harmonic oscillator approximation for the polar motion is valid when $m \sim l- c$, with $c$ an $O(1)$ constant, such that
\begin{equation}
    \mu \equiv \frac{m}{l+\frac{1}{2}} \to 1 \;, \quad \textrm{as } m\sim l \to \infty \; .
\end{equation}
For more generic scaling limits with $m \sim c \cdot l$ and $c \in (-1,1)$ 
\begin{equation}
    \mu \to c \;, \quad \textrm{as } m\sim l \to \infty \; ,
\end{equation}
and the turning points of the polar motion are sufficiently well-separated. In this case the harmonic oscillator approximation breaks down, but it can be replaced by the ordinary WKB approximation with first-order turning points. The polar dependence of the wavefunction is then controlled by the Hamilton-Jacobi function at leading order in the WKB approximation, with a subleading amplitude correction. One might expect the Penrose limit to geometrize this approximation, with the polar dependence arising from the $e^{i p v}$ factor as in \eqref{eq:Phase}. 

\begin{figure}[h]
    \centering
    \includegraphics[width=10cm]{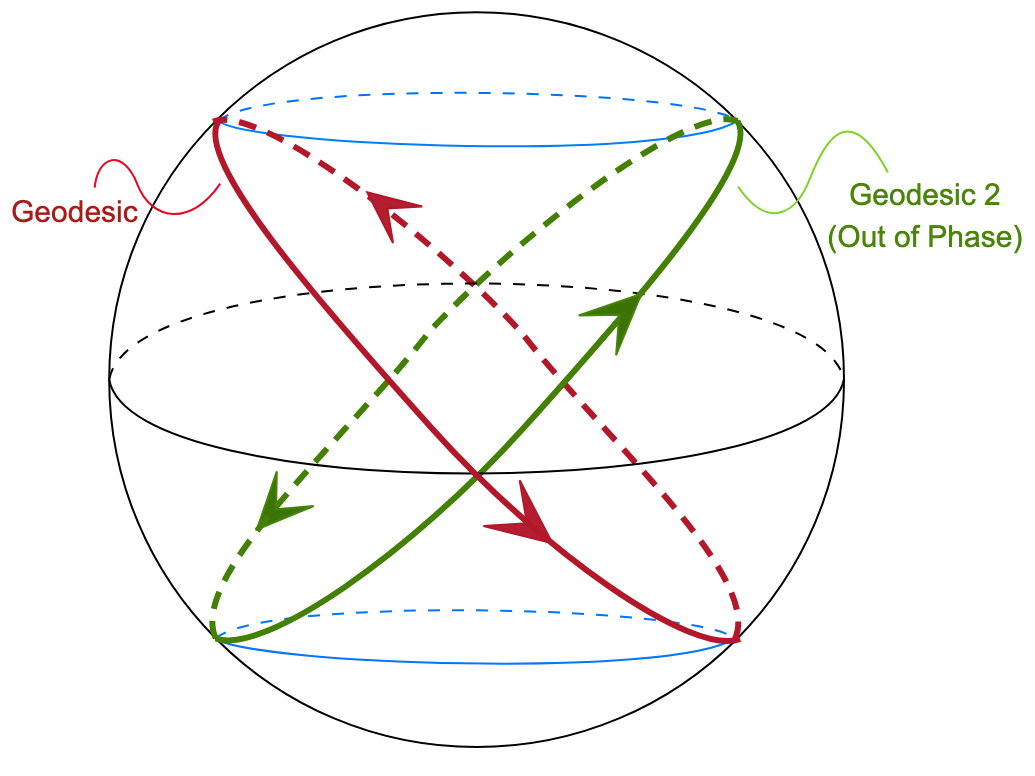}
    \caption{The geodesic in red is the tilted photon ring geodesic isolated by the Penrose limit. The green geodesic has polar motion that is exactly $\pi$ out of phase with the red geodesic -- related by $p_\theta \to -p_\theta$.}
    \label{fig:Tilted PR 2 geodesics}
\end{figure}

However, there is an important difference between the degenerate WKB approximation (involving a higher-order turning point) and the usual first-order turning point problem. When the turning points are first-order, the WKB approximation involves a solution of the form 
\begin{equation}\label{eq:WKBvsGeo}
\psi_{\text{WKB}}(\theta) \propto c_+\exp\left(+i\int^\theta_{\theta_0} p_\theta \; d\theta'\right) +c_-\exp\left(-i\int^\theta_{\theta_0} p_\theta \; d\theta'\right) \equiv c_+ e^{iS_+(\theta)}+c_-e^{iS_-(\theta)}\; ,
\end{equation}
and generically both terms are required in order to satisfy the appropriate boundary conditions. On the other hand, the geometric optics approximation (directly related to the Penrose limit through \eqref{eq:Phase}) takes an ansatz of the form
\begin{equation}\label{eq:geoVSwkb}
    \psi_{\text{Geodesic}, \pm} (\theta) \propto \exp\left(\pm i\int^\theta_{\theta_0} p_\theta \; d\theta'\right) 
\end{equation}
with a fixed sign. In other words, the solution \eqref{eq:geoVSwkb} is associated to a single geodesic when we pick a definite sign, while \eqref{eq:WKBvsGeo} is really associated to two separate geodesics corresponding to the two possible signs in \eqref{eq:HJfun}. This scenario is illustrated in figure \ref{fig:Tilted PR 2 geodesics}. Both \eqref{eq:WKBvsGeo} and \eqref{eq:geoVSwkb} can solve the wave equation locally, but they will generically satisfy different boundary conditions (or have different regularity properties in the case of polar motion).

This is of course not an issue for the standard WKB approach to calculating QNMs. Since the wave equations separate, it is possible to independently solve for the angular eigenfunctions (either exactly or using WKB) and their separation constants. These are then fed into the radial equation, which is based on a single geodesic (since it involves a second-order turning point). The Penrose limit instead attempts to do both approximations simultaneously. This could be viewed as an advantage, since the procedure does not require separability of the wave equation, but it does present certain complications. 

The basic issue is easy to explain in physical terms. When  the Penrose limit scales into a particular congruence, all geodesics not aligned with the congruence become infinitely blue-shifted by the singular scaling. The geometric optics wavefunctions based on those singular geodesics also do not have sensible scaling limits.  So if we take the Penrose limit using $S_+$ (so that the geodesic has positive $p_\theta$ at $\theta_0$) in order to reproduce the first phase in the angular WKB approximation \eqref{eq:WKBvsGeo}, then the geodesic associated to $S_-$ (with negative $p_\theta$ at $\theta_0$) will be singular. The factor of $e^{ipv}$ in the Penrose limit spacetime will therefore correctly reproduce the first term in \eqref{eq:WKBvsGeo} but not the second. In standard WKB both phases are necessary in order to determine the quantization condition for the angular separation constants, so this seems like a significant drawback.\footnote{The $t,\phi$ factors in the phase never require WKB approximation since they are protected by symmetry (i.e. the geometric optics approximation \eqref{eq:HJ equatorial} is exact for those coordinates).} However, both phases share the same amplitude correction, and as we will see, the Penrose limit of the titled photon ring does correctly reproduce it.

In the remainder of this section, we will discuss two different ways of taking the Penrose limit into the tilted photon ring. The first method, based on the $SO(3)$ symmetry of the problem, exhibits the SHO approximation to the polar wavefunction. It amounts to picking new coordinates $(\tilde{\theta},\tilde{\phi})$ for which $p_{\tilde{\theta}}=0$ on the tilted geodesic. The second method implements the standard Penrose limit in adapted coordinates and exhibits the first-order WKB approximation to the polar wavefunction. As we will explain, the first method singles out a particular linear combination of degenerate QNMs whose support is concentrated on the tilted photon ring, while the second method instead approximates individual spherical harmonics. The nontrivial difference in the polar dependence of the wavefunctions is encoded by the different relationships between the pp-wave coordinates and the Schwarzschild coordinates.

\subsection{Tilted Photon Ring Using $SO(3)$}\label{subsec:so(3)}

For the polar wavefunction, the difficulty in reproducing both terms in \eqref{eq:WKBvsGeo} from the Penrose limit arises because the support of a generic spherical harmonic is not concentrated along a single great circle. The generic $Y_l^m(\theta,\phi)$ is oscillatory in the region $\theta \in [\theta_+,\pi-\theta_+]$,  while only $Y_l^l(\theta,\phi)$ concentrates on a single geodesic (the equatorial photon ring). Since the Penrose limit only has access to a single geodesic, it can only faithfully reproduce $Y_l^l(\theta,\phi)$. Of course, because Schwarzschild has $SO(3)$ symmetry, all of the photon ring geodesics are related by rotations, and it is therefore possible to construct a particular linear combination of spherical harmonics that concentrates on any given photon ring trajectory. As discussed in
appendix \ref{App:tilted1}, if one performs a rotation to new angular coordinates $(\tilde{\theta},\tilde{\phi})$ so that the trajectory is equatorial (at fixed $\tilde{\theta}=\frac{\pi}{2})$ then this linear combination is given by $Y_l^l(\tilde{\theta}, \tilde{\phi})$, 
where $(\tilde{\theta}, \tilde{\phi})$ are given by \eqref{eq:rodriguez_rot}. 
This function can be expressed in terms of the original coordinates $(\theta, \phi)$ via the Wigner $D$ matrix\footnote{In Mathematica, the conventions for the Wigner $D$ matrix are WignerD$[\{l,-m,-m'\}$,$\alpha, \beta, \gamma$] for $D_{mm'}^{(l)}(\mathcal{R}(\alpha,\beta,\gamma))$, where $\alpha,\beta,\gamma$ are the Euler angles of the rotation matrix $\mathcal{R}$.}
\begin{equation} \label{eq:wignerD}
    Y_l^l(\tilde{\theta}, \tilde{\phi}) = \sum_{m = -l}^{l} [D^{(l)}_{l m} (\mathcal{R})]^* Y_{l}^m(\theta, \phi) \; ,
\end{equation}
where $\mathcal{R}$ is the matrix of rotation around the $x$-axis by the angle $\frac{\pi}{2} - \theta_L$ 
\begin{equation}
    \mathcal{R}_x(\pi/2-\theta_L) \; .
\end{equation}
Note that $\theta_L$ in the language of appendix \ref{App:SCHpenrose} corresponds to $\theta_+$, the turning point of the polar motion. $Y_l^l(\tilde{\theta}, \tilde{\phi})$ is a Gaussian peaked along the tilted geodesic given by \eqref{eq:angMom}.  Since the linear combination is composed of spherical harmonics with fixed $\ell$, and since the eikonal QNM spectrum is independent of the azimuthal number $m$, \eqref{eq:wignerD} can be combined with the appropriate phase factors in order to produce a QNM wavefunction with a definite frequency peaked about the photon orbit. The solution will not have a definite Hamilton-Jacobi function in the $(\theta,\phi)$ coordinates since it involves a superposition of waves with different angular momenta $L_z$ (as is evident in \eqref{eq:wignerD}), but in the $(\tilde{\theta}, \tilde{\phi})$ coordinates it has $p_{\tilde{\theta}}=0$ and $p_{\tilde{\phi}}=l$ (with $p_r=0$ and $p_t=-\omega_l$) and is nonsingular in the  tilted Penrose limit described in appendix \ref{App:tilted1}. This approach relies on the high degree of symmetry and does not generalize to Kerr, where the QNM spectrum depends on $m$ and the trajectories are not related by symmetry transformations.

In order to reproduce a wavefunction like \eqref{eq:WKBvsGeo} with definite angular momentum $m$ in the standard coordinates and nonzero $p_\theta$, we instead need to perform the Penrose limit in adapted coordinates for which $V=S(t,r,\theta,\phi)$. This limit is described in appendix \ref{App:tilted2}. As we will see, the factor of $e^{ipv}$  will only reproduce the first term in \eqref{eq:WKBvsGeo}, but we will be able to reproduce the amplitude correction using the pp-wave wave equation.
In order to make the comparison we first need to review the WKB approximation of the associated Legendre functions.

\subsection{WKB for Spherical Harmonics}
\begin{figure}[h]
\includegraphics[scale=.39]{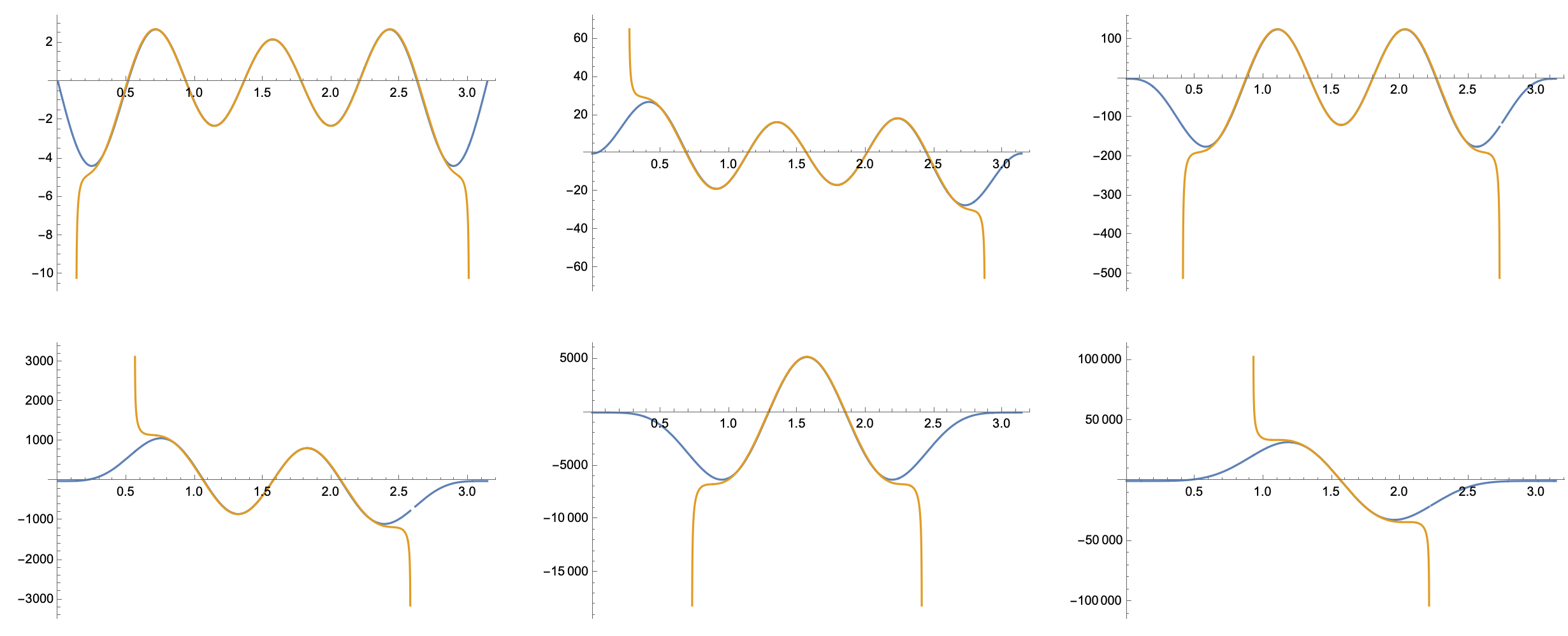}
\caption{Plots of $P_l^m(\cos(\theta))$ (in blue) and the WKB approximation to $S_{lm}(\theta)$ (in orange) for $l=7$ and $m=1,\dots, 6$.}\label{fig:Legendre_WKB}
\end{figure}

In this section, we review the construction of the WKB approximation of the Legendre polynomials \cite{landauer}.
 The wavefunction in the Schwarzschild geometry is 
\begin{equation}
    \Psi(t,r,\theta,\phi)=e^{-i\omega t}e^{im\phi}S_{lm\omega}(\theta)R(r) \; .
\end{equation}
Because the black hole is not spinning the frequency does not enter into the angular equation 
\begin{align}
    \br{\frac{1}{\sin{\theta}}\pd_\theta\pa{\sin{\theta}\pd_\theta} -\frac{m^2}{\sin^2\theta}+A_{\ell m\omega}}S_{\ell m\omega}(\theta)=0\;.
\end{align}
This is the associated Legendre ODE and it can be solved exactly in terms of the associated Legendre functions $P_l^m(\cos \theta)$. In particular the separation constants are given by $A_{lm\omega}=l(l+1)$ (in Kerr they will also depend on $m,\omega$). We would like to rederive these statements in the WKB approximation. To do this, we change variables $e^y=\tan{\frac{\theta}{2}}$, so that $\pd_y=\sin{\theta}\pd_\theta$. The angular equation is 
\begin{align}
    \br{ \pd_y^2  -m^2+A_{lm} \; \text{sech}^2y }S_{\ell m\omega}(y)=0\;.
\end{align}
$S_{lm\omega}(y)$ should vanish at the two boundaries $y=\pm \infty$ in order to assure regularity at the poles. This is a Schrodinger problem with a potential
\begin{equation}
    V_\theta(y)=m^2-A_{lm} \; \text{sech}^2y  \;.
\end{equation}
To perform the WKB approximation we assume $A_{lm}\sim m^2\gg1$ with  $\mu=\frac{m}{l+\frac{1}{2}}\in (-1,1)$ fixed.
The potential has a minimum at the equator $y=0$ with
\begin{equation}
    \frac12 V_{\theta}''(0)=A_{lm} \; , \qquad V_{\theta}(0)=m^2-A_{lm} \; .
\end{equation}
The zeroes of the potential, corresponding to classical turning points, occur at $\theta_+, \pi-\theta_+$ with
\begin{equation}
    \sin^2 \theta_+ = \frac{m^2}{A_{lm}} \;. 
\end{equation}
These are also the classical turning points for a great-circle geodesic on the 2-sphere with total angular momentum $L^2=A_{lm}$. Outside of these turning points the WKB approximation produces an exponentially decaying function
\begin{equation}
    S_{lm}(\theta)\sim \frac{1}{  V_\theta^{1/4}}\exp\left[-\int^\theta_{\pi-\theta_+}  \sqrt{m^2/\sin^2\theta'-A_{lm}}\; d\theta'\right] \; .
\end{equation}
In the region near the minimum of the potential we use the WKB approximation to write
\begin{equation}\label{eq:LegendrePhase}
    S_{lm }(\theta)\sim \frac{1}{ (-V_\theta)^{1/4}}\left[c_1e^{i\int_0^\theta \sqrt{-V_\theta}\frac{d\theta'}{\sin\theta'}}+c_2e^{-i\int_0^\theta \sqrt{-V_\theta}\frac{d\theta'}{\sin \theta'}}\right] \; .
\end{equation}
Near the turning points one uses the Airy approximation and obtains a matching condition \cite{landauer}
\begin{equation}
    \int_{\pi-\theta_+}^{\theta_+} d\theta \sqrt{A_{lm}-m^2/\sin^2\theta}=\pi\left(l+\frac12\right) \; ,
\end{equation}
for $l \in \mathbb{Z}$ . The integral can be evaluated explicitly and is equal to $\sqrt{A_{lm}}\pi$. One therefore concludes that
\begin{equation}
    A_{lm}=\left(l+\frac12\right)^2
\end{equation}
as expected, provided that $l>m$. The matching conditions also determine the relative coefficients in the oscillatory region \eqref{eq:LegendrePhase}, such that
\begin{equation} \label{eq:legendre_WKB}
    S_{lm}(\theta)=\sqrt{\frac{4(l+m)!}{\pi (2l+1)(l-m)!}} \frac{1}{\left(\sin^2\theta-\frac{m^2}{A_{lm}} \right)^{1/4}}\cos\left(\int_{\theta_+}^\theta \sqrt{A_{lm}-m^2/\sin^2\theta'}d\theta'-\frac{\pi}{4}\right) \; .
\end{equation}
The integral in the phase can be performed explicitly and is given by
\begin{equation}
    \pa{l+\frac{1}{2}} \arccos\left(\frac{\cos (\theta )}{\sqrt{1-\sin^2(\theta_+)}}\right)+m \arctan\left(\frac{\sin(\theta_+) \cos (\theta )}{\sqrt{\sin^2\theta -\sin^2(\theta_+)}}\right) + m \frac{\pi}{2}\;,
\end{equation}
with
\begin{equation}
    \sin(\theta_+) = \frac{m}{A_{lm}} = \frac{m}{l+\frac{1}{2}}\;.
\end{equation}
As is evident in figure \ref{fig:Legendre_WKB}, this approximation is very good even for small values of $l$. Note that the phase exactly matches the polar dependence of \eqref{eq:HJ tilted}, if we identify $\theta_+=\theta_L$, $l+\frac{1}{2}=L$, which gives $m=\sin(\theta_+) \,\l(l+\frac{1}{2}\r)=\sin(\theta_L) \,L=L_z$.

\subsection{General Penrose Limit}
In section \ref{subsec:so(3)}, we considered the Penrose limit of the geodesic with $L_z = \sin(\theta_L) L$ using special coordinates that relied on the $SO(3)$ symmetry of Schwarzschild. We now consider the Penrose limit in the standard coordinate system, discussed in detail in appendix \ref{App:tilted2}. The first step is to switch to adapted coordinates $(U,V,R,\Psi)$ defined by \eqref{eq:tilted_adapted}.  One then performs the scaling $(U,V,R,\Psi)=(u,\varepsilon^2v,\varepsilon \rho, \varepsilon \psi)$ so that 
\begin{equation} \label{eq:adapted_penrose}
    \begin{split}
        & t = 3 u \; , \\
        & r = 3M + \varepsilon \frac{\rho}{\sqrt{3}} \; , \\
        & \theta = \arccos \br{-\cos(\theta_L) \sin \pa{ \beta(u - \varepsilon \psi/\sqrt{3})}} \; ,\\
        & \phi = \frac{\beta}{\sin(\theta_L)} \l( \frac{1}{3} \varepsilon^2 v + \varepsilon \frac{\psi}{\sqrt{3}} \r) + \arctan \br{\sin(\theta_L) \tan \pa{ \beta (u-\varepsilon \psi/\sqrt{3})} } \; .
    \end{split}
\end{equation}
Scaling the metric by $\varepsilon^{-2}$ and taking $\varepsilon \to 0$ gives the metric
\begin{equation}\label{eq:tiltMetric}
    ds_{\mathrm{pp}}^2 = 2 du dv + \beta^2 \rho^2 du^2 + d\rho^2 + \cot^2(\theta_L) \cos^2(\beta u) d\psi^2 \;, \qquad \beta = \frac{b_0}{r_0^2} = \frac{1}{\sqrt{3} M} \; .
\end{equation}
Next we solve the scalar wave equation in this background. Since $\partial_v$ and $\partial_\psi$ are isometries, we take the ansatz
\begin{equation}
    f = e^{i p v} e^{i K \psi} X(\rho) F(u) \; .
\end{equation}
The wave equation then takes the form
\begin{equation}
    F(u) \left[ X''(\rho) + \beta^2 p^2 \rho^2 X(\rho) - i \beta p \tan(\beta u) X(\rho) - K^2 \tan^2(\theta_L) \sec^2(\beta u) X(\rho) \right] + 2 i p F'(u) X(\rho) = 0 \; . 
\end{equation}
This equation can be separated by assuming
\begin{equation}
    X''(\rho) + \beta^2 p^2 \rho^2 X(\rho) = C X(\rho) \; ,  
\end{equation}
which  is the familiar IHO equation for the radial direction. As discussed in section \ref{Sec:pp_wave_eq}, the correct boundary conditions are satisfied with $C$ such that
\begin{equation}
    X''(\rho) + \beta^2 p^2 \rho^2 X(\rho) = 2 i \beta p \left( n+ \frac{1}{2} \right) X(\rho) \; . 
\end{equation}
In that case $F(u)$ satisfies the first-order ODE
\begin{equation}
    F'(u) + F(u) \left[ \beta \left( n+ \frac{1}{2} \right) - \frac{1}{2} \beta \tan(\beta u) + i \frac{K^2}{2p} \tan^2(\theta_L) \sec^2(\beta u)\right] = 0 \; 
\end{equation}
which  can be integrated to give
\begin{equation}
    F(u) = e^{-\l(n+\frac{1}{2}\r) \beta u } \frac{1}{\sqrt{\cos(\beta u)}} \exp \br{ - i \frac{K^2}{2 \beta p} \tan^2(\theta_L) \tan(\beta u)} \; . 
\end{equation}
The full wavefunction is then given by
\begin{equation}\label{eq:fullwave}
    f = e^{i p v} X_n(\rho) e^{-\l(n+\frac{1}{2}\r) \beta u } \frac{1}{\sqrt{\cos(\beta u)}} G_K(u,\psi) 
\end{equation}
where
\begin{equation} \label{eq:Gk}
    G_K(u,\psi) = \exp \br{i K \l(\psi - \frac{K}{2 \beta p} \tan^2(\theta_L) \tan(\beta u) \r)} 
\end{equation}
and the $X_n(\rho)$ are the IHO wavefunctions as in section \ref{Sec:pp_wave_eq}.

\subsection{Relation to Schwarzschild Wavefunctions}\label{sec:tiltedWave}
The WKB mode solutions of the Schwarzschild wave equation are
\begin{equation}
    \Psi(x) = e^{-i \omega t} e^{i m \phi} R(r)S_{lm}(\theta) \; ,
\end{equation}
where the $S_{lm}(\theta)$ are given by \eqref{eq:legendre_WKB}. The near-ring approximation to the radial wavefunction is reproduced by $X_n(\rho)$ as in the equatorial Penrose limit, so we suppress the radial dependence in what follows. Separating out the two phase factors in $S_{lm}(\theta)$, we have
\begin{equation} \label{eq:2 phases}
    \begin{split}
        \Psi(x) &= A_{lm}(\theta) \left[ \exp \pa{-i \omega t + i m \phi + i \zeta_{lm}(\theta)} +  \exp \pa{-i \omega t + i m \phi - i \zeta_{lm}(\theta)} \right] \\
        &:= \Psi_+(x) + \Psi_-(x) \; ,
    \end{split}
\end{equation}
where $\zeta_{lm}(\theta)$ is the phase in \eqref{eq:legendre_WKB}. With the identifications
\begin{equation}
    \omega = (l+1/2) \Omega_R \;, \qquad m = \sin(\theta_L) (l+1/2) \; ,
\end{equation}
the phase of $\Psi_+$ is reproduced by the factor $e^{ipv}$ in the pp-wave
\begin{equation} \label{eq:single phase}
    \exp \pa{-i \omega t + i m \phi + i \zeta_{lm}(\theta)}= \exp(i \omega S(x)) = \exp( i \omega  \varepsilon^2 v) \to \exp(i p v)
\end{equation}
where $p = \omega \varepsilon^2 $ is held fixed in the limit.  
The amplitude correction in \eqref{eq:legendre_WKB} takes the form
\begin{equation}
    A_{lm}(\theta) = \frac{1}{(\sin^2\theta-\sin^2\theta_L )^{1/4}} 
    = \frac{1}{(\cos^2\theta_L-\cos^2\theta )^{1/4}} \; .
\end{equation}
Making the coordinate transformation \eqref{eq:adapted_penrose} and sending $\varepsilon\to 0$ we get
\begin{equation}
\begin{split}
    A_{lm}(\theta)
    &= \frac{1}{(\cos^2\theta_L-\cos^2\theta_L \sin^2(\beta (u-\varepsilon \psi/\sqrt{3}) )^{1/4}}\\
    &\to \frac{1}{(\cos^2\theta_L-\cos^2\theta_L \sin^2(\beta u) )^{1/4}}\\
    &= \frac{1}{\sqrt{\cos(\theta_L) \cos(\beta u)}} \; .
\end{split}
\end{equation}
So the Penrose limit of $\Psi_+(x)$ is
\begin{equation}
    \Psi_+(x) \to \frac{1}{\sqrt{\cos(\beta u)}} e^{i p v}
\end{equation}
which is reproduced by \eqref{eq:fullwave}.

On the other hand, $\Psi_-$ corresponds to the second (green) geodesic in figure \ref{fig:Tilted PR 2 geodesics}. As discussed below that figure, this geodesic is infinitely blueshifted when we take the Penrose limit onto the $S_+$ geodesic. $\Psi_-$ is therefore not simply related to quantities in the pp-wave spacetime.

Finally, the factor of $e^{-\l(n+\frac{1}{2}\r) \beta u}$ in \eqref{eq:fullwave} reproduces the Schwarzschild eikonal overtones  ${\omega \supset -i\l(n+\frac{1}{2}\r)\gamma_L}$, which are associated to the excited states $X_n(\rho)$ of the radial IHO.

So far we have accounted for the first four factors of \eqref{eq:fullwave} in terms of the scaling limit of eikonal QNM wavefunctions in the full Schwarzschild geometry. To recap: if we take a wavefunction with the conserved quantities $\omega,l,m$ and overtone number $n$, and we take the Penrose limit using the geodesic with conserved quantities $E = \omega_R, \, L=(l+1/2), \,  L_z=m$, then the wavefunction scales onto
\begin{equation}
    \Psi^{\omega,l,m,n}_+ \to e^{i p v} X_n(\rho) e^{-\l(n+\frac{1}{2}\r) \beta u } \frac{1}{\sqrt{\cos(\beta u)}} \;,
\end{equation}
i.e. \eqref{eq:fullwave} with $K=0$. 

When we took the Penrose limit of the equatorial photon ring, we found that solutions to the wave equation in the pp-wave background reproduced not only the exact geometric optics wavefunction, but also solutions that were ``close by.'' In particular, descendants of the IHO/SHO groundstates in the pp-wave lifted to solutions with $l-m\neq0$ (for the SHO) and overtone number $n\neq0$ (for the IHO). In general, the Penrose limit is expected to reproduce more than just the wavefunction $\Psi^{\omega,l,m,n}_+$. Scaling limits of wavefunctions with nearby parameters should also be sensible  and will correspond to ``excited'' states in the pp-wave. In the tilted pp-wave this is encoded in the remaining factor $G_K(u,\psi)$ in \eqref{eq:fullwave}.  In particular, if we take the the scaling limit of the wavefunction with $m'=m+1$ (equivalently $\sin(\theta_+')= \sin(\theta_+) + \frac{1}{l+\frac{1}{2}}$) using the Penrose limit of the geodesic with $L_z=m$, then we land on \eqref{eq:fullwave} with $G_K(u, \psi)$ turned on (i.e. $K \neq 0$).

\section{Highly-Damped Quasinormal Modes }\label{sec:highDamp}
The eikonal QNM spectrum describes the asymptotics of the black hole resonances at large real frequency. This approximation only captures the subset of QNMs whose real parts dominate their imaginary parts. 
Although the eikonal approximation
breaks down for QNMs with large imaginary parts,  there turns out to be a second asymptotic regime of the Schwarzschild QNM spectrum that can also be understood analytically. Rather than considering long-lived waves whose oscillation timescale is short compared to the curvature scale, one can instead take the highly-damped limit $|\omega_I| \to \infty$ onto waves with short decay timescales. This part of the spectrum also turns out to be universal. For scalars in Schwarzschild, it takes the form \cite{Hod:1998vk,Motl:2002hd,motl2003}
\begin{equation} \label{eq:highly_damped_qnm}
    \omega = \frac{\kappa}{2\pi} \log(3) - i \kappa \l( n + \frac{1}{2} \r)  \;, \qquad  \mathrm{as } \quad n \to \infty,
\end{equation}
where $\kappa = (4 M)^{-1}$ is the surface gravity of the horizon.
This formula has two important features which we will later interpret geometrically. First, the whole expression is independent of  angular momentum: both the $O(n)$ piece and the $O(1)$ piece are $l$-independent, and the angular momentum only shows up at sub-subleading order $O(n^{-1})$. Second, the imaginary parts are integer spaced in the Matsubara frequency. The fact that the overtones are equally spaced suggests a hidden symmetry of the spectrum, which can be realized geometrically in the Penrose limit as in section \ref{sec:plane wave}. 

\subsection{Review: Highly Damped Modes from Monodromy}\label{eq:dampedReview}
The approximate form of \eqref{eq:highly_damped_qnm}  was first noticed numerically \cite{Leaver:1985ax,Nollert:1993zz,Andersson_1993b} and later derived analytically \cite{Motl:2002hd,motl2003,Andersson_2004}.
For our purposes, the clearest derivation uses monodromy methods, which are reviewed in appendix \ref{App:monodromy}. The result of the analysis is a constraint on the highly damped QNM frequencies 
\begin{equation}\label{eq:Highly-damped condition}
    e^{2 \pi \omega /\kappa} = -3 \;.
\end{equation}
As reviewed in appendix \ref{App:monodromy}, this relation arises from comparing the monodromy of the radial QNM wavefunction around two paths in the complex $r$-plane. The first path encircles the event horizon and is controlled by the infalling boundary condition. The second contour is deformed away from the horizon and is controlled by the outgoing boundary condition at spatial infinity and the properties of the radial potential near the singularity. 
The condition \eqref{eq:Highly-damped condition} puts a constraint on the QNM frequency but does not fix it uniquely, since
\begin{equation}
    e^{2 \pi \omega /\kappa} = -3 \implies \omega = \left( \frac{\kappa}{2\pi} \log(3) - i \frac{\kappa}{2} \right) - i n \kappa  \;,
\end{equation}
where $n\geq 0$ because the derivation of \eqref{eq:Highly-damped condition} assumes that $\mathrm{Im}(\omega) < 0$. In particular, as discussed in appendix \ref{App:monodromy}, 
if we have an infalling solution at the horizon, then the shift in the frequency
\begin{equation}
    \omega \to \omega - i n \kappa
\end{equation}
 coupled with the change in the radial wavefunction leads to a new solution with  the same monodromy around the horizon.\footnote{To correctly compute the monodromy of the full QNM wavefunction one must switch to outgoing coordinates, or equivalently one can allow for complex $t$ so that $e^{-i\omega t}$ also has monodromy around the horizon} This in turn is sufficient to ensure that the new wavefunction is still a QNM in the limit of high-damping. In other words, if
\begin{equation}
    \Psi_0 =e^{-i \omega_0 t} \psi_{\omega_0}(r)
\end{equation}
is a solution that is infalling at the horizon, with a certain monodromy there, then 
\begin{equation} \label{eq:overtones_intro}
    \Psi_n = e^{-i \omega_0 t} e^{-n \kappa t} \psi_{\omega_0 - i n \kappa}(r)
\end{equation}
will also be an infalling solution with the same monodromy around the horizon. In general, $\psi_{\omega_0 - i n \kappa}(r)$ is not related  to $\psi_{\omega_0}(r)$ in a simple way throughout the spacetime. However, as we will show in section \ref{Sec:near-horizon wave eqn}, the highly damped overtones turn out to be related by an emergent symmetry of the near-horizon region, in perfect analogy with the IHO symmetry of the eikonal spectrum discussed in section \ref{subsec:near-ring dynamical symmetry}.  

In fact, since all QNMs behave as $\psi_\omega(r) \sim e^{-i \omega r_*}$ near the horizon, \eqref{eq:overtones_intro} takes the asymptotic form
\begin{equation}\label{eq:overtones}
    \Psi_n \sim e^{-n \kappa (t+r_*)} e^{-i \omega_0 t} \psi_{\omega_0}(r) = \l( e^{-\kappa v_S} \r)^n \Psi_0 \;, \qquad \textrm{as } r \to 2M \;,
\end{equation}
where $v_S$ is the ingoing Eddington-Finkelstein coordinate. This formula obviously resembles \eqref{eq:geometricOvertones}, and in this section we will try to make this analogy precise. For now, note that the geometric optics construction in section \ref{subsec:overtoneSym} generated overtones by multiplying the seed solution by a function of the quantity conserved along the underlying geodesic congruence. The analogous function in \eqref{eq:overtones} appears to be an analytic function $f(e^{-\kappa v_S})$ whose Taylor expansion corresponds to the individual overtones. Of course, $v_S$ is a constant precisely for radially infalling geodesics, and the presence of $\kappa$ in the formula suggests a geodesic close to the horizon. We are therefore led to consider a geodesic congruence based on the generators of the black hole's past horizon, as will be explained in section \ref{subsec:horizon geo_optics}.

\subsection{Wave Equation in the Near-Horizon Region} \label{Sec:near-horizon wave eqn}
As explained in appendix \ref{App:monodromy}, the conditions that single out the highly-damped QNMs \eqref{eq:highly_damped_qnm} can all be phrased using only the near-horizon region. One looks for solutions with large $\omega_I$ which are purely infalling at the horizon and which have a specific monodromy about $r=2M$. This monodromy condition is equivalent to the requirement that the mode is outgoing at spatial infinity. Importantly, the monodromy condition only involves geometric quantities associated to the near-horizon region.\footnote{This is no longer true in Reissner-Nordstrom and Kerr, where the monodromy condition involves the surface gravity of the inner horizon and the outer horizon.} 

\begin{figure}
    \centering
    \subfloat[\centering The radial wave potential \eqref{eq:radial_potential} as a function of $r_*$. $r_* \to -\infty$ as $r\to 2M$, and $r_* \to \infty$ as $r \to \infty$. We zoom into the near-horizon region, circled in red.]{{\includegraphics[width=0.45\textwidth]{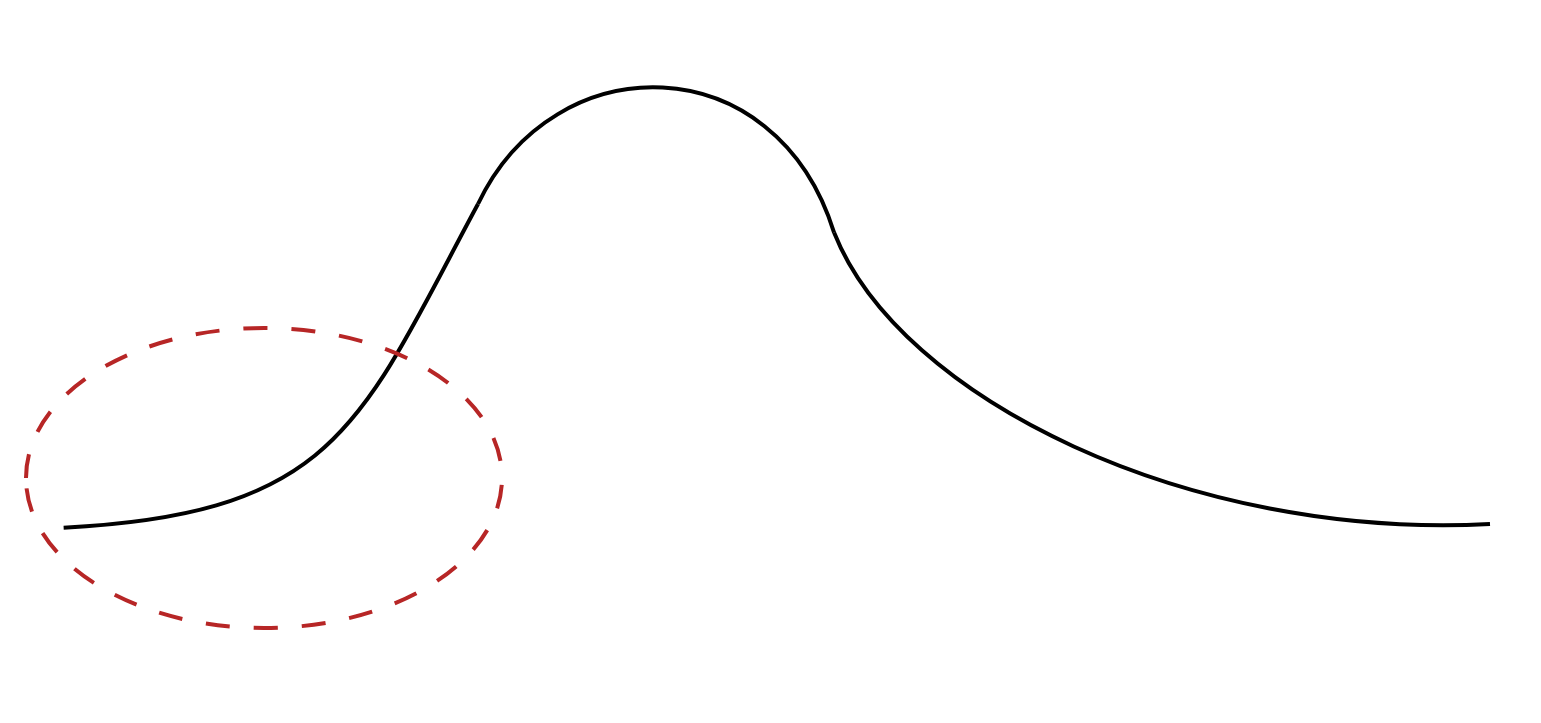} }}
    \quad
    \subfloat[In the near-horizon region, the potential looks exponential, and the wave equation reduces to the massive Rindler wave equation.]{{\includegraphics[width=0.40\textwidth]{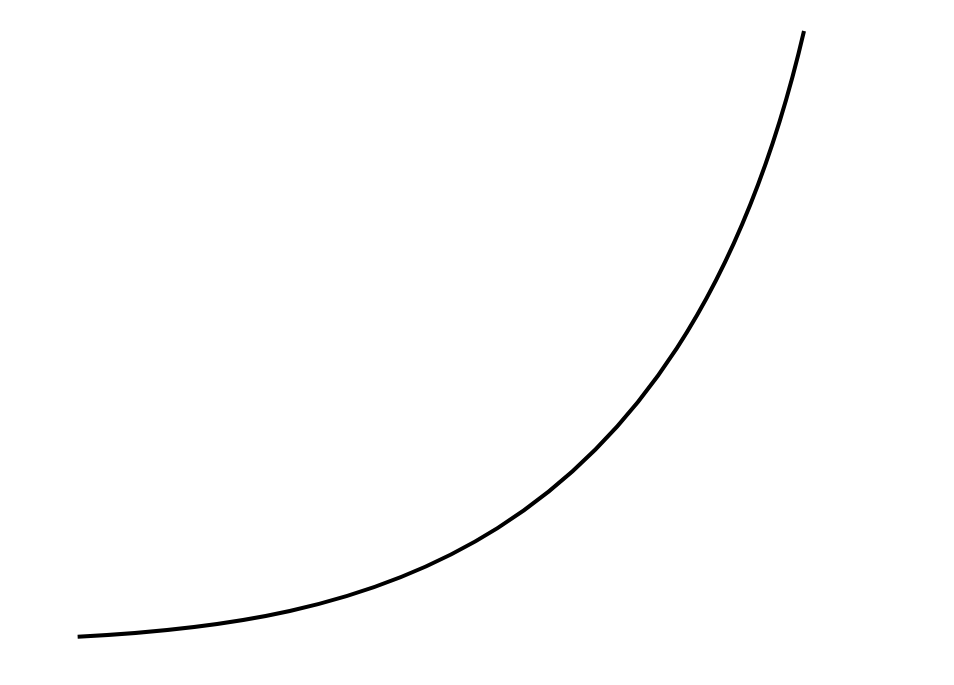} }}
    \caption{Near-horizon limit of the radial potential.}
    \label{fig:radial_pot}
\end{figure}

 This suggests that the analysis of the highly damped modes can be localized near the horizon in the same way that the analysis of the eikonal spectrum can be localized near the photon ring. In other words, in the limit of high-damping, we can solve the wave equation locally in the near-horizon region with infalling boundary conditions. Demanding a specific monodromy then restricts the allowed frequencies and guarantees that the local solution connects to a global solution that is purely outgoing at infinity. 
Near the horizon, we have
\begin{equation}
    r_* = r   + 2M \log\l(\frac{r}{2M}-1 \r) \approx 2M+ 2M \log\l(\frac{r}{2M}-1 \r) \;, \qquad \textrm{as } \quad |r-2M| \ll M  \;.
\end{equation}
Very close to the horizon the logarithm dominates so 
\begin{equation} \label{eq:r*(r) near-horizon}
    r \approx 2M \l(1+e^{r_*/2M} \r) = 2M \l(1+e^{2 \kappa r_*} \r) \;,
\end{equation}
such that $f(r) \approx e^{2 \kappa r_*}$. The radial wave potential \eqref{eq:radial_potential} then takes the approximate form
\begin{equation}
    V(r_*) \approx 4 \kappa^2 (l^2 + l + 1) e^{2 \kappa r_*} \equiv m_l^2 e^{2 \kappa r_*} \;
\end{equation}
and the radial wave equation \eqref{eq:radial} is  that of Liouville quantum mechanics
\begin{equation}\label{eq:Liouville}
    \left[-\frac{d^2}{dr_*^2} + m_l^2 e^{2 \kappa r_*}\right] \psi_{\omega l} = \omega^2\psi_{\omega l} \; .
\end{equation}
As we review below, \eqref{eq:Liouville} is the radial wave equation for a field of mass $m_l^2$ in Rindler space (with acceleration $a=\kappa$). The mass term can be thought of as arising from a dimensional reduction on the $S^2$ whose inverse radius is $2\kappa$. In Rindler space, the potential grows exponentially to the right because a massive field cannot make it out to $\mathcal{I}^+$. As illustrated in figure \ref{fig:radial_pot}, the black hole's radial potential eventually peaks several gravitational radii from the horizon due to the centrifugal barrier and then falls slowly to zero. However, the behavior of the potential beyond the barrier is essentially irrelevant for very low-energy excitations in the near-horizon region, and the boundary condition that is imposed there can be replaced with a condition on the monodromy of the solution about the horizon.

\subsubsection*{Rindler ``Overtones'' from Symmetries}
The Rindler metric and its massive wave operator are
\begin{equation}  
    ds^2 = e^{2 \kappa r_*} \l( -d t^2 + d r_*^2 \r) \; , \qquad     \Box -m_l^2= e^{-2\kappa r_*}\left(-\partial_t^2 + \partial_{r_*}^2\right)  - m_l^2 \; .
\end{equation}
Taking the ansatz
\begin{equation}
    \Psi_\omega = e^{-i\omega t}\psi(r_*) \;,
\end{equation}
one finds that the radial wavefunctions are modified Bessel functions
\begin{equation}\label{eq:modBessel}
    \psi(r_*) = c_1 \, I_{i\omega/\kappa} \l(m_l e^{\kappa r_*} /\kappa \r) + c_2 \, I_{-i\omega/\kappa} \l(m_l e^{\kappa r_*} /\kappa \r) \; .
\end{equation}
Since $I_\nu(z)\sim z^\nu$ as $z\to 0$, the solution that is infalling at the horizon is
\begin{equation}\label{eq:rindler_ingoing}
    \psi(r_*) \propto I_{-i\omega/\kappa}\l(m_l e^{\kappa r_*} /\kappa \r)  \implies \Psi_{\omega} = e^{-i \omega t} I_{-i\omega/\kappa}\l(m_l e^{\kappa r_*} /\kappa \r)  \;,
\end{equation}
but since this is a single condition it does not place a constraint on the allowed values of $\omega$. In the normal QNM problem we would impose a second boundary condition on the far-right end, but we have lost track of the far region in taking the Rindler limit and do not know how this near-solution is supposed to connect through the barrier. Instead we impose the second QNM boundary condition using the monodromy condition at the horizon as in \cite{motl2003}, which guarantees that the solution continues to an outgoing mode in the far region. As explained in appendix \ref{App:monodromy}, the condition states that in the vicinity of the past horizon (at fixed $u_S = t-r_*$), under $r-2M \to e^{2\pi i}(r-2M)$, the highly damped QNM wavefunctions must satisfy
\begin{equation}  
    \Psi^{\mathrm{QNM}}_\omega \to -3 \Psi^{\mathrm{QNM}}_\omega \;.
\end{equation}
This fixes 
\begin{equation}
    \omega_R=  \frac{\kappa}{2\pi} \log(3) \; , \qquad \omega_I=-  i \frac{\kappa}{2} -i\kappa n \; .
\end{equation}
The real part of this frequency is input for the near-horizon analysis and has no intrinsic significance in Rindler space. However, once we are given the real part, the QNM overtone spectrum can be generated   using the enhanced symmetries of Rindler that emerge in the near-horizon limit, as was the case for the photon ring.  
To proceed it is useful to introduce the  coordinates
\begin{equation} \label{eq:Kruskal coords}
    V = \frac{1}{\kappa} e^{\kappa(t+r_*)}  \;, \qquad U= \frac{1}{\kappa} e^{-\kappa(t-r_*)} \; .
\end{equation}
In Schwarzschild these are the lightlike Kruskal–Szekeres coordinates in the vicinity of the bifurcation surface, but in Rindler they are simply the advanced and retarded coordinates for two dimensional Minkowski space.
The three vector fields
\begin{equation}\label{eq:RindlerIso}
    P_+=\partial_V \; ,\qquad  P_-=\partial_U \; , \qquad   K = U\partial_U -V\partial_V = -\frac{1}{\kappa} \partial_t  \; ,
\end{equation}
generate the 2d Poincar\'e algebra
\begin{equation} \label{eq:Poincare}
    [K, P_-] = -P_- \;, \qquad [K, P_+] = P_+ \;, \qquad [P_-, P_+] = 0 \; .
\end{equation}
Of the three generators, only the boost $K$ preserves the Rindler patch.  This symmetry corresponds to global time translation in the exterior Schwarzschild geometry, while the other two vector fields $P_+$ and $P_-$ are emergent isometries in the Rindler limit.

Note that the first two commutation relations resemble the ladder structure present in the simple harmonic oscillator algebra  and $SL(2,\mathbb{R})$ if we identify $K$ with $H$ or $L_0$, and $P_{\pm}$ with $a_{\pm}$ or $L_{\pm}$. Indeed, these three algebras only differ in the commutation relation between the two ladder-type operators: it vanishes in Rindler, is a central term in the harmonic oscillator algebra, and is $2L_0$ in $SL(2,\mathbb{R})$. This  commutator does not play an important role in the construction of overtones so the three types of emergent symmetries all lead to equally spaced families of resonances. 

Note that in Schwarzschild coordinates, the vector fields \eqref{eq:RindlerIso} take the form
\begin{equation}\label{eq:horizon dynamical symmetries}
\begin{aligned}
    & P_+= \frac{1}{2} e^{- \kappa t} e^{-\kappa r} \pa{\frac{r}{2M} -1}^{-1/2} [ \partial_t + f(r) \partial_r ] \; , \\
    & P_-=\frac{1}{2} e^{\kappa t} e^{-\kappa r} \pa{\frac{r}{2M} -1}^{-1/2} [ -\partial_t + f(r) \partial_r ] \; , \\
    & K =-\frac{1}{\kappa}\partial_t \; .
\end{aligned}
\end{equation}
In other words, \eqref{eq:RindlerIso} are just the near-horizon limits of the global Kruskal–Szekeres vector fields $\partial_V,\partial_U, U\partial_U-V\partial_V$. These vector fields are defined everywhere in the Schwarzschild exterior, where they obey the 2d Poincar\'e algebra identically, but they are only relevant in the near-horizon region, where they  commute with the wave operator. Since any wavefunction in Schwarzschild that is infalling at the horizon exhibits the near-horizon behavior  \eqref{eq:rindler_ingoing},   \eqref{eq:horizon dynamical symmetries} can be used to produce descendants in the near-horizon region, as we now explain.

Since $K=-\frac{1}{\kappa}\partial_t$, the mode functions \eqref{eq:rindler_ingoing} diagonalize $K$  
\begin{equation}\label{eq:Keigenval} 
    K \Psi_\omega = \frac{i \omega}{\kappa} \Psi_\omega \;.
\end{equation}
The algebra \eqref{eq:Poincare} then indicates that $P_-$ sends $\omega \to \omega + i\kappa$, and $P_+$ sends $\omega \to \omega - i \kappa$. Indeed one can explicitly check by rewriting \eqref{eq:rindler_ingoing} 
\begin{equation}
  \Psi_\omega(U,V)=  \left(\frac{V}{U}\right)^{-i\omega/2\kappa}I_{-i\omega/\kappa}(m_l\sqrt{UV})
\end{equation}
that
\begin{equation} \label{eq:rindler_raising/lowering}
    P_- \Psi_\omega = \frac{m_l}{2} \Psi_{\omega + i \kappa} \;, \qquad P_+ \Psi_\omega = \frac{m_l}{2} \Psi_{\omega - i \kappa} \;.
\end{equation}
That means that if we start with a seed solution with frequency $\omega_0$, we can generate a family of solutions in the near-horizon region with
\begin{equation}\label{eq:RindlerResonance}
    \omega_n=\omega_0 -in\kappa\; .
\end{equation}
However, while these descendants will all satisfy the infalling boundary condition at the horizon, it is not guaranteed that the global completion of a Rindler descendant will obey the same boundary conditions as the seed $\Psi_{\omega_0}$ at spatial infinity.

The seed solution and its descendant are not multiplicatively related throughout the spacetime, but as noted in \eqref{eq:overtones}, near the horizon where $r_*\to -\infty$
\begin{equation}
    \Psi_{\omega - i \kappa}(r_*)\sim \Psi_{\omega }(r_*) \cdot e^{-\kappa t} e^{-\kappa r_*}=e^{-\kappa v_S} \Psi_{\omega }(r_*) \propto \frac{1}{V}\Psi_{\omega }(r_*)\; .
\end{equation}
Indeed, in the limit that $V\to 0$ we have
\begin{equation}
    P_+ \Psi =\frac{1}{V}\Psi + \text{higher order}\; .
\end{equation}
As in \eqref{eq:pp overtones} for the photon ring, the dynamical relation \eqref{eq:overtones} between overtones is a consequence of symmetry in the appropriate scaling limit.

All of the preceding statements are even simpler for the massless Rindler wave equation. In that case the potential vanishes identically, so a right moving wave does not reflect  and mix with the incoming wave. The solutions are superpositions of left and right moving waves and the Rindler plane waves $e^{-i\omega t}e^{\pm i\omega r_*}$ are simply
\begin{equation}
    \Psi_\omega^- =  V^{-i \omega/\kappa} \;, \qquad \Psi_\omega^+ = U^{i \omega/\kappa} \; .
\end{equation}
The left-movers $\Psi_\omega^-$ are annihilated by $P_-$, and $P_+$ raises the overtone $\omega \to \omega - i\kappa$. Similarly, $P_+$ annihilates the right-movers, and $P_-$ shifts their overtone number $\omega \to \omega + i \kappa$.  
As in the photon ring,
the infalling boundary condition can be phrased in terms of a ``highest weight'' condition: $P_-\Psi_\omega^-=0$.

\subsubsection*{Rindler Symmetries and Monodromy}
The infalling boundary condition \eqref{eq:rindler_ingoing} does not distinguish between scattering states, QNMs, or generic complex frequencies since the near-horizon analysis is blind to the boundary condition imposed at spatial infinity. In the near-horizon region the emergent Rindler isometries \eqref{eq:Poincare} preserve the infalling boundary condition and generate new solutions from old. However, there is no guarantee that the wavefunction  of a Rindler-descendant satisfies the same boundary condition as the seed solution at spatial infinity. The infalling solutions could represent the near-horizon behavior of infalling scattering states whose $\omega$ is real and continuous or QNMs whose frequencies are complex but discrete. Both cases diagonalize the boost $K$ as in \eqref{eq:Keigenval}, but only 
the ``resonance'' spectrum \eqref{eq:RindlerResonance} depends on the temperature. In particular, the two momentum operators $P_{\pm}$ do not map states with real $\omega$ to states with real $\omega$. Rather, for a mode with fixed $\omega_0$, $P_+$ generates a tower of decaying modes with increasingly negative imaginary parts, and $P_-$ generates a family of growing modes with large positive imaginary parts. \textit{The emergent symmetries are realized on the resonance spectrum, not on the scattering spectrum.}

To be concrete, consider the case when the seed wavefunction $\Psi_{\omega_0}$ is the near-horizon limit of a QNM wavefunction. In that case its extension to the full Schwarzschild geometry is purely outgoing at spatial infinity. The Rindler descendants $\Psi_{\omega_n}$ might extend to solutions with the same boundary conditions (if  $\omega_0-i n \kappa$ is also in the QNM spectrum) or they could contain both ingoing and outgoing waves far from the black hole (in which case they correspond to the analytic continuation of scattering states with complex frequencies). The latter case is clearly less interesting than the former. For instance, the near-horizon limit of an eikonal QNM would have the behavior \eqref{eq:rindler_ingoing} with $\omega_0=\left(l+\frac12\right)(3\sqrt{3}M)^{-1}-\frac{i}{2}(3\sqrt{3}M)^{-1}$. Acting with the Rindler raising operator would then produce modes which are not in the eikonal QNM spectrum since the imaginary part would not match. Indeed, the eikonal QNM overtones each have different monodromy around the horizon, while the action of the Rindler isometry preserves the monodromy of the original solution.

It is not hard to see that the emergent Rindler symmetries are only spectrum-generating when applied to the highly-damped QNMs. Since the Rindler vector fields are regular in the vicinity of the bifurcation surface, they do not introduce non-analytic coordinate  dependence when acting on wavefunctions. 
In terms of the lightlike Kruskal–Szekeres coordinates, the future  horizon corresponds to $U \to 0$ while the past horizon is at $V \to 0$.  
Any wavefunction that is purely infalling at the future horizon  locally behaves as 
\begin{equation}
\Psi^{\mathrm{infalling}}_\omega \sim V^{-i \omega/\kappa} \; .
\end{equation}
As we go around the horizon in the complex $r$ plane in the vicinity of the past horizon, ${r-2M \to e^{2\pi i}(r-2M)}$ at fixed $U$,  then $V \to e^{2 \pi i}V$ so the infalling solution picks up monodromy controlled by the frequency
\begin{equation} \label{eq:monodromy1}
\Psi^{\mathrm{infalling}}_\omega \to e^{2 \pi \omega/\kappa} \Psi^{\mathrm{infalling}}_\omega \;.
\end{equation}
The highly damped QNMs  all have a specific monodromy around the past horizon  
\begin{equation} \label{eq:monodromy2}
    \Psi^{\mathrm{QNM}}_\omega \to -3 \Psi^{\mathrm{QNM}}_\omega \;.
\end{equation}
Since the vector fields are analytic in $U$ and $V$, the Rindler isometries do not change the monodromy when acting on wavefunctions. They therefore generate a tower of highly-damped QNMs with $\omega_n = \omega_0 - i n \kappa$ provided that $\omega_0$ is the frequency of a highly damped QNM. Other overtone branches of the QNM spectrum have variable monodromy within the family, so acting with the Rindler isometries cannot produce new QNMs if the seed solution is not already a highly damped QNM.  Only when $\omega_0$ is of the form \eqref{eq:highly_damped_qnm} with $n$ large do the emergent Rindler symmetries act within the space of QNMs.

To recap, the equal spacing of the overtones is given by the Rindler acceleration of the near-horizon region, $\kappa$, which sets the relation between $t$ and the flat lightcone coordinates of the Rindler region \eqref{eq:Kruskal coords} that generates the 2d Poincar\'e algebra \eqref{eq:horizon dynamical symmetries}.

\subsubsection*{Angular Momentum Independence}
The spectrum \eqref{eq:highly_damped_qnm} is notably independent of angular momentum, and it is interesting to track how this arises in the near-horizon analysis. In the near-Rindler limit, the angular momentum of the wave shows up as a mass term, and the 2d mass is the coefficient of the exponential potential in the Liouville problem \eqref{eq:Liouville}. 

In the standard approach to \eqref{eq:Liouville} (for instance when it is treated as the zero-mode quantum mechanics of the Liouville quantum field theory), one requires normalizable solutions. This singles out a specific linear combination of the modified Bessel functions \eqref{eq:modBessel}. Since the exponential interaction vanishes so quickly to the left, the spectrum is still continuous and independent of the coefficient of the exponential as well as its rate. The normalizable solutions contain incoming and outgoing waves in the asymptotic region, so the only effect of the exponential function on the spectrum is to identify certain solutions of the free problem (the reflection identification). The resulting ``Liouville $S$-matrix'' is independent of the mass and the poles only depend on $\kappa$.

The treatment of \eqref{eq:Liouville} for the QNM problem is different. Instead of imposing normalizability, we impose the infalling condition at the horizon. Since this is only one condition, it does not constrain the allowed values of $\omega$ which can still be taken to be continuous. This reflects the fact that these infalling solutions could represent the near-horizon behavior of infalling scattering states whose $\omega$ is real and continuous or QNMs whose frequencies are complex but discrete. Only the resonance spectrum (in this case the highly-damped QNMs) is generated by the emergent symmetry $P_+$.

The independence of $\omega_0$ on the angular momentum is not fixed by the near-horizon analysis, but the $l-$independence of the overtones is guaranteed because the enhanced symmetries \eqref{eq:horizon dynamical symmetries} depend only on the surface gravity (in contrast with the emergent symmetries of the photon ring which  depend explicitly on the angular momentum). The angular momentum only serves to control the height of the centrifugal barrier, which has been effectively scaled out of the problem by zooming into the near-horizon region, as depicted in figure \ref{fig:radial_pot}.

\subsection{``Geometric Optics'' of the Past Horizon} \label{subsec:horizon geo_optics}

In the weakly-damped eikonal limit $\omega_R \to \infty$, the constant overtone spacing of the spectrum was a signature of the  hidden IHO symmetry of the wave equation in the near-ring region. This  hidden symmetry was realized geometrically by the isometries of the pp-wave geometry in the corresponding Penrose limit of the photon ring. 

As explained in the previous subsection, given any seed wavefunction that's purely infalling at the horizon, we can use the emergent symmetries \eqref{eq:horizon dynamical symmetries} to generate a tower of purely infalling wavefunctions
\begin{equation} \label{eq:tower}
    \Psi_{\omega_n} \propto (P_+)^n \Psi_{\omega_0} \;, \qquad \qquad \omega_n = \omega_0 - i n \kappa \;.
\end{equation}
The constant overtone spacing of the highly-damped QNM spectrum is seen as a signature of the emergent Poincar\'e symmetry of the wave equation in the near-horizon region.

\begin{figure}[h]
    \centering
    \includegraphics[width=9cm]{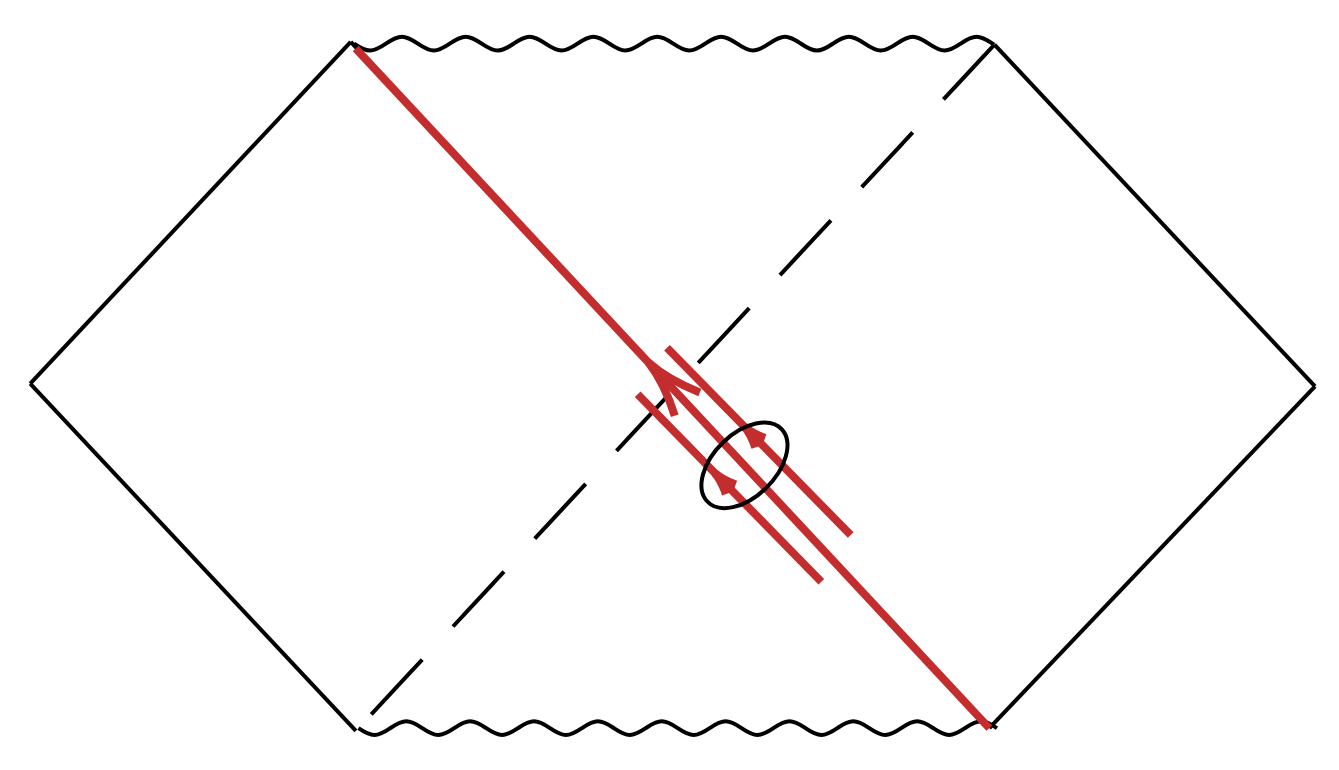}
    \caption{A bundle of radially infalling geodesics, centered around the past horizon. }
    \label{fig:past horizon bundle}
\end{figure}

This tower of solutions satisfies the outgoing QNM boundary condition at spatial infinity iff $|\text{Im}\, \omega_0|\gg1$. This is not the usual geometric optics limit in which one expects to make a connection with geodesics, but section \ref{Sec:near-horizon wave eqn} demonstrated that the overtone spectrum can be derived by zooming into the near-horizon region. In the rest of this section, we will seek an analogous geometric construction of the highly-damped QNMs using the bundle of radially infalling geodesics centered around the past horizon, as  depicted in figure \ref{fig:past horizon bundle}.

\subsubsection*{Photon Ring vs. Horizon}\label{subsec:prvshorizon}

There are two interesting classes of trapped null geodesics in any black hole spacetime: those that lie in the photon ring and those that generate the horizon. 
The past horizon has a coordinate invariant definition as the boundary of the future of $\mathcal{I}^-$, but the photon ring is usually defined  as the set of null geodesics with constant Boyer-Lindquist radius. That is obviously not a coordinate invariant definition. The naive geometric characterization would be in terms of the set of bound null geodesics, but that would also include the horizon generators. One sensible definition that differentiates between the two involves boundaries in the space of geodesics in the maximally extended Schwarzschild geometry.

\begin{figure}[h]
    \centering
        \includegraphics[width=9cm]{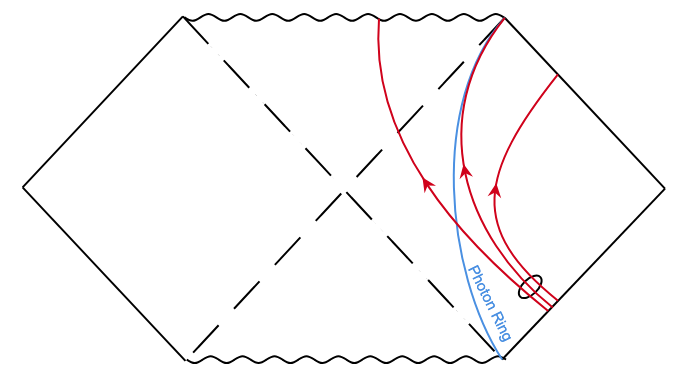}
    \caption{The photon ring is represented by the blue line at $r=3M$, extending all the way from $i^-$ to $i^+$. Null geodesics sent in from $\mathcal{I}^-$ are labeled in red. The geodesic with $b \equiv L/E$ tuned to its critical value $b_0= 3\sqrt{3}M$ will asymptote to the photon ring, and it represents the boundary between geodesics that escape back to $\mathcal{I}^+$ and those that fall into the black hole.}
    \label{fig:photon ring penrose diagram}
\end{figure}

Null geodesics sent in from $\mathcal{I}^-$ either scatter off of the black hole and end up on $\mathcal{I}^+$, or they fall through the horizon and end on the singularity. The photon ring geodesics represent the boundary between these two possibilities, as depicted in figure \ref{fig:photon ring penrose diagram}. A geodesic congruence based on the photon ring with finite radial extent includes geodesics of both types, and the divergence of those geodesics controls the eikonal QNM spectrum.

Likewise, we can view the past horizon of the black hole as the boundary between those null geodesics that start on $\mathcal{I}^-$ and end on the future singularity, and the ones that start at the past singularity and escape to $\mathcal{I}^+$ on the other asymptotically flat end, as depicted in figure \ref{fig:past horizon bundle}. A congruence based on a horizon generator with spread transverse to the horizon includes geodesics of both types, and as we will see their divergence apparently controls the highly-damped QNM spectrum.

This educated guess passes a number of sanity checks. The QNMs \eqref{eq:highly_damped_qnm} have small real part (zero to first order in $n^{-1}$) which matches the infinite redshift between the horizon and a $\partial_t$ observer. The geodesics in figure \ref{fig:past horizon bundle}  are independent of angular momentum at leading order, and the modes in \eqref{eq:highly_damped_qnm} are $l$-independent. A small perturbation near the horizon falls immediately into the black hole, and the modes in \eqref{eq:highly_damped_qnm} have very short lifetimes controlled by the surface gravity. The highly-damped overtone wavefunctions \eqref{eq:overtones} are controlled by a quantity conserved along the geodesics in figure \ref{fig:past horizon bundle} just as in the photon ring.

\subsubsection*{Highly Damped Geometric Optics} \label{Sec:horizon geo optics}
In the region outside the horizon,  the bundle of geodesics depicted in figure \ref{fig:past horizon bundle} move according to $v_S = t+r_* = \mathrm{const}$ for a thin band of $v_S$. In outgoing Eddington–Finkelstein coordinates with
\begin{equation} \label{eq:Schw advanced metric}
    ds^2 = -f(r) du_S^2 - 2 du_S dr + r^2 d\Omega^2 \;, \qquad  u_S = t - r_* \; ,
\end{equation}
these geodesics can be parameterized\footnote{This parametrization is valid for $\lambda < 0$.  $\lambda \to 0$ corresponds to the geodesic crossing the future horizon into the black hole region, where the $u_S$ coordinate breaks down.} 
\begin{equation} \label{eq:radially infalling bundle}
    \begin{split}
        &u_S(\lambda) = - \frac{1}{\kappa} \ln(- \lambda) + 2 E \lambda \;, \\
        &r(\lambda) = 2M - E \lambda \;.
    \end{split}
\end{equation}
Such a geodesic has $v_S=4M(1+\log \frac{E}{2M})$, so as $E\to 0$ the geodesic moves onto the past horizon at $v_S=-\infty$. The geodesics \eqref{eq:radially infalling bundle} have the Hamilton-Jacobi function 
\begin{equation}
    S = \int p_\mu dx^\mu = -E (t+r_*) = -E (u_S + 2 r_*) = -E v_S \;.
\end{equation}
If we use the geometric optics approximation to construct a solution to the wave equation out of this bundle, the phase will be given by
\begin{equation}\label{eq:dampedphase}
    \psi_0(x) \sim e^{i S(x)} = e^{-i E v_S} \;.
\end{equation}
This approximate form of the solution is valid in the vicinity of the radially infalling geodesic and has purely infalling boundary conditions at the future horizon. This can be seen from figure \ref{fig:past horizon bundle}, or from the fact that $\psi(x) \sim e^{-i E (t+r_*)}$.

A generic ingoing geodesic will produce the correct QNM boundary condition at the horizon, but not at infinity since it is coming in from $\mathcal{I}^-$. To fix this we could take  $E\to 0$ (equivalently $v_S\to -\infty$) so that the geodesic moves onto the past horizon (trapped and not coming in from $\mathcal{I}^-$).\footnote{There is another heuristic way to motivate the $E\to0$ limit to the past horizon. As explained in appendix \ref{App:monodromy}, the monodromy condition \eqref{eq:monodromy2} is most naturally understood as the monodromy of the radial wavefunction $\phi(r)$ \eqref{eq:monodromy_app2} around the horizon $r-2M \to e^{2\pi i}(r-2M)$ at fixed $u_S$, which is the monodromy around the past horizon.} However, in that limit the geometric optics approximation breaks down since $E\lesssim R^{-1}$, where $R$ is the characteristic curvature length-scale in that region. 

Of course, for the highly damped modes \eqref{eq:highly_damped_qnm} the phase is not really the important part of the wavefunction: the amplitude decay dominates the behavior. In other words, we do not expect to generate QNMs with small real part and small imaginary part using the congruence in figure \ref{fig:past horizon bundle}.
The usual geometric optics approximation works when the wave's oscillation timescale is short compared to $R$. The solutions \eqref{eq:highly_damped_qnm} instead have a \textit{decay timescale} short with respect to $R$ so the situation is nonstandard and it is not obvious that there should be any relation to a geodesic. Nevertheless,  if we begin with the ansatz \eqref{eq:dampedphase} (even though it is not a good approximation) and then start building overtones using the procedure in section \ref{subsec:overtoneSym}, the approximation eventually becomes more accurate and limits onto the correct behavior. 

With these caveats in mind, we begin with the ansatz based on the geodesics \eqref{eq:radially infalling bundle}
\begin{equation} \label{eq:seed}
    \psi_0(x) \sim e^{- i E v_S} \; .
\end{equation}
Along infalling geodesics the coordinate $v_S$ is constant, which means that 
\begin{equation}
    p^\mu \partial_\mu f(v_S) = 0 \;
\end{equation}
for any function $f(v_S)$. Multiplying the seed  $\psi_0$ by $f(v_S)$ therefore commutes with the geometric optics amplitude equation \eqref{eq:amp}, provided that the new waveform still obeys the correct boundary condition. We saw in section \ref{Sec:near-horizon wave eqn} that $f(v_S)$ should admit a Laurent expansion at $V=e^{\kappa v_S}=0$ in order to preserve the outgoing boundary condition (equivalently to preserve the monodromy around the horizon).
If we choose
\begin{equation} \label{eq:f geo_optics_horizon}
    f(v_S) = e^{- \kappa n v_S}\sim V^{-n} \;,
\end{equation}
then the new amplitude is 
\begin{equation}
    \psi_n(x) = e^{-i(E - i n \kappa) v_S} \equiv e^{-i \omega_n v_S} \;,
\end{equation}
for $\omega_n = E - i n \kappa$. These wavefunctions are bad approximations for small $n$ but get better as the imaginary part of the frequency becomes large if $E= \frac{\kappa}{2\pi} \log(3) - i \frac{\kappa}{2} \ll n$.  

In the near-horizon region, multiplying by $e^{-\kappa v_S}$ is the same as acting with the Rindler momentum operator $P_+$. 
Following the discussion in section \ref{subsec:overtoneSym}, we can identify the conserved quantity along the geodesic \eqref{eq:f geo_optics_horizon} with the emergent symmetry of the wave equation \eqref{eq:RindlerIso}
\begin{equation}
    P_+^\mu p_\mu \propto e^{- \kappa v_S} \propto V^{-1} \; .
\end{equation}
Finally, we remark on the $l-$independence of the spectrum from the geometric optics perspective. For a geodesic with angular momentum, the Hamilton-Jacobi function is given by
\begin{equation} \label{eq:HJ schw L}
    S = -Et - \int \frac{1}{f(r)} \sqrt{E^2 - \frac{L^2 f(r)}{r^2}} dr \; .
\end{equation}
The near-horizon behavior of these geodesics is given by the Rindler Hamilton-Jacobi function for massive geodesics, \eqref{eq:HJ rindler}. The fact that \eqref{eq:HJ rindler} is independent of the mass uplifts to the angular momentum independence of the Schwarzschild Hamilton-Jacobi function near the horizon, which can be seen in \eqref{eq:HJ schw L} since $L^2f(r)/r^2 \to 0$ as $r \to 2M$.

\subsubsection*{Penrose Limit into the Past Horizon} \label{App:past horizon penrose}

To conclude this section,   we describe the Penrose limit of the geodesic \eqref{eq:radially infalling bundle} that moves onto the past horizon in the limit $E\to 0$. The limit yields Rindler space (a trivial pp-wave with zero curvature) and the emergent symmetries of this flat space geometrize the overtone-generating symmetries discussed in section \ref{Sec:near-horizon wave eqn}.

We start with the bundle of radially infalling geodesics described by \eqref{eq:radially infalling bundle} in the Schwarzschild geometry. If we fix the affine time parametrization for the entire bundle to match \eqref{eq:radially infalling bundle}, then
\begin{equation}
    u_S = -\frac{1}{\kappa} \ln (-\lambda) + \frac{1}{\kappa} e^{\kappa v_S-1} \lambda \;, \qquad r = 2M - 2M e^{\kappa v_S-1} \lambda \;.
\end{equation}
This satisfies $u_S + 2 r_* = v_S$. At any fixed $v_S$, it gives the corresponding radially infalling geodesic moving with affine parameter $\lambda$, with $E = 2M e^{\kappa v_S-1}$.
The bundle has the Hamilton-Jacobi function $S = -E(u_S + 2 r_*) = -E v_S = -2M e^{\kappa v_S-1} v_S$. In order to perform the Penrose limit, we set $\lambda = -u$ and $S = \tilde{\varepsilon}^2 v$ (the minus sign in $u$ to have $u>0$). This gives 
\begin{equation}
    -v_S e^{\kappa v_S} = 2 \kappa e \tilde{\varepsilon}^2 v \;,
\end{equation}
whose solution is 
\begin{equation}
    \kappa v_S =  W(-2 \kappa^2 e \tilde{\varepsilon}^2 v) \;,
\end{equation}
where $W$ is the Lambert function. For small $\varepsilon$, this has the approximate form
\begin{equation}
    \kappa v_S \approx \log(2 \kappa^2 e \tilde{\varepsilon}^2 v) - \log( -\log(2 \kappa^2 e \tilde{\varepsilon}^2 v) ) \;.
\end{equation}
Then, we get
\begin{equation}
    \begin{aligned}
        & u_S = -\frac{1}{\kappa} \ln (u) +\frac{2 \kappa  \tilde{\varepsilon}^2 v}{\log(2 \kappa^2 e \tilde{\varepsilon}^2 v)} u \approx -\frac{1}{\kappa} \ln (u) +\frac{ \tilde{\varepsilon}^2}{\log(\tilde{\varepsilon}^2)} 2 \kappa v u \; , \\
        & r = 2M - 2M \frac{2 \kappa^2  \tilde{\varepsilon}^2 v}{\log(2 \kappa^2 e \tilde{\varepsilon}^2 v)} u \approx 2M - \frac{ \tilde{\varepsilon}^2}{\log(\tilde{\varepsilon}^2)} \kappa v u \; .
    \end{aligned}
\end{equation}
 Redefining the small parameter  
\begin{equation}
    \varepsilon^2 \equiv - \frac{\tilde{\varepsilon}^2}{\log(\tilde{\varepsilon}^2)} \;,
\end{equation}
we take the Penrose limit the usual way by performing the coordinate transformation
\begin{equation} \label{eq:past horizon coord penrose}
    \begin{aligned}
        & u_S = -\frac{1}{\kappa} \ln (u) - \varepsilon^2 2 \kappa v u \; ,\\
        & r = 2M + \varepsilon^2 \kappa v u \; ,\\
        & \theta = \frac{\pi}{2} + \frac{\varepsilon \psi_1}{2M} \; ,\\
        & \phi = \frac{\varepsilon \psi_2}{2M} \; ,
    \end{aligned}
\end{equation}
 scaling the metric by $\varepsilon^{-2}$, and sending $\varepsilon \to 0$. The result is flat space in lightcone coordinates,
\begin{equation} \label{eq:past horizon flat space}
    ds_{\textrm{pp}}^2 = 2du dv + d\psi_1^2 + d\psi_2^2 \;.
\end{equation}
The angular directions are decompactified in the limit.

\section{Future Directions}\label{sec:future}
In this paper we investigated how the universal features of the near-ring scaling limits for black holes are encoded in pp-wave geometries  via the Penrose limit. We conclude in this section with some potential follow-up questions. 

\subsubsection*{Higher-Order WKB}
We have seen that  the exact pp-wave geometry captures the leading order WKB approximation of the Schwarzschild eikonal QNMs. The WKB approximation has been pushed to higher orders (see \cite{Iyer:1986np} for early work). These improved approximations involve corrections to the IHO potential among other ingredients. Meanwhile, the paper \cite{Blau:2006ar} extended the Penrose limit, viewed as a covariant expansion about the geodesic, to higher orders. Roughly speaking, this amounts to retaining the subleading $O(\varepsilon)$ terms in \eqref{eq:PR_Pen_metric} in a covariant way. It would be interesting to connect the subleading terms in both approaches in order to reproduce the higher-order WKB approximation from the corrected pp-wave geometry. Can one geometrize the full expansion covariantly?

\subsubsection*{Eikonal Kerr Spectrum}
The near-ring limit introduced in \cite{Hadar:2022xag}  works for Schwarzschild as well as Kerr, but there are important differences between the two cases. In Schwarzschild there is a single Lyapunov exponent due to the $SO(3)$ symmetry. Since there is only one exponent, it can be expressed in terms of the geometric quantities of the black hole rather than the momenta of the geodesic. This is no longer true in Kerr, where the elliptic integrals related to polar motion enter directly into the definition of the Lyapunov exponent, with sensitive dependence on the particular photon orbit. Rather than a single set of vector fields \eqref{eq:symmetries_Schw_coords} representing the emergent symmetries \eqref{eq:symm_dynamical}, there is a family of distinct vector fields fibered over the different photon ring orbits in phase space. The Penrose limit seems like the perfect formalism to handle this situation, with a separate spacetime associated to each photon orbit and correspondingly distinct isometries.

Fransen initiated the investigation of the Penrose limit of the Kerr photon ring in \cite{Fransen:2023eqj}. The pp-wave that he obtains is unsurprisingly  more complicated than our tilted pp-wave metric \eqref{eq:tiltMetric}. Instead of directly treating the wave equation in that geometry, Fransen effectively performs the geometric optics approximation again within the pp-wave background in order to estimate the wave spectrum, which is then compared to the known eikonal spectrum of Kerr.

We are optimistic that a more direct, symmetry based approach to the wave equation in the Kerr pp-wave is possible. According to the general analysis in section \ref{sec:Tilt}, it may not be possible to recover both phases in the WKB approximation to the spheroidal harmonics from this wave equation, but the amplitude corrections might be obtained as in section \ref{sec:tiltedWave}. Since the wave equation separates in Kerr, it should separate in the limiting pp-wave as well, and we hope that the geometric formulas from the WKB approximation to the spectrum \cite{Yang:2012he} can be more easily recovered. This might involve eigenvalue perturbation theory in the pp-wave for the spheroidal separation constants, which amounts to averaging over polar motion as in \cite{Yang:2012he}. In particular, the time-dependence and the off-diagonal entries in Eq. (4.20) of \cite{Fransen:2023eqj} are completely determined by $\cos^2(\theta(u))$,  which is effectively averaged over in the WKB approximation.

\subsubsection*{Fast Damping with Charge and Spin}
Numerical and analytic  investigations have demonstrated that the highly damped QNMs of Kerr \cite{Berti:kerr, Berti:kerr_numerical, Keshet:2007nv,Keshet:2007be} and Reissner-Nordstrom \cite{Andersson_2004, motl2003} exhibit a rich structure not present for Schwarzschild. In particular, the uniform overtone spacing, when it exists, is not simply controlled by the surface gravity of the outer horizon. 

Mathematically, this occurs because the radial  wave potential has a different analytic structure (more poles and zeroes)
when the black hole has an inner horizon. This effects the monodromy technique in important ways. In particular, the monodromy of the QNMs about the outer horizon gets related to the monodromy at the inner horizon, and both surface gravities appear in the analog of \eqref{eq:Highly-damped condition}. 

\begin{figure}[h]
    \centering
    \subfloat[\centering A plot of the radial potential \eqref{eq:RN_pot} for massless geodesics in the Reissner-Nordstrom geometry.]{{\includegraphics[width=0.45\textwidth]{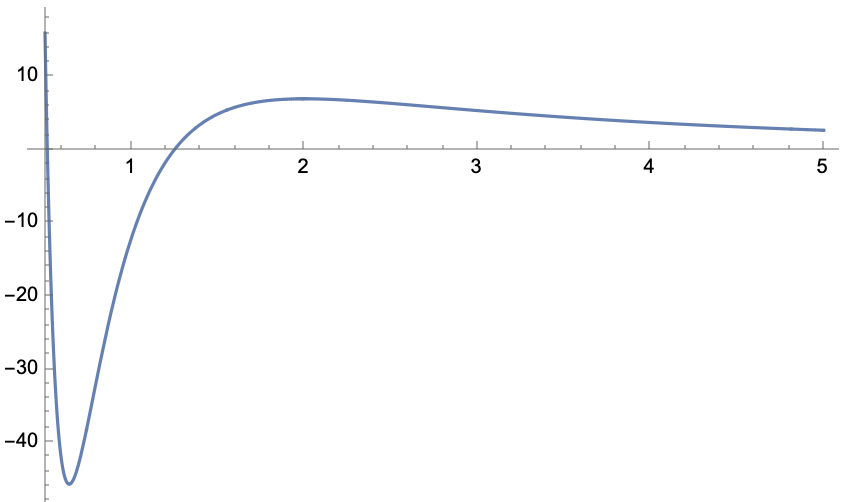} }}
    \quad
    \subfloat[\centering A plot of the radial potential \eqref{eq:Schw_geo_pot} for massless geodesics in the Schwarzschild geometry.]{{\includegraphics[width=0.40\textwidth]{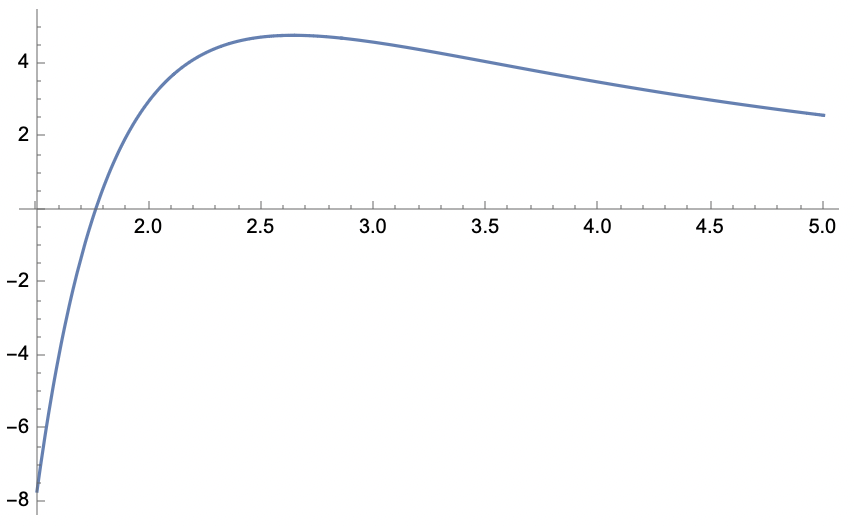} }}
    \caption{Radial potentials for Reissner-Nordstrom and Schwarzschild.}
    \label{fig:RN Schw_pot}
\end{figure}

Physically, we believe the simple geodesic picture presented in section \ref{subsec:horizon geo_optics} is unlikely to hold in the more general setting since the qualitative behavior of infalling geodesics differs significantly when there is charge. The geodesic potential for the charged black hole is 
\begin{equation} \label{eq:RN_pot}
    V_g(r) = \frac{L^2}{r^2} \l(1-\frac{2M}{r} + \frac{Q^2}{r^2}\r) \; .
\end{equation}
For small enough $r$, the third term dominates the second and the potential becomes repulsive again. It therefore has a local minimum, and geodesics can bounce into the other asymptotic regions located to the future in the Penrose diagram.
The infalling geodesics with small angular momentum hugging the past horizon therefore have different behavior than in Schwarzschild. It would be interesting if a modification of section \ref{sec:highDamp} taking these effects into account could be related to the highly damped Reissner-Nordstrom QNMs. For instance, it is sometimes possible to deform the contours employed in the monodromy method onto approximately geodesic trajectories in the complexified geometry, as was done for Kerr in \cite{Keshet:2007be}.

\subsubsection*{Quantum Considerations}
The considerations in this paper were purely classical. However, the classical QNMs of black holes and black branes have an interesting quantum mechanical interpretation within known examples of the gauge/gravity duality. In this dictionary the QNMs map to the Ruelle resonances of the approximately thermal black hole microstate.

The pp-waves typically have one-dimensional null lines for conformal boundaries \cite{Berenstein:2002sa} (the precise statement involves more machinery \cite{Marolf:2002ye,Geroch:1972un} for the vacuum pp-waves since the metrics are not conformally flat).
In the spirit of BMN \cite{Berenstein:2002jq}, it would be interesting to know if there is a $0+1$ dimensional quantum mechanical system associated to this boundary that describes quantum gravity in the photon ring pp-wave background. The Ruelle spectrum of this system would presumably relate to the eikonal QNMs isolated by the Penrose limit. In holographic contexts, the pp-waves typically arise as limits of previously understood dualities. Since we do not know the quantum mechanical description of Schwarzschild or Kerr, a bottom up approach is required.

Ultimately one would like to understand the microscopic embedding of the 4d Kerr black hole in string theory. We do not have a tractable worldsheet description in the Kerr background, but string theory is known to be exactly solvable in pp-wave backgrounds \cite{Horowitz:1989bv,Horowitz:1990sr, Kiritsis:2002kz,Metsaev:2002re}. 
Most bottom-up analyses of Kerr as a quantum system (for example \cite{Guica:2008mu,Bredberg:2009pv,Kapec:2023ruw} in the near-extremal case) work with the Einstein gravity approximation, but one might hope that the Penrose limit of the Kerr photon ring could say something stringy about the dual to Kerr (for example, it could capture a universal set of states in the full stringy dual with a specific scaling of the quantum numbers). Recent interesting bottom-up  work along these lines includes \cite{Dodelson:2020lal,Dodelson:2023nnr}.

\section*{Acknowledgments}
We would like to thank Andrew Strominger for collaboration at an early stage of the project, and Matthew Dodelson for useful discussions. This work is supported by  DOE grant de-sc/0007870.

\appendix
\section{Penrose Limit of the Schwarzschild Photon Ring}
\label{App:SCHpenrose}

In this section, we derive the Penrose limit for bound photon orbits  in the Schwarzschild geometry. The first step is to solve the geodesic equation in terms of the conserved quantities.

\subsection*{Geodesic Equation}
The  Schwarzschild $SO(3)$ Killing vectors are given by
\begin{equation}
    \begin{aligned}
        & l_z = \partial_\phi \; ,\\
        & l_y = \cos (\phi) \partial_\theta - \cot(\theta) \sin(\phi) \partial_\phi \; ,\\
        & l_x = -\sin(\phi) \partial_\theta - \cot(\theta) \cos(\phi) \partial_\phi \; .
    \end{aligned}
\end{equation}
If we define the conserved quantities $L_i = (l_i)_\mu p^\mu$ then   the total angular momentum  is
\begin{equation}
    L = \sqrt{L_x^2 + L_y^2 + L_z^2} = r^2 \sqrt{\l( \frac{d \theta}{d \lambda} \r)^2 + \sin^2(\theta) \l( \frac{d \phi}{d \lambda} \r)^2} \; .
\end{equation}
Instead of using the conserved quantities $L_z$ and $L^2$ (which are in involution), we can instead consider $L$ and the angle $\theta_L$ between the angular momentum vector $\vec{L}$ and the $x-y$ plane. $\theta_L$ corresponds to the turning point of the polar motion, such that $\theta \in [\theta_L, \pi-\theta_L]$ (this is often denoted as $\theta_+$). Inverting the relations  
\begin{equation}\label{eq:Lz}
    \begin{aligned}
        & L \sin(\theta_L) = L_z = r^2 \sin^2 (\theta) \frac{d\phi}{d\lambda}\; , \\
        & L \cos(\theta_L) = \sqrt{L_x^2+L_y^2} = r^2 \sqrt{ \l( \frac{d \theta}{d \lambda} \r)^2 + \sin^2(\theta) \cos^2(\theta) \l( \frac{d \phi}{d \lambda} \r)^2} \; ,
    \end{aligned}
\end{equation}
determines the angular components of $p^\mu$ in terms of the conserved quantities $L$, $\theta_L$
\begin{equation}\label{eq:angMom}
\begin{aligned}
    & \frac{d \theta}{d \lambda} = \pm \frac{L}{r^2} \sqrt{1 - \sin^2(\theta_L) \csc^2(\theta)} \; ,\\
    & \frac{d \phi}{d\lambda} = \frac{L}{r^2} \sin(\theta_L) \csc^2(\theta) \; .
\end{aligned}
\end{equation} 
The other 2 components of $p^\mu$ are then fixed in terms of the conserved quantities 
\begin{equation}\label{eq:trMom}
    \begin{aligned}
        \frac{dt}{d\lambda} = \frac{E}{f(r)} \;, \quad  \frac{d r}{d \lambda} = \pm \sqrt{E^2 - L^2f(r)r^{-2}} \; .
    \end{aligned}
\end{equation}

\subsection{Tilted Photon Ring Using $SO(3)$ Symmetry} \label{App:tilted1}
The covariant information retained in the Penrose limit is the geodesic deviation, so by the $SO(3)$ symmetry of Schwarzschild all photon ring geodesics will have the same Penrose limit (they all have the same Lyapunov exponent). The nontrivial information about the geodesic is instead contained in the relationship between the pp-wave coordinates and Schwarzschild coordinates. 

Consider a photon ring geodesic with $\theta_L \neq \pi/2$ so that the motion is not equatorial. We can use the rotational symmetry of the problem  to transform to new Schwarzschild coordinates $(t, r, \theta, \phi) \to (t, r, \tilde{\theta}, \tilde{\phi})$ where $\tilde{\theta}_L = \pi/2$. In other words, we choose new coordinates in which the geodesic is equatorial. One way to accomplish this is to rotate the original coordinates $(\theta, \phi)$ by an angle $\pi/2 - \theta_L$ around the $x$-axis, to get $(\tilde{\theta}, \tilde{\phi})$.
The answer could be obtained by the Rodriguez rotation formula, which gives
\begin{equation} \label{eq:rodriguez_rot}
\begin{split}
    & \tilde{\theta} = \mathrm{arccos}\br{ \sin(\theta_L) \cos(\theta) + \cos(\theta_L) \sin(\theta) \sin(\phi) }\; , \\
    & \tilde{\phi} = \mathrm{arctan} \br{\sin(\theta_L) \tan(\phi) - \cos(\theta_L) \frac{\cot(\theta)}{\cos(\phi)}} \; .
\end{split}
\end{equation}
Alternatively, we can obtain the coordinates $(\theta, \phi)$ by rotating the coordinates $(\tilde{\theta}, \tilde{\phi})$ by angle $\pi/2-\theta_L$ in the opposite direction around the $x$-axis. This gives
\begin{equation} \label{eq:rodriguez_rot_inv}
\begin{split}
    & \theta = \mathrm{arccos}\br{ \sin(\theta_L) \cos(\tilde{\theta}) - \cos(\theta_L) \sin(\tilde{\theta}) \sin(\tilde{\phi}) } \; , \\
    & \phi = \mathrm{arctan} \br{\sin(\theta_L) \tan(\tilde{\phi}) + \cos(\theta_L) \frac{\cot(\tilde{\theta})}{\cos(\tilde{\phi})}} \; .
\end{split}
\end{equation}
Since the geodesic is equatorial in the $(\tilde{\theta}, \tilde{\phi})$ coordinates, they will be related to the pp-wave coordinates via \eqref{eq:coord_transf}
\begin{equation} 
    \begin{aligned}
        & \tilde{\theta} = \frac{\pi}{2} + \varepsilon \frac{\psi}{3M}\;, \\
        & \tilde{\phi} =  3 \Omega_R u + \varepsilon^2 \Omega_R v \;.
    \end{aligned}
\end{equation}
Combining this with \eqref{eq:rodriguez_rot_inv} gives the relation between $(\theta, \phi)$ and the Penrose limit coordinates as
\begin{equation}
    \begin{split}
        & t = 3 u \; ,\\
        & r = 3M + \varepsilon \frac{\rho}{\sqrt{3}}\; , \\
        & \theta = \mathrm{arccos}\br{ \sin(\theta_L) \cos\l( \frac{\pi}{2} + \varepsilon \frac{\psi}{3M} \r) - \cos(\theta_L) \sin\l( \frac{\pi}{2} + \varepsilon \frac{\psi}{3M} \r) \sin\l( 3 \Omega_R u + \varepsilon^2 \Omega_R v \r) } \; , \\
        & \phi = \mathrm{arctan} \br{\sin(\theta_L) \tan\l( 3 \Omega_R u + \varepsilon^2 \Omega_R v \r) + \cos(\theta_L) \frac{\cot\l( \frac{\pi}{2} + \varepsilon \frac{\psi}{3M} \r)}{\cos\l( 3 \Omega_R u + \varepsilon^2 \Omega_R v \r)}} \; .
    \end{split}
\end{equation}
Note that this relation can be inverted to obtain $\psi, v$ in terms of $\theta, \phi$, by observing that
\begin{equation}
    \begin{split}
        & u = \frac{1}{3} t \;,\\
        & \rho = \frac{1}{\varepsilon} \sqrt{3} \l(r-3M \r) \;,\\
        & \psi = \frac{1}{\varepsilon} 3M \l( \tilde{\theta} - \frac{\pi}{2} \r) = \frac{1}{\varepsilon} 3M \l(\mathrm{arccos}\br{ \sin(\theta_L) \cos(\theta) + \cos(\theta_L) \sin(\theta) \sin(\phi) } - \frac{\pi}{2} \r)\;,  \\
        & v= \frac{1}{\varepsilon^2} \l( \frac{\tilde{\phi}}{\Omega_R} - t \r) = \frac{1}{\varepsilon^2} \l( \frac{1}{\Omega_R} \mathrm{arctan} \br{\sin(\theta_L) \tan(\phi) - \cos(\theta_L) \frac{\cot(\theta)}{\cos(\phi)}} - t \r) \; .
    \end{split}
\end{equation}

\subsection{Tilted Photon Ring Using Adapted Coordinates}\label{App:tilted2}
There is an alternate way to take the Penrose limit of the tilted photon ring   that does not make use of symmetry and therefore generalizes more naturally to Kerr. Throughout this section 
\begin{equation} \label{eq:beta}
    \beta \equiv \frac{b_0}{r_0^2} = \frac{1}{\sqrt{3} M} \; ,
\end{equation}
where we rescaled the affine parameter to set $E=1$ so  that $L=b_0$. 

\textbf{Solving Geodesic Equation} We can explicitly solve for $\theta$ motion along the geodesic,
\begin{equation}
    \frac{d\theta}{d\lambda} = \beta \sqrt{1-\sin^2(\theta_L) \csc^2(\theta)} \implies -\textrm{arcsin} \l( \frac{\cos(\theta)}{\cos(\theta_L)} \r) = \beta (\lambda - \lambda_0)
\end{equation}
with initial conditions $\theta_0 \equiv \theta(\lambda_0) = \frac{\pi}{2}$.
The sign of the square root flips when the motion reaches a polar turning point. We get
\begin{equation}\label{eq:theta(lambda)}
    \cos(\theta) = -\cos(\theta_L) \sin \pa{\beta(\lambda - \lambda_0)} \; .
\end{equation}
We can use this to solve the $\phi$ motion explicitly:
\begin{equation} \label{eq:phi(lambda)}
    \frac{d\phi}{d\lambda} = \beta \sin(\theta_L) \frac{1}{1-\cos^2(\theta_L) \sin^2 \pa{\beta (\lambda-\lambda_0)}} \implies \phi = \phi_0 + \arctan \br{\sin(\theta_L) \tan\pa{\beta (\lambda-\lambda_0)}}
\end{equation}
where we keep track of the $\phi$ initial condition via $\phi(\lambda_0) = \phi_0$. 

We can also solve for $\phi(\theta)$ along the geodesic, since the geodesic equation gives
\begin{equation}
    \frac{d\phi}{d\theta} = \frac{\sin(\theta_L) \csc^2(\theta)}{\sqrt{1-\sin^2(\theta_L) \csc^2(\theta)}}
\end{equation}
which we can solve to get
\begin{equation} \label{eq:phi(theta)}
    \phi(\theta) = \phi_0 - \arctan \br{ \frac{\sin(\theta_L) \cos(\theta)}{\sqrt{\sin^2(\theta) - \sin^2(\theta_L)}} } \; .
\end{equation}
Indeed, if we use \eqref{eq:theta(lambda)} and \eqref{eq:phi(lambda)}, we find that \eqref{eq:phi(theta)} is satisfied along the geodesic. 

Note that if we plug \eqref{eq:phi(theta)}, with $\phi_0 = 0$, into \eqref{eq:rodriguez_rot}, we verify that $\cos(\tilde{\theta}) = 0$ since the geodesic is equatorial in these coordinates. Alternatively, if we plug \eqref{eq:theta(lambda)}, \eqref{eq:phi(lambda)} into \eqref{eq:rodriguez_rot}, and use $\theta_0 = \frac{\pi}{2}$, $\phi_0 = 0$, then we verify that $\cos(\tilde{\theta}) = 0$ and that $\tan(\tilde{\phi}) = \tan \pa{\beta \lambda}$ as expected in these coordinates.

Finally, we can write the $t$ motion as
\begin{equation} \label{eq:t(lambda)}
    t(\lambda) = 3 (\lambda - \lambda_0) + t_0
\end{equation}
with the initial condition $t(\lambda_0) = t_0$. 

\textbf{Hamilton Jacobi Function} We next consider the Hamilton Jacobi function for a  geodesic with motion in $t,\theta,\phi$ but fixed $r = 3M$. According to \eqref{eq:angMom} we have 
\begin{equation} \label{eq:HJ tilted}
\begin{split}
    S &= \int p_\mu dx^\mu \\
    &=-E t + L \int \sqrt{1-\sin^2(\theta_L) \csc^2(\theta')} d\theta' + L \sin(\theta_L) \phi \\
    &= -t + b_0 \arccos\left(\frac{\cos (\theta )}{\cos(\theta_L)}\right) + b_0 \sin(\theta_L) \br{ \arctan \left(\frac{\sin(\theta_L) \cos (\theta )}{\sqrt{\sin^2(\theta) -\sin^2(\theta_L)}}\right) + \phi} + S_0 
\end{split}
\end{equation}
where $S_0$ is an overall additive constant, and we used $E=1, L=b_0$ in the last line. Note that the term in the brackets is constant along the geodesic, as can be verified from \eqref{eq:phi(theta)}. Moreover, the sum of the first two terms is also constant along the geodesic, as can be verified from \eqref{eq:theta(lambda)} and \eqref{eq:t(lambda)}. Therefore $S(x) = S(x_0)$ is constant along the geodesic as required.

\textbf{Adapted Coordinates}   
We can solve for the adapted coordinate $V$ by solving ${S(x_0) = V}$, where $x_0 \equiv x(\lambda_0)$. The geodesic motion is given by
\begin{equation}
    \begin{split}
        & t(\lambda) = t_0 + 3 (\lambda - \lambda_0) \; ,\\
        & \theta(\lambda) = \arccos \br{- \cos(\theta_L) \sin\pa{\beta(\lambda-\lambda_0)} } \; , \\
        & \phi(\lambda) = \phi_0 + \arctan \br{\sin(\theta_L) \tan\pa{\beta (\lambda-\lambda_0)}} \; .
    \end{split}
\end{equation}
In order to switch to adapted coordinates, we parametrize the initial conditions in terms of the two coordinates $V,\Psi$: $t_0(V, \Psi), \theta_0(V, \Psi), \phi_0(V,\Psi)$, subject to the constraint $S(t_0, \theta_0, \phi_0) = V$ \cite{Blau:notes}. This can be accomplished by writing
\begin{equation} \label{eq:adapted_initial_coords}
    \lambda_0 = \frac{\Psi}{\sqrt{3}} \;, \quad t_0 = \sqrt{3} \Psi \;, \quad \phi_0 = \frac{1}{b_0 \sin(\theta_L)} \pa{V + \sqrt{3} \Psi} = \frac{\beta}{\sin(\theta_L)} \pa{\frac{1}{3} V + \frac{\Psi}{\sqrt{3}}} \; ,
\end{equation}
where the $\sqrt{3}$ normalization of $\Psi$ is adopted for later convenience. The diffeomorphism to the adapted coordinates is therefore
\begin{equation} \label{eq:tilted_adapted}
    \begin{split}
        & t = 3 U \; , \\
        & r = 3M + \frac{R}{\sqrt{3}} \; , \\
        & \theta = \arccos \br{-\cos(\theta_L) \sin \pa{ \beta(U - \Psi/\sqrt{3})}} \; ,\\
        & \phi = \frac{\beta}{\sin(\theta_L)} \l( \frac{1}{3} V + \frac{\Psi}{\sqrt{3}} \r) + \arctan \br{\sin(\theta_L) \tan \pa{ \beta (U-\Psi/\sqrt{3})} } \; .
    \end{split}
\end{equation}

\textbf{Penrose Limit} To take the Penrose limit, we set $U = u$, $V = \varepsilon^2 v$, $R = \varepsilon \rho$, and $\Psi = \varepsilon \psi$, and then take the scaling limit
\begin{equation}
    ds^2_{\textrm{pp}} = \lim_{\varepsilon\to 0} \varepsilon^{-2} ds^2 = 2 du dv + \beta^2 \rho^2 du^2 + d\rho^2 + \cot^2(\theta_L) \cos^2(\beta u) d\psi^2 \; .
\end{equation}
It is straightforward to verify that this metric satisfies the vacuum Einstein equation. It is a plane wave presented in a mix of Rosen and Brinkmann coordinates. The $\psi$ dependence is in Rosen form $ds_{\textrm{pp}}^2 \supset g_{\psi \psi}(u) d\psi^2$, while the $\rho$ dependence is in Brinkmann form $ds_{\textrm{pp}}^2 \supset A_{\rho \rho} \rho^2 du^2 + d\rho^2$. This occurs because we constructed adapted coordinates only in the $r=3M$ hypersurface, so we get a pp-wave in Rosen form for the 3-dimensional metric at fixed $r=3M$ ($\rho = 0$). On the other hand, the relation between $r$ and $R$ is determined by the pseudo-orthonormal frame, discussed in footnote\footref{fn:coframe}, which  gives the pp-wave in Brinkmann form.

\section{Monodromy and Highly-Damped QNMs} \label{App:monodromy}
In this appendix, we briefly review the monodromy methods that were used to derive the highly-damped Schwarzschild QNMs \cite{motl2003, Andersson_2004}. The radial   equation \eqref{eq:radial} in terms of the variable $r$ is
\begin{equation}
    \br{ \partial_r^2 + \frac{f'(r)}{f(r)} \partial_r + \frac{1}{f(r)^2} \left( \omega^2 - V_l(r)\right) } \psi_{\omega l}(r) = 0 \;.
\end{equation}
This a second-order ODE with a regular singular point at the horizon $r=2M$. Solving the indicial equation, we get the two linearly independent solutions near the horizon\footnote{These simply correspond to the non-analytic part of the asymptotic solutions $\psi(r) \sim e^{ \pm i \omega r_*}$ in the vicinity of the horizon. The subleading terms in the expansion that multiply \eqref{eq:indicial} are single-valued about $r=2M$.} 
\begin{equation}\label{eq:indicial}
    \psi(r) \sim (r-2M)^{\pm i \omega / 2\kappa}\;,  \qquad \textrm{as } r \to 2M \;.
\end{equation}
Since the horizon is a singular point of the ODE, solutions are multi-valued in its vicinity. In particular, if we analytically continue the solutions into the complex $r$ plane, we can ask how the solutions change under a counter-clockwise rotation around the horizon 
\begin{equation}  \label{eq:horizon_monodromy}
    (r-2M) \to e^{2\pi i} (r-2M) \;.
\end{equation}
Under this rotation the solutions \eqref{eq:indicial} pick up monodromy $ e^{\mp \frac{\pi \omega}{\kappa}}$. However, we must keep in mind that the full QNM wavefunction is actually of the form $\Psi\sim e^{-i\omega t}\psi(r)$, and $t$ is also a singular coordinate on the horizon ($t \to -\infty$ on the past horizon). 
 It is therefore more natural to consider time-dependence in terms of the outgoing coordinate $u_S$, which is finite and analytic around the past horizon. Expressed in terms of this variable
\begin{equation}\label{eq:uvst}
    \Psi(x^\mu) \propto e^{-i \omega t} \psi(r) = e^{-i \omega u_S} e^{-i \omega r_*} \psi(r) \equiv e^{-i \omega u_S} \phi(r) \;.
\end{equation}
This radial wavefunction $\phi(r)$ has the asymptotic behavior 
\begin{equation} \label{eq:solns_horizon}
    \phi(r) \sim 1, (r-2M)^{-i \omega /\kappa} \;,  \qquad \textrm{as } r \to 2M \;.
\end{equation}
The solutions \eqref{eq:solns_horizon} are chosen to diagonalize the monodromy matrix: they do not mix under \eqref{eq:horizon_monodromy}. In particular, if we consider the wavefunction that's purely infalling at the horizon so that
\begin{equation}
    \phi(r) \sim (r-2M)^{-i \omega /\kappa} \;,  \qquad \textrm{as } r \to 2M \;,
\end{equation}
then the solution picks up the monodromy  
\begin{equation}
    \phi(r) \to e^{2 \pi \omega/ \kappa}  \phi(r)\;
\end{equation}
under \eqref{eq:horizon_monodromy}. On the other hand, \cite{motl2003, Andersson_2004} compute the monodromy of $\phi(r)$ assuming the purely outgoing boundary condition at $r \to \infty$ as well as $\omega_I \to -\infty$, and they find that the monodromy of the solution should be $-3$, i.e.
\begin{equation} \label{eq:monodromy_app2}
    \phi(r) \to -3 \cdot \phi(r) \;.
\end{equation}
A highly-damped QNM wavefunction $\phi(r)$ will obey both boundary conditions, so its monodromy can be obtained both ways -- and since a globally well-defined solution should have a definite monodromy, this imposes the condition
\begin{equation}
    e^{2\pi \omega/\kappa}  = -3 \;,
\end{equation}
which implies \eqref{eq:highly_damped_qnm}.

An alternative way to think about \eqref{eq:uvst} is in terms of the analytically continued complex Schwarzschild metric. In the vicinity of the past horizon, the coordinate function $t$ is singular, but the complexified coordinate can be continued around the horizon \cite{Fidkowski_2004}. It is however multi-valued, just like the coordinate $r_*$ : it shifts $t \to t + i \frac{\beta}{2} = t + i \frac{\pi}{\kappa}$ when we go around the horizon in the complex $r$ plane, holding $u_S$ fixed.\footnote{Since $u_S = t-r_*$ is well-defined in the vicinity of the past horizon, the shift $r_* \to r_* + i \frac{\beta}{2}$ fixes the shift in $t$.} If we work with the monodromy of $\psi(r)$ rather than $\phi(r)$, then we must keep track of the additional non-analyticity due to the factor of $e^{-i \omega t}$, which compensates to produce the same monodromy of the full wavefunction
\begin{equation}
    \Psi(x^\mu) \to e^{2\pi \omega/\kappa} \Psi(x^\mu) \;, \qquad \textrm{as } (r-2M) \to e^{2\pi i} (r-2M) \;. 
\end{equation}
In other words, it is really the counterclockwise monodromy of the full wavefunction around $V=0$ with ${V= e^{\kappa v_S}\to e^{2\pi i} V}$ (at fixed $u_S$) that is fixed to $-3$ for the highly-damped QNMs.

\section{Rindler Wave Equation}\label{App:rindler}
In this section we review properties of solutions to the wave equation in the Rindler patch of 2d flat space. The metric in tortoise coordinates takes the form
\begin{equation} \label{eq:rind metric chi}
    ds^2 = e^{2 a \chi} \l( -d \eta^2 + d \chi^2 \r) \;.
\end{equation}
It is sometimes more useful to work with the metric  
\begin{equation} \label{eq:rind metric rho}
    ds^2 = -a^2 \rho^2 d \eta^2 + d\rho^2 \; , \qquad \qquad   \rho = a^{-1} e^{a\chi} \; .
\end{equation}
The relation to flat space lightcone coordinates is 
\begin{equation} \label{eq:rind LC coords}
    v = t_{\textrm{flat}}+x = \frac{1}{a} e^{a(\chi+\eta)} = \rho e^{a\eta} \;, \qquad u=t_{\textrm{flat}}-x = -\frac{1}{a} e^{a(\chi-\eta)} = -\rho e^{-a\eta}
\end{equation}
where $ds^2 = -du dv$. In these global coordinates the future horizon is located at $u=0$ and the past horizon is located at $v=0$.
In Rindler coordinates this corresponds to $\rho \to 0$ ($\chi \to -\infty$) with $\eta \to \pm \infty$.  Spatial infinity corresponds to $\rho \to \infty$ ($\chi \to \infty$) with $\eta$ fixed.

\subsection*{Massless Solutions}
In the $(\eta, \chi)$ coordinates, solutions to the massless wave equation $\Box \Psi = 0$ are Rindler plane waves 
\begin{equation}
    \Psi_\omega \propto e^{-i\omega(\eta\pm \chi)} \propto u^{i\omega /a},v^{-i \omega /a} \;.
\end{equation}
Since there is no potential for the massless field in either global coordinates or in Rindler coordinates, there is no scattering and hence no meaningful notion of a resonance. Indeed, since in global coordinates the wave equation implies that $\Psi$ separates into left and right-moving parts $\Psi(u,v)=f(u)+g(v)$, there is no smooth solution that is purely outgoing as $\chi \to \infty$ and purely infalling at the horizon $\chi \to -\infty$. For instance, a solution that is purely infalling at the horizon, given by $v^{-i \omega/a}$,  will also be purely incoming (rather than outgoing) at spatial infinity.

Nevertheless, the Matsubara overtones $\omega =\pm \, i n a $ do have special properties in this simplified problem. 
If we fix boundary conditions at $\chi \to \infty$ in the Rindler region (for instance, by connecting the near-horizon solution to the solution in the full black hole geometry), then the solution will generically be a superposition of incoming and outgoing waves at the horizon
\begin{equation}
    \Psi_\omega = (A/a) e^{-i\omega(\eta-\chi)} + (B/a) e^{-i \omega(\eta + \chi)} = A u^{i \omega/a} + B v^{-i\omega/a} \;.
\end{equation}
Generically, $A,B \neq 0$, and so the solution is not purely infalling at the horizon. However, if $\omega = -i n a$, for positive integer $n$, then at the horizon
\begin{equation} \label{eq:rindler_massless_bc}
    \Psi = A u^n + B v^{-n}  \xrightarrow{u \to 0} B v^{-n} \;.
\end{equation}
This is a mode that is smooth and purely decaying along the future horizon (as we evolve along the future horizon by $\partial_v$, the mode decays). It is however not a QNM since it is not outgoing at spatial infinity: there is no potential for the wave to scatter off of so it always retains the same orientation.\footnote{Solutions of this type are QNM's in AdS$_2$, which is conformally related to \eqref{eq:rind metric rho}. In that case one imposes Dirichlet boundary conditions at the conformal boundary at $\rho=1$,  which requires $B=-A$ \cite{DoubleCone}.} Indeed, it is not even clear that a solution of the form $e^{-\frac{n}{a}(\eta + \chi)}$ should be considered infalling since it is purely decaying with no propagating wavefronts.

\subsection*{Massive Solutions}
As described in section \ref{Sec:near-horizon wave eqn}, the near-horizon limit of the massless Schwarzschild wave equation  is the massive wave equation in the Rindler patch. Here we review the massive Rindler wave equation in more detail. 
The equation 
\begin{equation}
    \Box \Psi = m^2 \Psi 
\end{equation}
in the $(\eta, \rho)$ coordinates takes the form
\begin{equation}
    \l( -a^{-2} \partial_\eta^2 + \rho \partial_\rho + \rho^2 \partial_\rho^2 \r) \Psi = \rho^2 m^2 \Psi \;.
\end{equation}
Using the $\partial_\eta$ isometry, we can write $\Psi_\omega = e^{-i \omega \eta} \psi_\omega(\rho)$, which gives the radial equation
\begin{equation} \label{eq:rindler radial rho}
    \rho^2 \psi_\omega''(\rho) + \rho \psi_\omega'(\rho) + (\omega/a)^2 \psi_\omega(\rho) - \rho^2 m^2 \psi_\omega(\rho) =0 \;.
\end{equation}
The solutions are linear combinations of modified Bessel functions 
\begin{equation} \label{eq:massive_rindler_solns}
    \psi(\rho) \propto K_{i\omega/a} \l( m \rho \r) \;, I_{i\omega/a} \l( m \rho \r) \;
\end{equation}
which have the asymptotic behavior as $\rho \to \infty$
\begin{equation}
         K_{i \omega/a} \l(  m \rho \r) \sim \frac{e^{-m \rho}}{\sqrt{m \rho}} \;, \qquad \qquad
         I_{i \omega/a} \l(  m \rho \r) \sim \frac{e^{m \rho}}{\sqrt{m \rho}} \;.
\end{equation}
For non-integer order $\nu$ the modified Bessel function of the second kind is
\begin{equation}\label{eq:mBessel12}
    K_\nu(z)= \frac{\pi}{2}\frac{I_{-\nu}(z)-I_\nu(z)}{\sin \pi \nu}
\end{equation}
so an equivalent basis of solutions is $I_{\pm i\omega/a}(m\rho)$. As $z\to 0$ the modified Bessel functions of the first kind have the asymptotic behavior $I_\nu(z)\sim z^\nu$, so $I_{-i\omega/a}$ is infalling at the horizon and $I_{i\omega/a}$ is outgoing. 
If we require normalizability then $K_{i \omega/a}$ is the relevant solution. According to \eqref{eq:mBessel12} this will contain a superposition of infalling and outgoing waves at the Rindler horizon. $I_\nu$ has the Frobenius series
\begin{equation}
    I_\nu(z)=\left(\frac{z}{2}\right)^\nu\sum_{k=0}^\infty\frac{\left(\frac{z}{2}\right)^{2k}}{k!\Gamma(k+\nu+1)}
\end{equation}
so according to \eqref{eq:mBessel12} as $\rho \to 0$ for generic $\omega$, there is an asymptotic expansion
\begin{equation}\label{eq:Kasymp}
    K_{i \omega/a} (m \rho) \propto (m/2)^{-i \omega/a} \Gamma (i\omega/a) \rho^{-i \omega /a}+(m/2)^{i \omega/a} \Gamma (-i\omega/a) \rho^{i \omega /a}   \;.
\end{equation}

\subsubsection*{Integer Order}

For integer order $I_n(z)=I_{-n}(z)$ and $I_n(z)$ becomes an entire function with the asymptotic $I_n(z)\sim z^{|n|}$ as $z\to 0$. In this case the second independent solution is defined by the limit $\nu \to n$ of the expression $\eqref{eq:mBessel12}$. However one cannot naively apply the asymptotic \eqref{eq:Kasymp} in this case. 

It is occasionally stated in the literature that \eqref{eq:Kasymp} leads to the S-matrix
\begin{equation}
    S(\omega) = (m/2)^{-2i \omega /a} \frac{\Gamma(i \omega /a)}{\Gamma(-i \omega /a)}
\end{equation}
with poles at $\omega = i n a$ for positive integer $n$, and that this therefore signals the presence of quasinormal modes. This identification is the typical way to read off the spectrum of resonances through poles of a scattering amplitude, and the thermal interpretation of Rindler space (and its relation to the black hole problem)  make it natural to look for QNMs in this ODE. Naively when the frequency approaches $\omega = i na$ the coefficient of the infalling component of \eqref{eq:Kasymp} dominates the coefficient of the outgoing component and the wave appears to satisfy QNM boundary conditions. However, if we consider the solution at precisely $\omega = i na$, then the two roots of the indicial equation for the expansion about the horizon differ by an integer and \eqref{eq:Kasymp} does not correctly represent the asymptotics. Instead, the two linearly independent solutions are $I_{|n|}(z)$ and
\begin{equation}
    K_n(z)= (-1)^{n+1}\log \left(\frac{z}{2}\right)I_{|n|}(z) + \frac12 \left(\frac{z}{2}\right)^{-n} + \text{less singular}
\end{equation}
These modified Bessel functions at integer order behave hear the horizon as
\begin{equation} \label{eq:Bessel_integer}
    \begin{split}
        &I_{\pm |n|} (m \rho) \sim (m \rho)^{|n|} \;, \\
        &K_{\pm |n|} (m \rho) \sim (m \rho)^{-|n|} + (m \rho)^{|n|} \log(m \rho) \;.
    \end{split}
\end{equation}
So even though the infalling coefficient in \eqref{eq:Kasymp} becomes large as $\omega\to ina$, $K_{i\omega/a}(m\rho)$ is not given by a purely infalling mode at the horizon in this limit. 
The massive Rindler wave equation does not admit QNMs: there are no solutions that are exponentially decaying as $\rho \to \infty$ and that are smooth, non-vanishing and infalling at the future horizon. 

However, the pure Matsubara frequencies are not the ones which are relevant to the highly damped QNM problem in Schwarzschild. Importantly, the spectrum \eqref{eq:highly_damped_qnm} is equally spaced in the Matsubara frequency, but it includes a small subleading part
\begin{equation}
    \omega_0= \frac{\kappa}{2\pi}\log 3 -i\frac{\kappa}{2}
\end{equation}
which prevents the logarithmic branching \eqref{eq:Bessel_integer}.
It is also not clear that normalizability (exponential decay), which singles out $K_{i\omega/a}$, is the correct boundary condition for the problem since the full solution connects to a wavefunction on the other side of the potential barrier. 
In this case the massive Rindler wave equation is simply a tool to understand the near-horizon behavior of massless waves on the full black hole geometry, and since we are not considering solutions with $\omega=ian$ the  $I_{\pm i \omega/a}$ are linearly independent. Generic waves in the near-horizon region behave as  
\begin{equation} \label{eq:rindler_massive_bc}
    \psi_\omega(\rho) = A(\omega) I_{-i\omega/a}(m \rho) + B(\omega) I_{i \omega/a}(m\rho) \; ,
\end{equation}
where $A(\omega)$ and $B(\omega)$ are fixed by matching the near-horizon solution to the solution in the far region.  When $A,B$ are nonzero the wave has infalling and outgoing components near the horizon
\begin{equation} \label{eq:rindler massive horizon}
    \psi_\omega(\rho) \propto A(\omega) (m \rho)^{-i \omega /a} + B(\omega) (m \rho)^{i \omega /a} \;, \qquad \rho \to 0 \; .
\end{equation} 
If the solution is the near-horizon limit of a QNM wavefunction it is necessary that $B=0$ so  
\begin{equation}
    \psi_\omega(\rho) \propto (m \rho)^{- i\omega /a} \;, \qquad \rho \to 0 \;.
\end{equation}
Including the time dependence gives
\begin{equation}
    e^{-i \omega \eta} \psi_\omega(\rho) \propto v^{-i \omega /a} \;, \qquad u \to 0 \; .
\end{equation}
$I_{-i\omega/a}(m\rho)$ grows exponentially as $\rho \to \infty$, but it can connect to a regular solution in the full black hole geometry that is purely outgoing at spatial infinity.
This second boundary condition at the asymptotically flat end selects a discrete set of frequencies but is input for the near-horizon analysis. As described in section \ref{Sec:near-horizon wave eqn}, given a seed solution corresponding to a highly damped QNM, the emergent symmetries of Rindler space generate the family of overtones \eqref{eq:highly_damped_qnm}.

\subsection*{Geometric Optics in Rindler Space}

Massive Rindler geodesics satisfy
\begin{equation}
    \frac{d\rho}{d\lambda} =\pm \sqrt{\frac{E^2}{a^2 \rho^2} - m^2} \;, \qquad \frac{d\eta}{d\lambda} = \frac{E}{a^2 \rho^2} \; 
\end{equation}
so the Hamilton Jacobi function is  
\begin{equation}
    S = -E\eta \pm \int \sqrt{\frac{E^2}{a^2 \rho^2} - m^2} \, d\rho \; .
\end{equation}
The behavior of the Hamilton-Jacobi function near the horizon is therefore given by
\begin{equation} \label{eq:HJ rindler}
    S \sim -E \left( \eta \pm \frac{1}{a} \log(\rho) \right) \;, \qquad \textrm{as } \rho \to 0 \; .
\end{equation}
This reproduces the  asymptotic form of the Bessel functions in the limit \eqref{eq:Kasymp}. The behavior of the wavefunction near the horizon is independent of the Rindler mass because in that limit the geodesic momentum becomes independent of the mass $\frac{d\rho}{d\lambda}\sim \frac{E}{a\rho}$.

\bibliographystyle{JHEP}
\bibliography{refs}

\end{document}